\documentclass[aps,prd,amsmath,amsfonts,nofootinbib,preprintnumbers, superscriptaddress,eqsecnum,secnumarabic]{revtex4-1}

\usepackage{color}

\pdfoutput=1

\usepackage{graphicx}
\usepackage[caption=false]{subfig}
\AtBeginDocument{\usepackage{booktabs}}

\newcommand{\overbar}[1]{\mkern 1.5mu\overline{\mkern-1.5mu#1\mkern-1.5mu}\mkern 1.5mu}

\newcommand{\bea}{\begin{eqnarray}}
\newcommand{\eea}{\end{eqnarray}}

\newcommand{\eq}[1]{Eq.~\ref{#1}}

\begin{document} 

\preprint{PSI-PR-18-08, ZU-TH-28/18}

\title{Low- and high-energy phenomenology of a doubly charged scalar}

\author{Andreas Crivellin}
\affiliation{Paul Scherrer Institut, CH--5232 Villigen PSI, Switzerland}
\author{Margherita Ghezzi}
\affiliation{Paul Scherrer Institut, CH--5232 Villigen PSI, Switzerland}
\author{Luca Panizzi}
\affiliation{Dipartimento di Fisica, Universit\`a di Pisa and INFN, Sezione di Pisa, Largo Pontecorvo 3, I-56127, Pisa, Italy}
\affiliation{School of Physics and Astronomy, University of Southampton, Highfield, Southampton SO17 1BJ, UK}
\author{Giovanni Marco Pruna}
\affiliation{Laboratori Nazionali di Frascati, via E. Fermi 40, I--00044 Frascati, Italy}
\author{Adrian Signer}
\affiliation{Paul Scherrer Institut, CH--5232 Villigen PSI, Switzerland}
\affiliation{Physik-Institut, Universit\"at Z\"urich, Winterthurerstrasse 190, CH-8057 Z\"urich, Switzerland}


\begin{abstract}
\noindent
We explore the phenomenology of an $SU(2)$-singlet doubly charged scalar at the high and low energy frontier. Such a particle is predicted in different new physics models, like left-right symmetric models or the Zee--Babu model. Nonetheless, since its interactions with Standard Model (SM) leptons are gauge invariant, it can be consistently studied as a UV complete SM extension. Its signatures range from same-sign di-lepton pairs to flavour changing decays of charged leptons to muonium-antimuonium oscillations. In this article, we use a systematic effective-field-theory approach for studying the low-energy observables and comparing them consistently to collider bounds. For this purpose, experimental searches for doubly charged scalars at the Large Hadron Collider are reinterpreted, including large width effects, and projections for exclusion and discovery reaches in the high-luminosity phase are provided. The sensitivities of the future International Linear Collider and Compact Linear Collider for the doubly charged scalar are presented with focus on di-lepton final states and resonant production. Theoretically and phenomenologically motivated benchmark scenarios are considered showing the different impact of low- and high-energy observables. We find that future low- and high-energy experiments display strong complementarity in studying the parameter space of the model.

\vskip 2cm
\tableofcontents
\end{abstract}

\maketitle

\section{Introduction}
\label{sec_1}
\noindent
Doubly charged scalars were initially proposed in the context of Left--Right models~\cite{Pati:1974yy,Mohapatra:1974gc,Senjanovic:1975rk} where they can be identified either with components of the associated bi-doublet of $SU(2)_{L}$ and $SU(2)_{R}$ or with elements of the $SU(2)_{L,R}$ triplets. It is well known that the doubly charged scalar embedded either in the $SU(2)_L$ triplet or in the $SU(2)_{L,R}$ bi-doublet does not allow for interactions with Standard Model (SM) fields that are gauge invariant on their own. Instead, if the doubly charged scalar is a component of the $SU(2)_R$ triplet it is always possible to introduce renormalisable interactions with the SM fermions of this single particle. Consequently, the addition of a $SU(2)_L$-singlet doubly charged scalar to the SM degrees of freedom represents an intriguing phenomenological option that can be considered both in the context of a minimal ultraviolet (UV) complete extension or as the low-energy limit of a more complicated UV complete theory after the decoupling of the other states.

In this paper, both options are considered and the phenomenology of a doubly charged scalar that couples to the right-handed charged leptons is explored in detail. Specific beyond-the-SM (BSM) extensions with doubly charged scalars have been studied in the literature before: low-energy observables~\cite{Konetschny:1977bn,Beall:1981ze,Mohapatra:1981pm,Halprin:1982wm}, neutrino mass generation in extended see-saw scenarios~\cite{Mohapatra:1979ia,Cheng:1980qt,Magg:1980ut,Schechter:1980gr}, and both collider~\cite{Azuelos:2004mwa,Nebot:2007bc,Chen:2007dc,Han:2007bk,Chen:2008qb,Akeroyd:2009nu,Akeroyd:2011zza,Akeroyd:2011ir,Sugiyama:2012yw,delAguila:2013yaa,Babu:2013ega,Chaudhuri:2013xoa,delAguila:2013hla,Alloul:2013raa,Chun:2013vma,Dutta:2014dba,Herrero-Garcia:2014hfa,Mitra:2016wpr,Dev:2018upe} and exotic signatures~\cite{Lazarides:1980nt,Mohapatra:1980qe} were considered. Moreover, scenarios motivated by the Zee--Babu mechanism for neutrino mass generation were also investigated~\cite{Zee:1985id,Ma:2006km,Babu:1988ki,Babu:2002uu,Gustafsson:2012vj,Gustafsson:2014vpa}. In our analysis, we adopt an even more comprehensive approach. Concerning the low-energy analysis, we exploit the aforementioned fact that the Yukawa insertion of a $SU(2)_L$-singlet doubly charged scalar is always renormalisable. Therefore, we match such a UV-complete theory on a low-energy effective field theory (EFT) where both the doubly charged scalar and the SM degrees of freedom are integrated out. For the high-energy aspects of our analysis, we study the full theory by additionally treating the width of the doubly charged scalar as a free parameter to account for large couplings or possible exotic decay modes.

The most general interactions of the doubly charged scalar are described by the Lagrangian
\begin{eqnarray}
\label{Lagrangian}
\mathcal{L}_{\mathrm{UV}} &=& \mathcal{L}_{\mathrm{SM}} +
\left(D_{\mu}S^{++}\right)^{\dagger}\left(D^{\mu}S^{++}\right)
 +\left(\lambda_{ab}\, \overline{\left(\ell_R\right)}_{\,a}^{\,c} \left(\ell_R\right)_{b}^{\phantom{c}}\, S^{++}
+ \mathrm{h.c.}\right)+\nonumber\\
&+&\lambda_2 \left(H^\dagger H\right)\left(S^{--} S^{++}\right)
+\lambda_4\left(S^{--} S^{++}\right)^2+\left[\dots\right],
\end{eqnarray}
where $a$ and $b$ are flavour indices and $\lambda_{ab}$ is a symmetric complex coupling matrix in the flavour space. This Lagrangian introduces 16 parameters: the mass of the doubly charged scalar $m_S$, six complex Yukawa parameters $\lambda _{ab}$, a coupling to the Higgs sector $\lambda_2$, the $\lambda_4$ quartic self-coupling and the $S$ width $\Gamma_S$. No specific assumption on the origin of $m_S$ is made, therefore $\lambda_2$ and $m_S$ are understood to be unconstrained by the electroweak-symmetry-breaking (EWSB) mechanism. Any form of new physics contributing to the value of $m_S$ and $\Gamma_S$ is intended to be represented by the ellipsis. 

The minimal SM extension introduced by~\eq{Lagrangian} breaks the lepton number by two units and explicitly violates charged lepton flavour. This implies ``smoking gun'' signatures such as lepton-flavour violating (LFV)~decays~\cite{Babu:2002uu,AristizabalSierra:2006gb,Akeroyd:2006bb,Calibbi:2017uvl,Dev:2018upe,Chen:2006vn} at low energy as well as same-sign lepton pairs appearing in high-energy collisions~\cite{King:2014uha,Geib:2015tvt,Dev:2018sel}. In this paper, the interplay of such signatures at different energy scales is examined 
using a systematic EFT approach improved by the renormalisation-group evolution (RGE) of its operators~\cite{Crivellin:2017rmk,Jenkins:2017dyc}. In addition, a detailed collider study is performed. On the one hand, phenomenological scenarios motivated by the anarchic pattern displayed by the Pontecorvo--Maki--Nakagawa--Sakata (PMNS) matrix are studied, hence assuming the couplings $\lambda_{ab}$ to the first and second generations to be the most sizeable ones. On the other hand, a cautious exploration of the phenomenology of $\tau$ final states is performed, and benchmark scenarios involving mainly couplings to the third generation are studied.

The scope of the Large Hadron Collider (LHC) to probe the resonant production of doubly charged scalars in same-sign leptonic final states is discussed in detail. In hadronic machines doubly charged scalars are dominantly produced through Drell-Yan processes where an off-shell photon or a $Z$ boson propagate in the s-channel, \emph{i.e.} $q\bar{q}\rightarrow \gamma^*(Z^*)\rightarrow S^{++}S^{--}$. However, photon-initiated sub-channels can also be important and give sizeable effects~\cite{Babu:2016rcr}, hence they are included in the analysis. Doubly charged scalars subsequently decay into pairs of same-sign leptons and, possibly, other exotic particles which can contribute to the width of the $S$. In this document, the narrow width approximation (NWA) and sizeable width effects are analysed through the recasting of a the CMS experiment that explores final states with same-sign leptons. Projections for the future high-luminosity (HL) stage of the LHC are also presented.

Concerning the impact of future linear colliders (LCs), such as the International Linear Collider (ILC)~\cite{Behnke:2013xla,Baer:2013cma} or the Compact Linear Collider (CLIC)~\cite{Aicheler:2012bya,Linssen:2012hp}, on doubly charged scalars searches, one should note that such machines are extremely sensitive to the exchange of an $S$ in the $t$-channel~\cite{Nomura:2017abh}. Moreover, if the mass of the doubly charged scalar is within the energy reach of the collider, on-shell production of a single $S$ associated with uncorrelated same-sign leptons is possible as well. In this paper, the capability of both sub-TeV and (multi-)TeV linear colliders to detect this particle is analysed in the light of several benchmark scenarios. Also in this case, sizeable width effects are considered and beam polarisation, initial-state radiation and beamstrahlung effects are taken into account.

The article is organised as follows: in Section~\ref{sec_2} we analyse the impact of doubly charged scalars on low-energy observables by means of a systematic EFT approach. In Section~\ref{sec_LHC} we study the current status of the model in \eq{Lagrangian} at the LHC and the prospects for searches at the HL phase. In Section~\ref{sec_ILC} we examine the scope of future LCs to probe this BSM particle in combination with current and future low-energy and collider constraints. In Section~\ref{sec_Pheno} we explore the phenomenology of several scenarios motivated by different underlying assumptions. In Section~\ref{sec_4} we present our conclusion.

\section{Low-energy phenomenology and allowed parameter space}
\label{sec_2}
\noindent
In this section we review the impact of the Lagrangian (\ref{Lagrangian}) on low-energy observables and discuss the resulting
limits on the couplings $\lambda_{ab}$. 

Doubly charged scalars contribute at tree level to three-body decays of charged leptons. The current limits on the branching ratios (BRs) for such decays are listed in the left column of Table~\ref{tab:exp}. The results of \cite{Amhis:2016xyh} for the $\tau$ decays are based on the measurements of the B-factories BELLE~\cite{Hayasaka:2010np} and BaBar~\cite{Lees:2010ez} but also on LHCb~\cite{Aaij:2014azz} and ATLAS results~\cite{Aad:2016wce}.

In addition, loop diagrams involving doubly charged scalars contribute to radiative lepton decays at the one-loop level. Furthermore, the QED RGE effects from the scale $m_S$ to experimental scale generate operators involving quarks which then contribute to $\mu\to e$ conversion in nuclei. The current limits of these processes are given in the right column of Table~\ref{tab:exp}.

\begin{table}[htbp!]
\begin{center}
\begin{minipage}{.45\textwidth}
\centering
\begin{tabular}{ll}
\toprule
\multicolumn{2}{c}{Three-body decays}\\
\midrule
${\rm{BR}}\left[{{\mu^\mp} \to {e^\mp}{e^\pm}{e^\mp}} \right]
\leq 1.0 \times10^{-12}$ & \cite{Bellgardt:1987du} \\[4pt]
${\rm{BR}}\left[ {{\tau^\mp} \to {\mu^\mp}{\mu^\pm }{\mu^\mp}}\right]
\leq 1.2\times 10^{-8}$ & \cite{Amhis:2016xyh} \\[4pt]
${\rm{BR}}\left[ {{\tau^\mp} \to {e^\mp}{e^\pm }{e^\mp}}\right]
\leq 1.4\times 10^{-8}$ & \cite{Amhis:2016xyh} \\[4pt]
${\rm{BR}}\left[{{\tau^\mp} \to {e^\mp}{\mu^\pm}{\mu^\mp}} \right]
\leq 1.6\times 10^{-8}$ & \cite{Amhis:2016xyh} \\[4pt]
${\rm{BR}}\left[{{\tau^\mp} \to {\mu^\mp}{e^\pm}{\mu^\mp}} \right]
\leq 9.8\times 10^{-9}$ & \cite{Amhis:2016xyh} \\[4pt]
${\rm{BR}}\left[{{\tau^\mp} \to {\mu^\mp}{e^\pm }{e^\mp}} \right]
\leq 1.1\times 10^{-8}$ & \cite{Amhis:2016xyh} \\[4pt]
${\rm{BR}}\left[{{\tau^\mp} \to {e^\mp}{\mu^\pm}{e^\mp}} \right]
\leq 8.4\times 10^{-9}$ & \cite{Amhis:2016xyh} \\
\bottomrule
\end{tabular} 
\end{minipage}
\begin{minipage}{.45\textwidth}
\centering
\begin{tabular}{ll}
\toprule
\multicolumn{2}{c}{Radiative decays}\\
\midrule
${\rm{BR}}\left[ \mu \to e\gamma \right] \leq 4.2\times 10^{-13}$
&\cite{TheMEG:2016wtm}\\[4pt]
${\rm{BR}}\left[ \tau \to \mu\gamma \right] \leq 4.4\times 10^{-8}$
&\cite{Aubert:2009ag}\\[4pt]
${\rm{BR}}\left[ \tau \to e\gamma \right] \leq 3.3\times 10^{-8}$
& \cite{Aubert:2009ag}\\
\bottomrule\\[5pt]
\toprule
\multicolumn{2}{c}{$\mu$-$e$ conversion}\\
\midrule
${\rm BR}_{\mu \to e}^{\rm Au} \leq 7 \times 10^{-13}$
& \cite{Bertl:2006up} \\
\bottomrule
\end{tabular}
\end{minipage}
\caption{Current experimental limits on charged LFV processes. \label{tab:exp}}
\end{center}
\end{table}

Lepton flavour violating hadronic tau decays like $\tau^\mp\to \mu^\mp P$ 
(where $P$ is a pseudoscalar meson), are not generated in our setup as these processes require
an axial coupling to quarks. Even though a doubly charged scalar can lead to decays like $\tau^\mp\to \mu^\mp K^+K^-$ and
$\tau^\mp\to \mu^\mp \pi^+\pi^-$ through the quark vector operator
which are generated via the RGE, we will not consider these strongly phase space suppressed 3-body
decays. Finally, the limits on $J/\psi \to \ell \ell'$ or  $Y \to \ell \ell'$ decays are much too weak to help in constraining the model due to the huge $J/\psi$ and $Y$ decay width.

Also muonium-antimuonium oscillations are generated at tree level~\cite{Chang:1989uk}. Here the current bound is~\cite{Willmann:1998gd}
\begin{equation}
{\cal P}(\overbar{M}-M)=8.3\times 10^{-11}/S_B
\label{exp:mmbar}
\end{equation}
where for our interactions consisting of right-handed currents~\cite{Hou:1995np,Horikawa:1995ae}, we
have for the correction factor $S_B=0.35$ (see table II of~\cite{Willmann:1998gd}).

Most processes mentioned above have excellent perspective for future experimental improvements. For $\mu\to 3e$~\cite{Blondel:2013ia,Berger:2014vba,Perrevoort:2018cqi} and $\mu \to e$ conversion in nuclei~\cite{Carey:2008zz,Cui:2009zz,Kutschke:2011ux} the sensitivity will be increased by several orders of magnitude. Also $\mu\to e \gamma$ will be improved by an order of magnitude~\cite{Baldini:2018nnn} and BELLE~II will improve on all $\tau$ decays by approximately one order of magnitude~\cite{Abe:2010gxa}.

The physical scale of the processes listed above is much below the electroweak symmetry breaking (EWSB) scale or $m_S$. Hence, they are best described by an effective theory valid below the EWSB scale. According to~\cite{Appelquist:1974tg}, a Lagrangian extended with dimension-six operators which are invariant under
$U(1)_{\rm QED} \times SU(3)_{\rm QCD}$ and contain the fermion fields
$f\in\{u,d,c,s,b,e,\mu,\tau\}$, as well as the QED and QCD gauge fields, is adopted to parameterise the interactions induced by the doubly charged scalar at the EWSB scale. Concretely, it reads
\begin{align}
\mathcal{L}_{\rm eff}&=\mathcal{L}_{\rm QED}
+ \mathcal{L}_{\rm QCD} +\frac{1}{m_S^2}\sum_{i}C_i Q_i, 
\label{Leff}
\end{align}
where the explicit form of those dimension-six operator that potentially induce charged LFV processes is presented in
Table~\ref{tab:app1}.
\begin{table}[htbp!]
\centering
\renewcommand{\arraystretch}{1.5}
\begin{tabular}{||c|c||c|c||} 
\hline \hline
\multicolumn{4}{||c||}{Dipole}\\
\hline
$Q_{e\gamma}$ & \multicolumn{3}{c||}{
$e m_{[pr]} (\bar l_p\sigma^{\mu\nu}P_Ll_r)F_{\mu\nu}+ {\rm H.c.}$}\\
\hline \hline
\multicolumn{2}{||c||}{Scalar/Tensorial} & 
\multicolumn{2}{c||}{Vectorial} 
\\
\hline
$Q_{S}$  & $(\bar l_p P_L l_r) (\bar l_s P_L l_t)+ {\rm H.c.}$&  
$Q_{VLL}$  &$(\bar l_p \gamma^\mu P_L l_r) (\bar l_s \gamma_\mu P_L l_t)$
\\
\  & \ &   
$Q_{VRL}$ & $(\bar l_p \gamma^\mu P_R l_r) (\bar l_s \gamma_\mu P_L l_t)$
\\
\  & \ &    
$Q_{VRR}$  & $(\bar l_p \gamma^\mu P_R l_r) (\bar l_s \gamma_\mu P_R l_t)$ 
\\
\hline
$Q_{Slq(1)}$  & $(\bar l_p P_L l_r) (\bar q_s P_L q_t)+ {\rm H.c.}$&  
$Q_{VlqLL}$  &$(\bar l_p \gamma^\mu P_L l_r) (\bar q_s \gamma_\mu P_L q_t)$
\\
$Q_{Slq(2)}$ & $(\bar l_p P_L l_r) (\bar q_s P_R q_t)+ {\rm H.c.}$ &   
$Q_{VlqLR}$ & $(\bar l_p \gamma^\mu P_L l_r) (\bar q_s \gamma_\mu P_R q_t)$
\\
$Q_{Tlq}$ & $(\bar l_p \sigma^{\mu \nu} P_L l_r) (\bar q_s \sigma_{\mu \nu} P_L q_t)+ {\rm H.c.}$ &    
$Q_{VlqRL}$  & $(\bar l_p \gamma^\mu P_R l_r) (\bar q_s \gamma_\mu P_L q_t)$
\\
\ & \ &    
$Q_{VlqRR}$  & $(\bar l_p \gamma^\mu P_R l_r) (\bar q_s \gamma_\mu P_R q_t)$
\\
\hline \hline
\end{tabular}
\caption{Dimension-six operators giving rise to effective leptonic
	transitions below the EWSB scale allowed by Lorentz and $U(1)_{\rm EM}$ gauge invariance. For our case, only the operators $Q_{e\gamma}$, $Q_{VRR}$, $Q_{VRL}$, $Q_{VlqRR}$ and $Q_{VlqRL}$ play a role. }
\label{tab:app1}
\end{table}
Here,  the indices $p$, $r$, $s$ and $t$ identify the flavour structure of the operator while $l$ and $q$ indicate lepton and quark fields, respectively. Furthermore, we introduce the notation $m_{[pr]} \equiv \max\{m_p,m_r\}$. The convention for the chirality projectors is fixed to $P_{L/R} = \left(\mathbb{I}\mp \gamma^5\right)/2$, and $F^{\mu\nu}$ is the
field-strength tensor of the photon. The sum in (\ref{Leff}) runs over all operators of Table~\ref{tab:app1} and over all family indices, even including equivalent terms multiple times.  In fact pure four-fermion leptonic operators with same chirality in the bilinear structures are invariant if the flavour indices of the bilinears are exchanged and this implies
some equalities among coefficients, \emph{i.e.} $C_{X}^{prst}=C_{X}^{stpr}$ with $X\in\{S,VLL,VRR\}$. Moreover, a further equality holds among coefficients of $Q_{VLL}$ and $Q_{VRR}$ operators due to Fierz relations: $C_{X}^{prst}=C_{X}^{ptsr}$ with
$X\in\{VLL,VRR\}$. In the following, these equalities are understood. Thus, the Lagrangian $\mathcal{L}_{\rm eff}$ contains
terms like $C_{VRR}^{1122} Q_{VRR}^{1122} + C_{VRR}^{2211} Q_{VRR}^{2211} +
C_{VRR}^{2112} Q_{VRR}^{2112} + C_{VRR}^{1221} Q_{VRR}^{1221} = 4\,
C_{VRR}^{1122} Q_{VRR}^{1122}$ and $C_{VRR}^{1112} Q_{VRR}^{1112} + C_{VRR}^{1211}
Q_{VRR}^{1211} = 2\, C_{VRR}^{1112} Q_{VRR}^{1112}$.

In order to link the UV-complete theory (\ref{Lagrangian}) to the EFT (\ref{Leff}) we perform the matching at the EWSB scale, implicitly assuming $m_W \sim m_S$. This matching produces the dipole operator and a four-fermion operator at the scale $m_W$.
\begin{figure}[htbp!]
\begin{center}
\subfloat[]{
\includegraphics[width=0.26\textwidth]{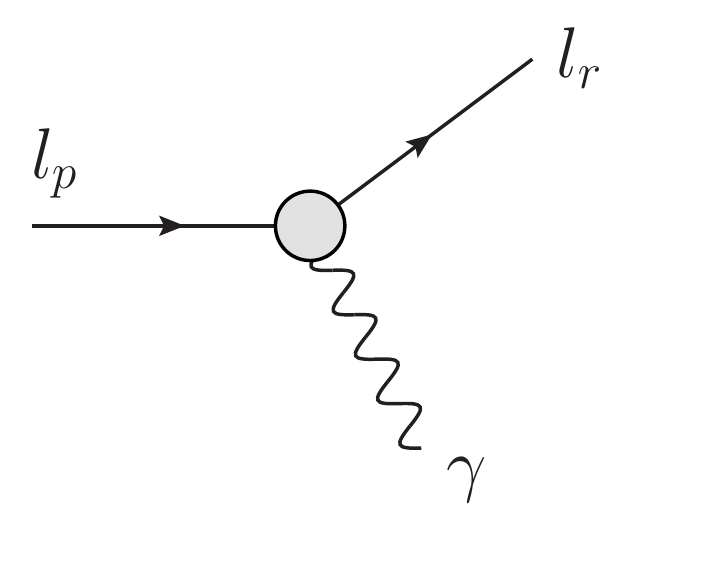}
\label{fig:feyna}
}
\hspace{0.1\textwidth}
\subfloat[]{
\includegraphics[width=0.26\textwidth]{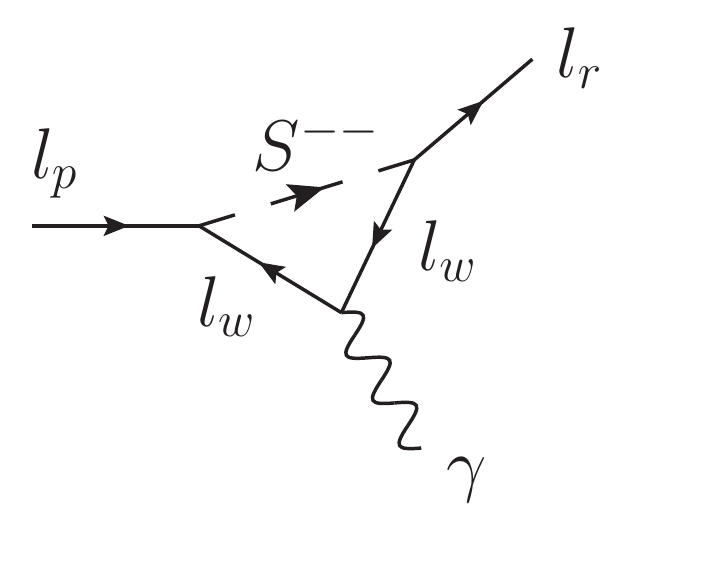}
\includegraphics[width=0.26\textwidth]{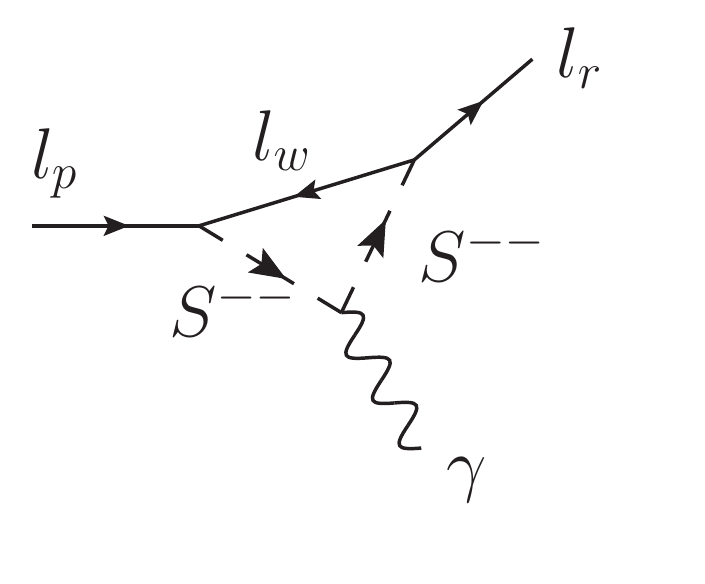}
\label{fig:feynb}
}\\
\subfloat[]{
\includegraphics[width=0.24\textwidth]{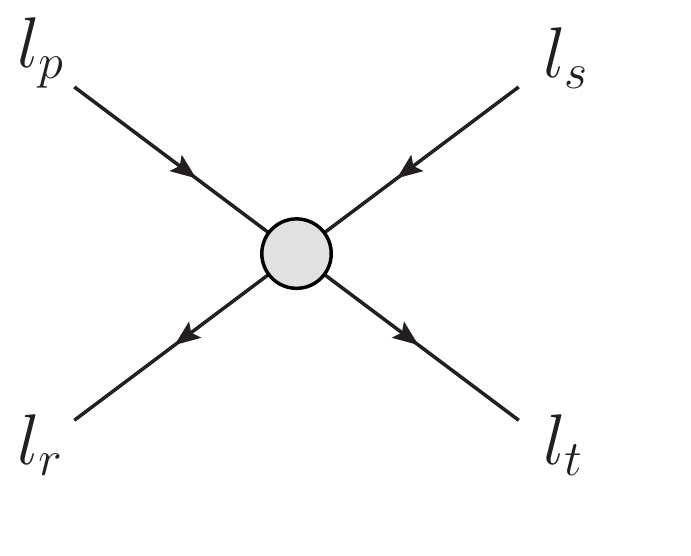}
\label{fig:feync}
}
\hspace{0.1\textwidth}
\subfloat[]{
\includegraphics[width=0.24\textwidth]{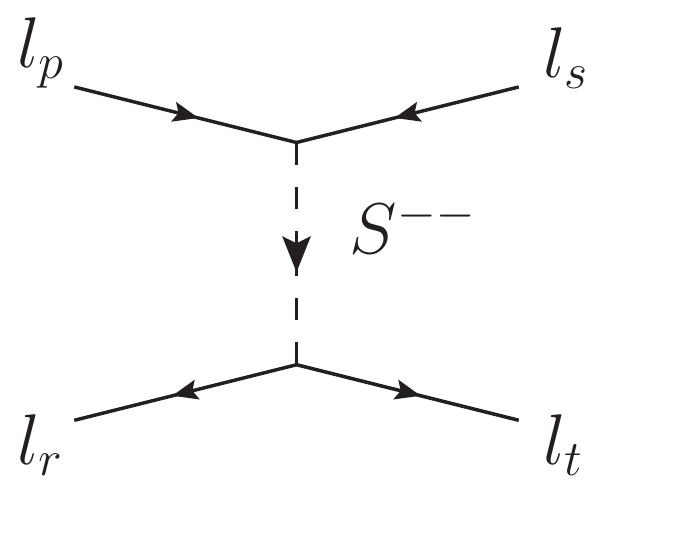}
\label{fig:feynd}
}
\end{center}
\caption{Feynman diagrams representing the UV-complete contributions
that match to the dipole and four-fermion operators. Diagrams in
Figure~\ref{fig:feynb} match into the diagram in Figure~\ref{fig:feyna}
(dipole interaction) and the diagram in  Figure~\ref{fig:feynd} matches
into the diagram in Figure~\ref{fig:feync} (contact interaction).}
\label{fig:feyn}
\end{figure}
First, as depicted in Figure~\ref{fig:feyn}, the hard part of the dipole interaction induced by the doubly charged scalar (at one loop) can be matched on the effective $Q_{e\gamma}$ operator using a straightforward application of the method of regions~\cite{Beneke:1997zp}. Second, the tree-level four-lepton interaction mediated by the doubly charged
scalar can be trivially matched on the corresponding contact interaction $Q_{VRR}$. No other Wilson coefficient of Table~\ref{tab:app1} is generated at the tree level in the UV-complete theory. In agreement with other studies of doubly charged scalars~\cite{Halprin:1982wm, Chang:1989uk, Raidal:1997hq, Cirigliano:2004mv, Akeroyd:2006bb, Giffels:2008ar, Akeroyd:2009nu, Fukuyama:2009xk}, the following matching at the EWSB scale is found:
\begin{eqnarray}
C_{VRR}^{prst}\left(m_W\right)&=&\frac{ \lambda_{rt}^{\phantom{*}}\lambda_{ps}^*}{2}, \\
C^{pr}_{e\gamma}\left(m_W\right)&=&\frac{1}{24 \pi^2} \frac{m_r}{m_{[pr]}}
\sum_{w=1}^3(\lambda_{rw}^{\phantom{*}}\lambda_{pw}^*).
\label{matchCey}
\end{eqnarray}

Now, we can use RGEs to determine the Wilson coefficients at the low scale~\cite{Crivellin:2017rmk} relevant for the processes. This evolution generates non-vanishing Wilson coefficients for the operators
\begin{align}
\{Q_{e\gamma}, Q_{VRR}, Q_{VRL}, Q_{VlqRR}, Q_{VlqRL}\}
\subset \mathcal{L}_{\rm eff}
\label{oplist}
\end{align}
As a final step, we express the BRs for the processes in terms of the Wilson coefficients given at the physical scale of the process. For the decay $l_p\to l_r\gamma$ we get
\begin{align}
\mathrm{BR}[l_p^+\to l_r^+\gamma] =
\frac{\alpha\, m_p^5}{m_S^4 \Gamma_p} \left(
\left|C_{e\gamma}^{rp}(m_p)\right|^2 +
\left|C_{e\gamma}^{pr}(m_p)\right|^2 \right)
\simeq \frac{\alpha\, m_p^5}{m_S^4 \Gamma_p}
\left|C_{e\gamma}^{rp}(m_p)\right|^2 ,
\label{brmeg}
\end{align}
where $\Gamma_p$ and $m_p$ are the decay width and mass of $l_p$, respectively. Note that the Wilson coefficient $C_{e\gamma}^{pr}$ in (\ref{brmeg}) is suppressed by $m_r/m_p\ll 1$ and will be neglected in what follows.  For the LFV decays of a lepton into three leptons the BRs can be written as
\begin{align}
&\mathrm{BR}[l_p^+\to l_r^+ l_s^- l_t^+] =
\frac{m_p^5}{s_{rt}\, 6\, m_S^4  \Gamma_p} \bigg[
\frac{1}{2 (4\pi)^3}  \Big(8 |C_{VRR}^{prst}|^2  + \frac{\delta_{st}}{2}|C_{VRL}^{prst}|^2 +
  \frac{\delta_{sr}}{2}|C_{VRL}^{ptsr}|^2\Big)   \nonumber \\
 & \qquad + \ \frac{\, \alpha^2}{\pi}   \left(\delta_{st}|C_{e\gamma}^{rp}|^2 +
   \delta_{sr}|C_{e\gamma}^{tp}|^2\right)   \Big(4 \, \log(\frac{m_p}{m_s})
   - 6 + \frac{1}{2}\, \delta_{sr}\, \delta_{st}\Big)   \nonumber \\
  & \qquad - \ \frac{\alpha}{8 \pi^2}\, \text{Re}\Big(
 \delta_{st}\, C_{e\gamma}^{rp}      (4\, C_{VRR}^{prst} + C_{VRL}^{prst})
  + \delta_{sr}\, C_{e\gamma}^{tp}      (4\, C_{VRR}^{prst} + C_{VRL}^{ptsr})      \Big) \bigg]\, ,
\label{brmu3e} 
\end{align}
where the symmetry factor is $s_{rt}=1+\delta_{rt}$ and we have only included the operators appearing in (\ref{oplist}). All Wilson coefficients in (\ref{brmu3e}) are to be evaluated at the scale $m_p$. In fact, the RGE effects for these decays can be very large. If $\lambda_{ps}$ and $\lambda_{rt}$ are suppressed with respect to the other couplings, the naive tree-level expressions are completely inadequate. Furthermore, in this case the dipole contribution and
interference terms can be numerically significant.

Turning to $\mu$-$e$ conversion in nuclei we can express the conversion rate normalised to the capture rate as 
\begin{align}
{\rm BR}_{\mu \to e}^{\rm N} =
\frac{m_\mu^5}{4\, m_S^4  \Gamma_\mathrm{capt}^N}\bigg|
e(m_\mu) C_{e\gamma}^{12}(m_\mu)\, D_N +
4 \Big( \tilde{C}^{(p)}_{VR}(m_\mu)\, V_N^{(p)} + p\to n\Big)\bigg|^2
\label{brconversion}
\end{align}
with
\begin{align}
\tilde{C}^{(p/n)}_{VR} = \sum_{q=u,d}\Big(
C_{VlqRR}^{12qq} + C_{VlqRL}^{12qq} \Big) f_{V\, p/n}^{(q)}
\end{align}
and $f_{V\, p}^{(u)}=2$, $f_{V\, n}^{(u)}=1$, $f_{V\, p}^{(d)}=1$, $f_{V\, n}^{(d)}=2$. The quantities $D_N$ and $V_N^{(p/n)}$ are related to overlap integrals~\cite{Czarnecki:1998iz} between the lepton wave functions and the nucleon densities. They depend on the nature of the target $N$ and for gold we use the numerical values~\cite{Kitano:2002mt}
\begin{align}
D_\mathrm{Au} &= 0.189 & V_\mathrm{Au}^{(p)} &= 0.0974 & V_\mathrm{Au}^{(n)} &= 0.146\,.
\end{align}
The capture rate $\Gamma_\mathrm{capt}^\mathrm{Au} = 8.7\times 10^{-15}\, \mathrm{MeV}$ is taken from~\cite{Suzuki:1987jf}.

In (\ref{brconversion}) we use the RGE of the effective Lagrangian $\mathcal{L}_{\rm eff}$, (\ref{Leff}), down to the scale
$m_\mu$. Strictly speaking, at a scale below $\mu_N\sim 1$~GeV, $\mathcal{L}_{\rm eff}$ is not suitable any longer to describe
processes involving hadrons and a matching to an effective Lagrangian with QCD bound states is required. However, because we are only dealing with QED corrections we use the perturbative RGE down to the scale $m_\mu$. Since the vector operator is protected by the Ward identity we expect that the QED effects are the dominant contribution due to the evolution from $\mu_N$ to the physical scale $m_\mu$.

The RGE effects are crucial for muon conversion in nuclei since the operators $C_{VlqRR}^{12qq}$ and $C_{VlqRL}^{12qq}$ are not generated through the matching at the EWSB scale. However, they are generated at the lower scale through the RGE. Hence, a meaningful description of this process hinges on the inclusion of RGE effects.

Finally, we turn to muonium-antimuonium oscillations. Expressing the oscillation probability ${\cal P}(\overbar{M}-M)$ through the effective Lagrangian we get~\cite{Feinberg:1961zza, Chang:1989uk}
\begin{align}
{\cal P}(\overbar{M}-M) =
\frac{ 72 (8\pi)^4 \, \alpha^6\, m_e^6}{m_S^4\, m_\mu^{10}\,G_F^4}
\, \Big|C_{VRR}^{2121}(m_\mu)\Big|^2 \, ,
\end{align}
where $G_F$ is the Fermi constant.

\begin{figure}[htbp!]
\begin{center}
\includegraphics[width=0.49\textwidth]{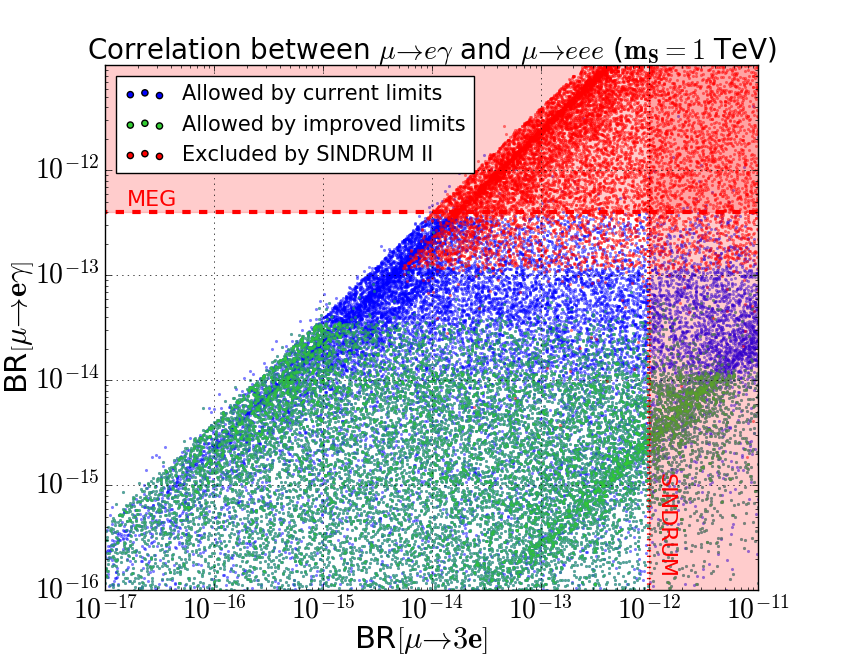}
\includegraphics[width=0.49\textwidth]{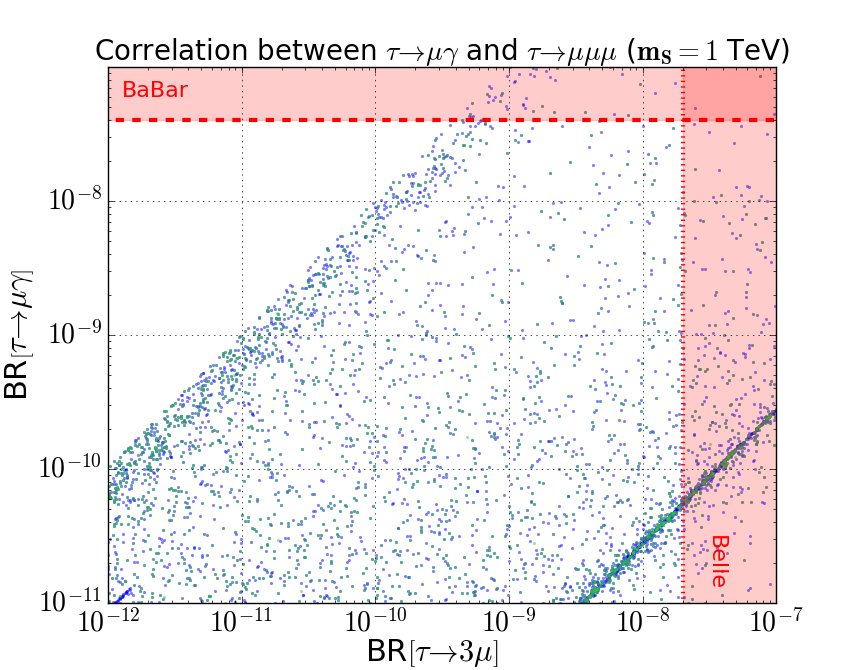}
\caption{
Correlations between BR$\left[\mu\to e\gamma\right]$(BR$\left[\tau\to\mu\gamma\right]$) and BR$\left[\mu\to3e\right]$(BR$\left[\tau\to3\mu\right]$) plotted in the left(right) panel.
The blue points are allowed by all other current experimental limits and the green points are still allowed in a future scenario where all bounds are improved by a factor of 10. Furthermore, the red points in the left panel are compatible with the current experiments with the exception of $\mu\to e$ conversion in nuclei. In both plots, a mass of $m_S=1$ TeV is considered and the $\lambda$-matrix is scanned over $100000$ random points with logarithmic scaling of the six-dimensional parameter space. Direct bounds on the observables are plotted in red-dashed lines. }
\label{phenoscatter}
\end{center}
\end{figure}

Before moving to the next section, we want to illustrate how the bounds resulting from the processes discussed in this section can be combined to analyse the parameter space of our model. In Figure~\ref{phenoscatter} we plot the correlations between BR$[\mu\to e\gamma]$ and BR$[\mu\to3e]$ (left panel) and between BR$[\tau\to\mu\gamma]$ and BR$[\tau\to3\mu]$ (right panel). The picture is that a band is populated by points, most densely at stripes at the edge, while there are only thinly scattered points outside this band. These stripes originate from \eq{brmeg} and \eq{brmu3e} when the interaction is mostly dipole (upper-left band) or 4-fermion contact-interaction (lower-right band) dominated. The scattered points outside the band correspond to points with fine-tuned cancellations between the dipole and the 4-fermion contributions.


\section{Bounds from LHC and projections for high-luminosity}
\label{sec_LHC}
\noindent
A comprehensive analysis of different production channels of doubly charged scalars at the LHC has been performed in \cite{King:2014uha}, where the cross sections for pair production through Drell--Yan (DY) processes, $Z$ boson fusion as well as single production of $S$ through $W$ boson fusion were computed for different values of the doubly charged scalar mass and for the $WWS$ coupling. A recasting of experimental searches at $7$ TeV was performed as well (see also \cite{Geib:2015tvt} for an extrapolated recasting at $13$ TeV using $7$ TeV data). 

This part of the analysis will consider the production of doubly charged scalars and has two purposes: 1) recast current limits of experimental analysis by including not only the DY topologies but also processes initiated by photons~\cite{Babu:2016rcr} which play a relevant role in the determination of the signal; 2) investigate the effect of the $S$ width ($\Gamma_S$) in the determination of the final state kinematics. We will limit our analysis to decays into leptons, including flavour-changing final states.

All the numerical results of this sections have been obtained at leading order using a dedicated model implemented in the {\tt UFO}~\cite{Degrande:2011ua} format; simulations have been performed within {\tt MadGraph5\_aMC@NLO}~\cite{Alwall:2014hca} considering the {\sc LUXqed17\_plus\_PDF4LHC15\_nnlo\_100} PDF set~\cite{Butterworth:2015oua,Manohar:2016nzj,Manohar:2017eqh}, which contains the photon contribution, with renormalisation and factorisation scales set to $2m_S$. {\tt PYTHIA\_v8}~\cite{Sjostrand:2014zea} has been used for parton showering and hadronisation, while the fast detector simulation has been run through {\tt Delphes\_v3}~\cite{deFavereau:2013fsa}. The recasting of experimental results has been obtained within the {\tt MadAnalysis5}~\cite{Conte:2012fm} framework.

\begin{figure}[htbp!]
\centering
\begin{tabular}{c|c|c}
\toprule
& narrow-width approximation & finite width \\
\midrule
quark-initiated & 
\centering\raisebox{-0.45\height}{\includegraphics[width=.24\textwidth]{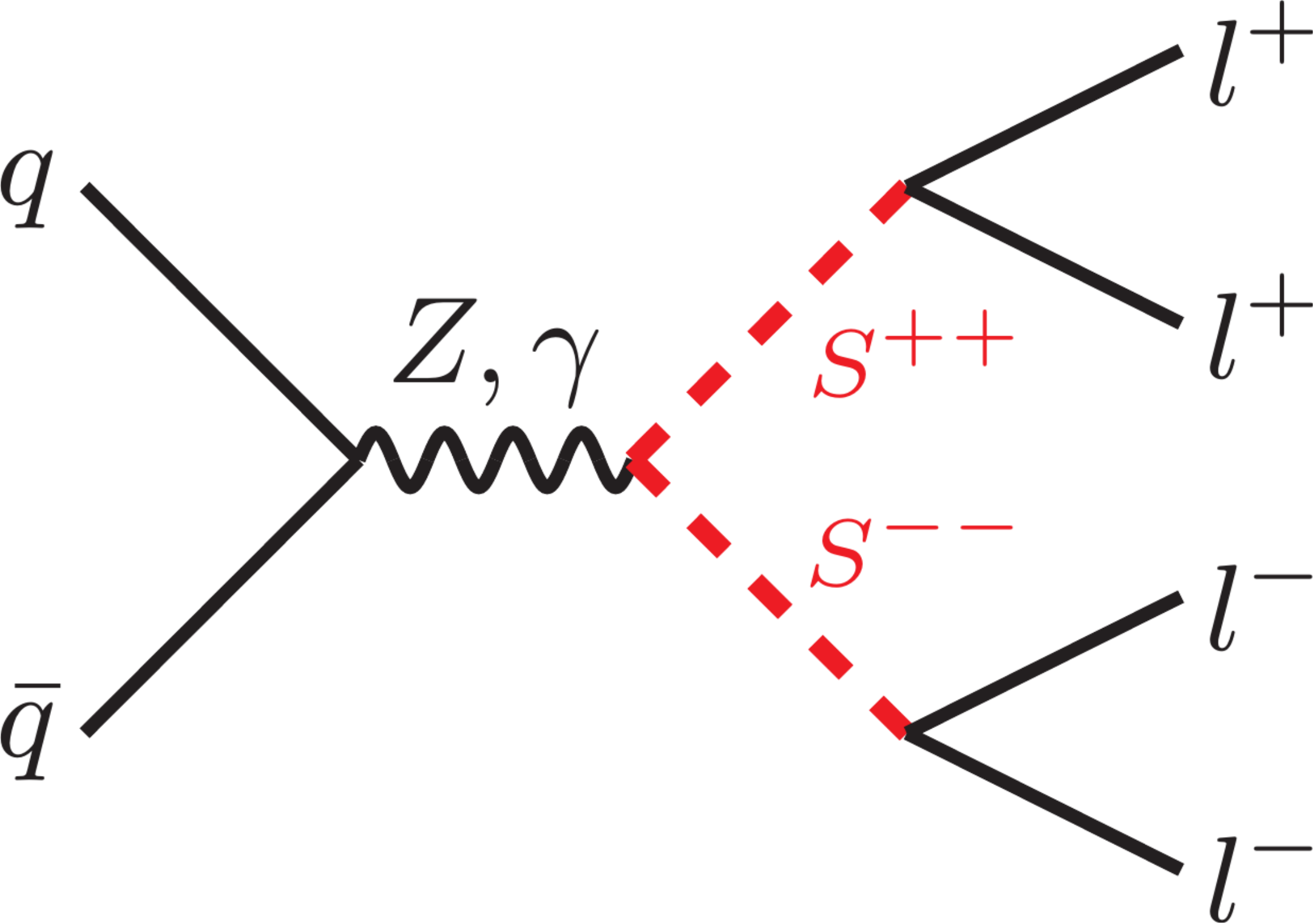}} & 
\raisebox{-0.45\height}{\includegraphics[width=.26\textwidth]{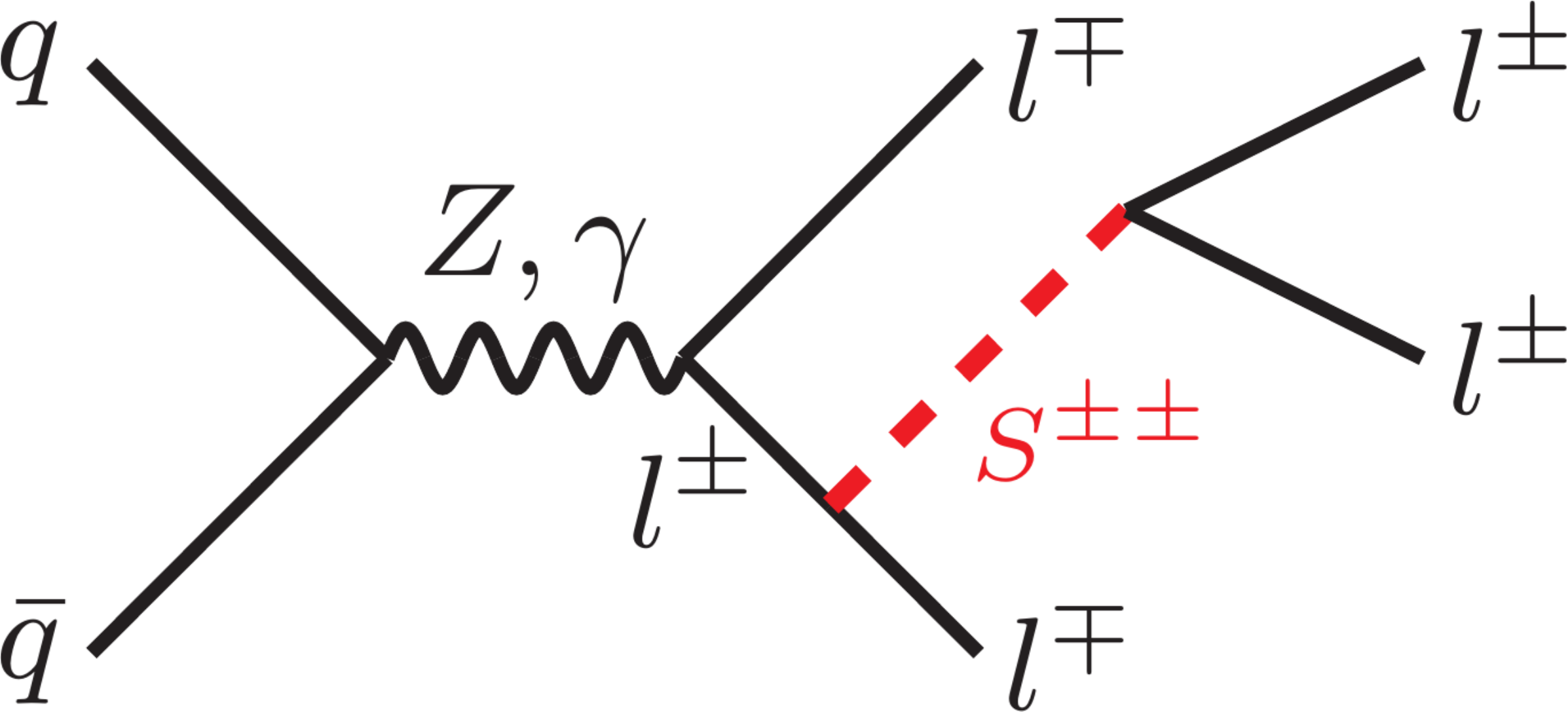}} \\[30pt]
\midrule
photon-initiated &
\centering\raisebox{-0.45\height}{\includegraphics[width=.22\textwidth]{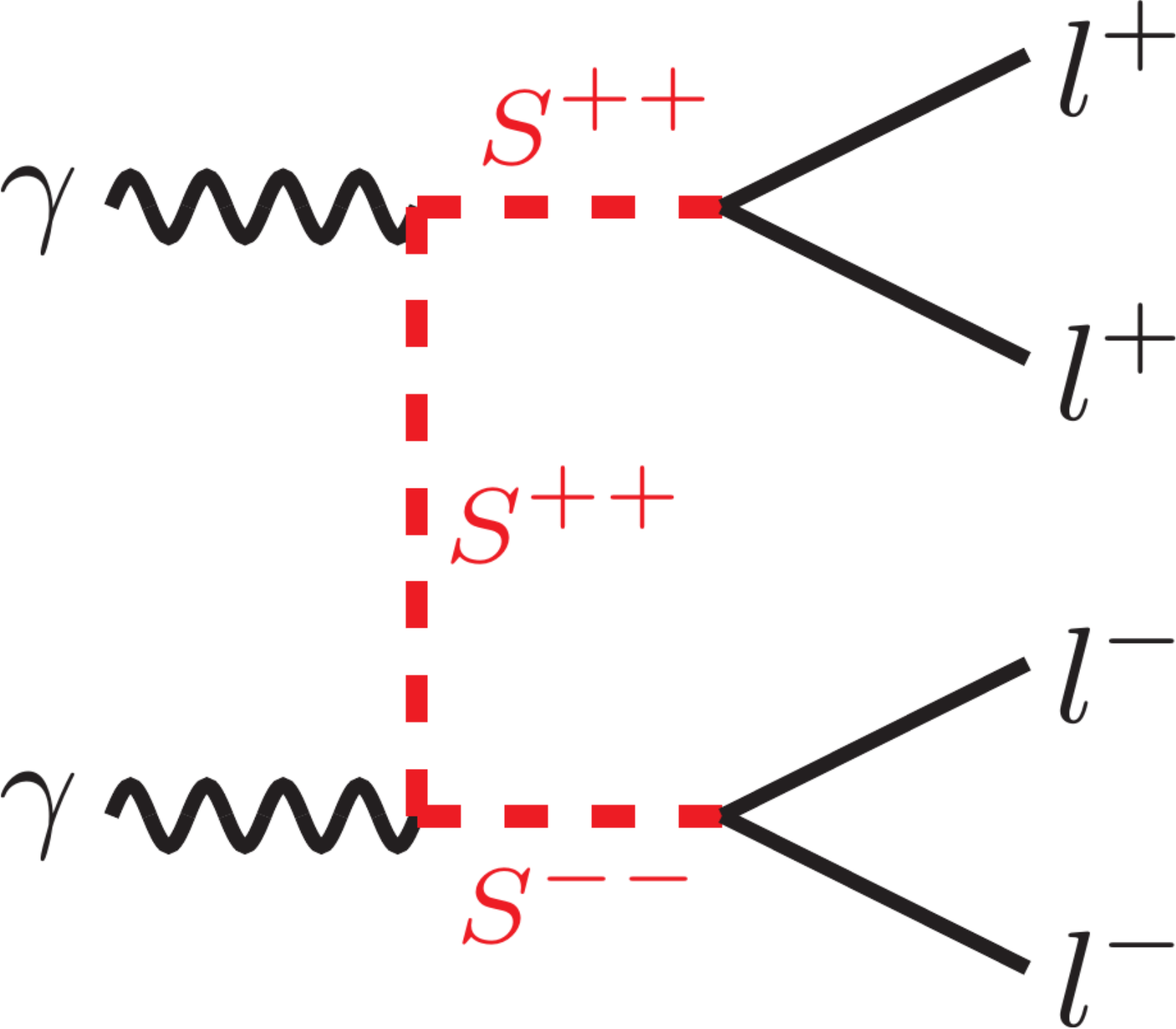}} \hskip 10pt
\centering\raisebox{-0.45\height}{\includegraphics[width=.22\textwidth]{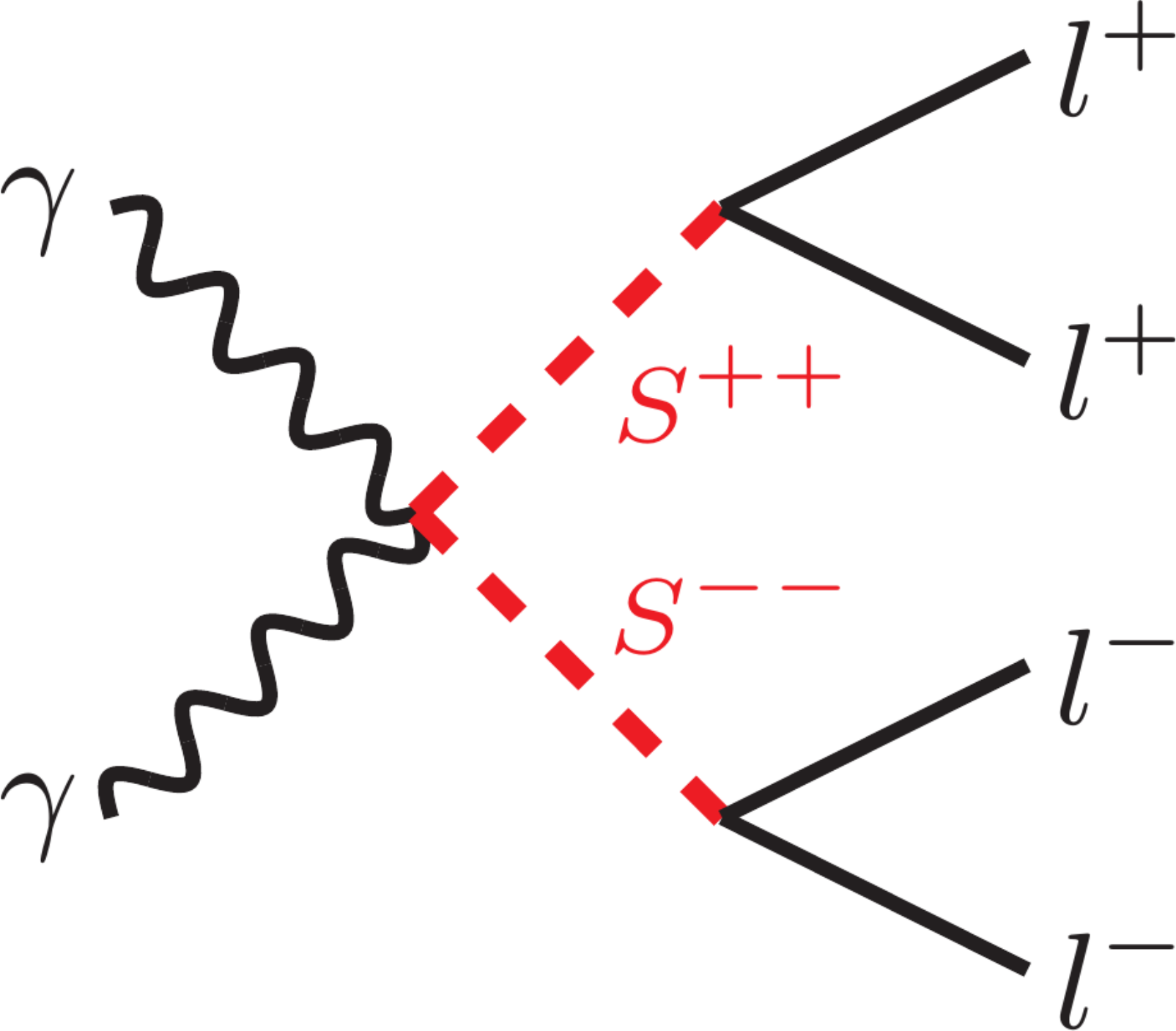}} & 
\raisebox{-0.45\height}{\includegraphics[width=.22\textwidth]{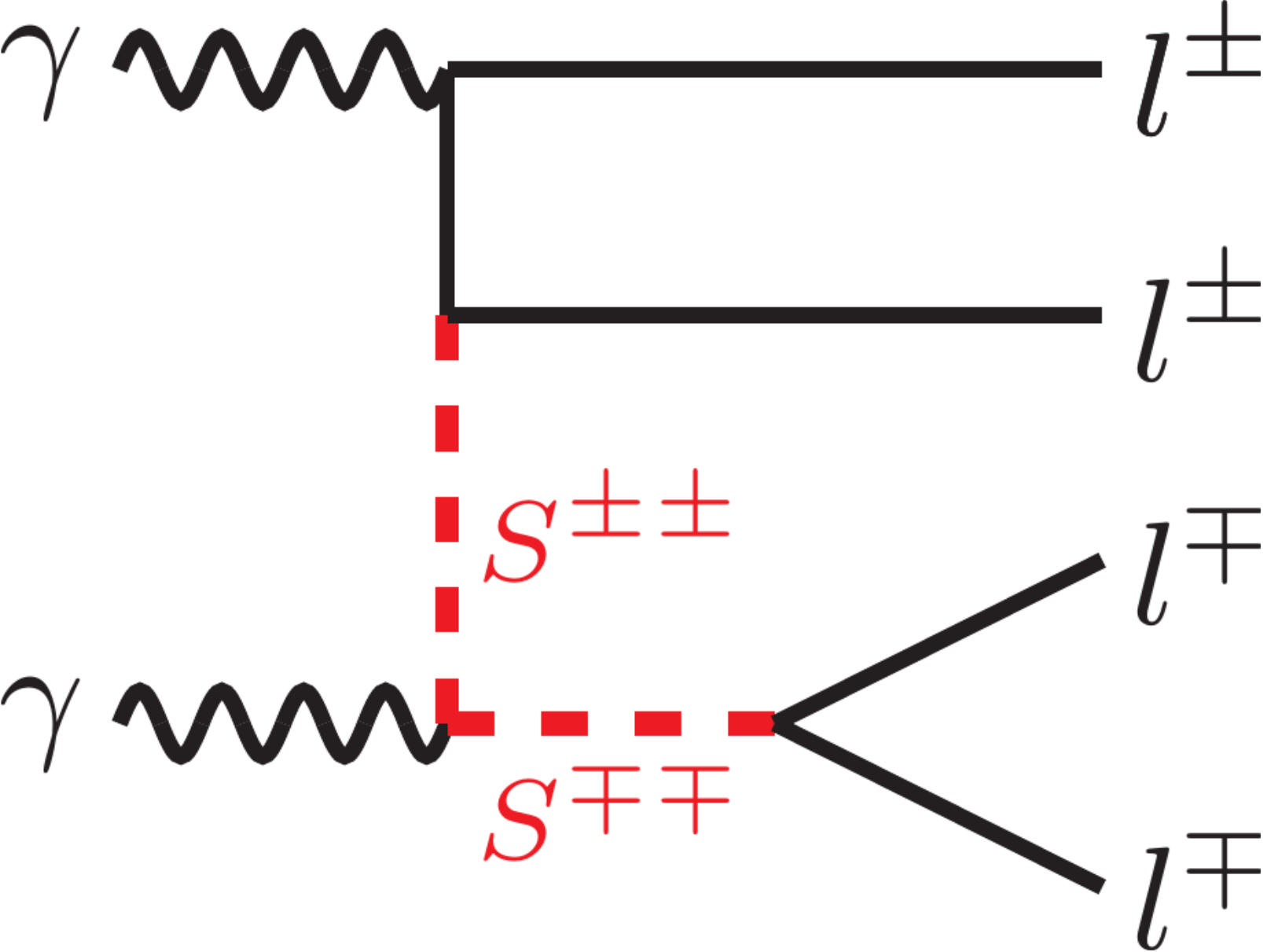}} \\[35pt]
\bottomrule
\end{tabular}
\caption{\label{fig:FWtopologies} Representative topologies for the process $pp\to2l^+2l^-$, $q\bar q$-initiated (\emph{i.e.} DY) and $\gamma\gamma$-initiated. The topologies in the last column are neglected in the NWA but can become relatively important if the $S$ width is large with respect to its mass.}
\end{figure}

Note that if $\Gamma_S$ is not narrow, it is not possible to factorise $S$ production and decay. Consequently, off-shell effects and topologies neglected by construction in the NWA, represented in the last column Figure~\ref{fig:FWtopologies}, can become relevant in scenarios where the $S$ has a finite width.

To evaluate the impact of a finite $S$ width on the determination of the cross section and at the same time ensure model-independence, \textit{the total width $\Gamma_S$ is considered as a free parameter}. The values of the $S$ couplings to SM leptons are then bounded from above by the fact that the sum of the corresponding partial widths must be smaller than $\Gamma_S$, for consistency. The partial width corresponding to a coupling $\lambda_{ab}$ and mass $m_S$ is given by
\begin{eqnarray}
 \Gamma_S^{\rm part}(m_S,\lambda_{ab},m_a,m_b) &=& \frac{\lambda_{ab}^2 (m_S^2 - m_a^2 - m_b^2)}{(1+\delta_{ab})16 \pi m_S} f^{\frac{1}{2}}(1,\frac{m_a}{m_S},\frac{m_b}{m_S})\stackrel{m_{a,b}\ll m_S}{\longrightarrow} \frac{\lambda_{ab}^2 m_S}{(1+\delta_{ab})16 \pi } \;, \qquad 
\end{eqnarray}
where $f^{\frac{1}{2}}(a,b,c)=\sqrt{a^4+b^4+c^4-2a^2b^2-2a^2c^2-2b^2c^2}$. The consistency requirement translates therefore into $\sum \Gamma_S^{\rm part}\leq \Gamma_S$.

In the context of a minimal extension of the SM where $S$ is the only new scalar and where the gauge sector of the SM is not modified, the coupling of $S$ to $Z$ boson is uniquely determined by the electric charge of $S$ and given by $g_{ZSS} = 2 \frac{g}{c_W} s_W^2$, since we assume that it is a singlet under $SU(2)_L$. Hence, \textit{$g_{ZSS}$ is not a free parameter in our analysis}. The relevance of this consideration comes from the fact that the $g_{ZSS}$ coupling only appears in a subset of the signal topologies leading to the four-lepton final state (namely, those in which the $S$ is produced through $Z$ boson exchange in the $s$-channel), and therefore determines the relative importance of such contributions with respect to those for which $g_{ZSS}$ does not appear, such as DY production via photon exchange, production initiated by photons, or radiation of $S$ from leptons, all represented in Figure~\ref{fig:FWtopologies}. While $g_{ZSS}$ is fixed by the $S$ representation, the coupling of $S$ to photon is always determined by its electric charge and therefore does not pose further issues.

\begin{figure}[htbp!]
\centering
\includegraphics[width=.48\textwidth]{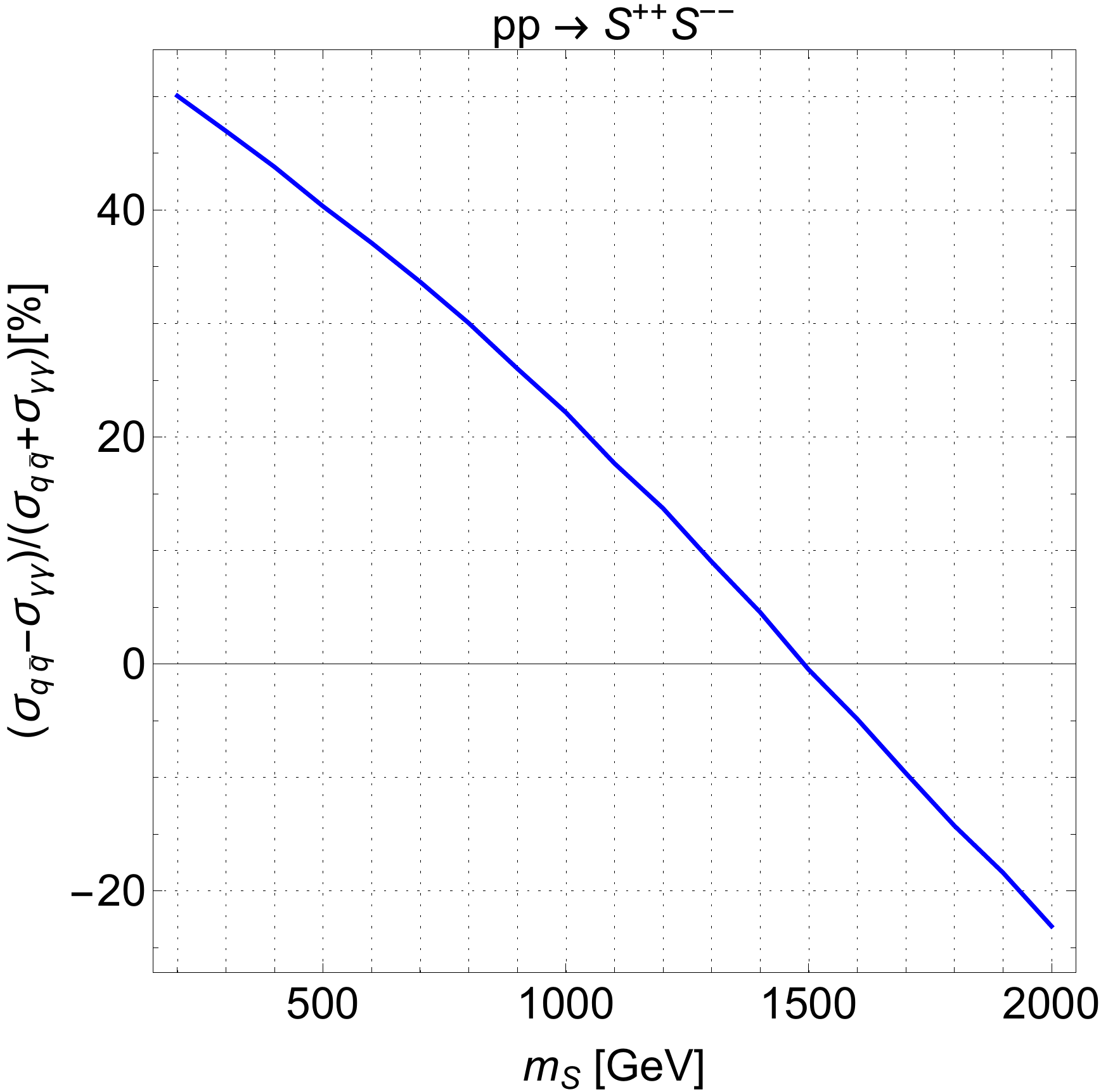}
\includegraphics[width=.48\textwidth]{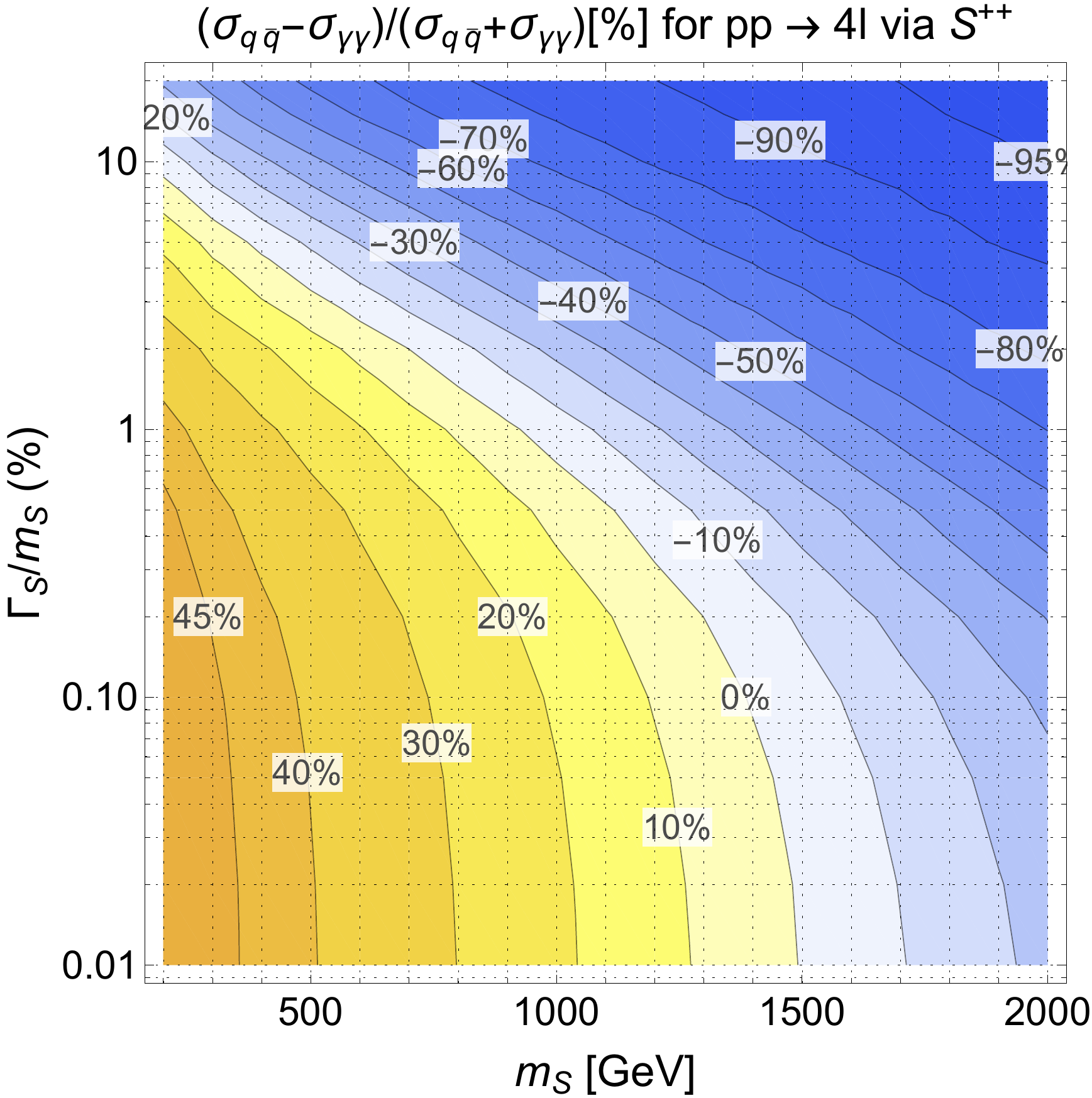}\hfill
\caption{\label{fig:QQvsGG} Relative weight of the $q\bar q$- and $\gamma\gamma$-initiated contributions over the total cross section. Left panel: $PP\to S^{++}S^{--}$; right panel: $PP\to 2l^+ 2l^-$ via propagation of $S$.}
\end{figure}

The difference of the weights of the $q \bar q$- and $\gamma\gamma$-initiated contributions in the determination of the total cross section, defined as $\eta=\left(\sigma_{q\bar q} - \sigma_{\gamma\gamma}\right)/\left(\sigma_{q\bar q} + \sigma_{\gamma\gamma}\right)$, depends only on $m_S$ and $\Gamma_S$ since (under the assumptions above) the Yukawa couplings can be factorised. The weights of the two processes relative to the total cross section can be then derived as $w_{q\bar q}=(1+\eta)/2$ and $w_{\gamma\gamma}=(1-\eta)/2$. In the left panel of Figure~\ref{fig:QQvsGG} the ratio $\eta$ is shown for the 2$\to$2 process $pp\to S^{++}S^{--}$, to emphasize the role of the different topologies in the NWA. The DY process gives a dominant contribution for low masses while the photon-initiated one dominates for large masses. The photon contribution becomes more relevant because the DY topologies require the s-channel propagation of $Z$ and $\gamma$, which receives larger suppression as the mass of $S$ increases; the $S$ pair production initiated by photons, on the other hand, involves the t-channel propagation of $S$ and a four-leg vertex (see Figure~\ref{fig:FWtopologies}), and the only suppression for large masses is due to the phase space. In the right panel of Figure~\ref{fig:QQvsGG} the same ratio is shown for the 2$\to$4 process $pp\to 2l^+2l^-$ containing at least one $S$ propagator, to determine the effects due to the $S$ width. In a region spanning from low mass and large width to high mass and small width, the two processes contribute equally to the total cross section. At any fixed mass of $S$, the photon contribution becomes more important for increasing values of the $S$ width. Given the large difference between $m_S$ and the mass of any SM lepton, this result is valid with excellent approximation for any 4-lepton final states generated via propagation of $S$. 

It is also important to numerically determine the relative importance of contributions which are usually neglected in the NWA. In the NWA, the cross section can be written as
\begin{equation}
\label{eq:NWA}
 \sigma_{pp\to l^+_a l^+_b l^-_c l^-_d}(m_S,{\Gamma_S\over m_S}\to0,\lambda_{ij})=\sigma_{pp\to S S}(m_S) {\rm BR}\left[S\to2l^+\right] {\rm BR}\left[S\to2l^-\right]\;,
\end{equation}
which is by construction independent of the width of $S$ and can be decomposed as before into two components corresponding to the quark- and photon-initiated topologies (which have not been explicitly written in \eqref{eq:NWA}).

\begin{figure}[htbp!]
\centering
\includegraphics[width=.33\textwidth]{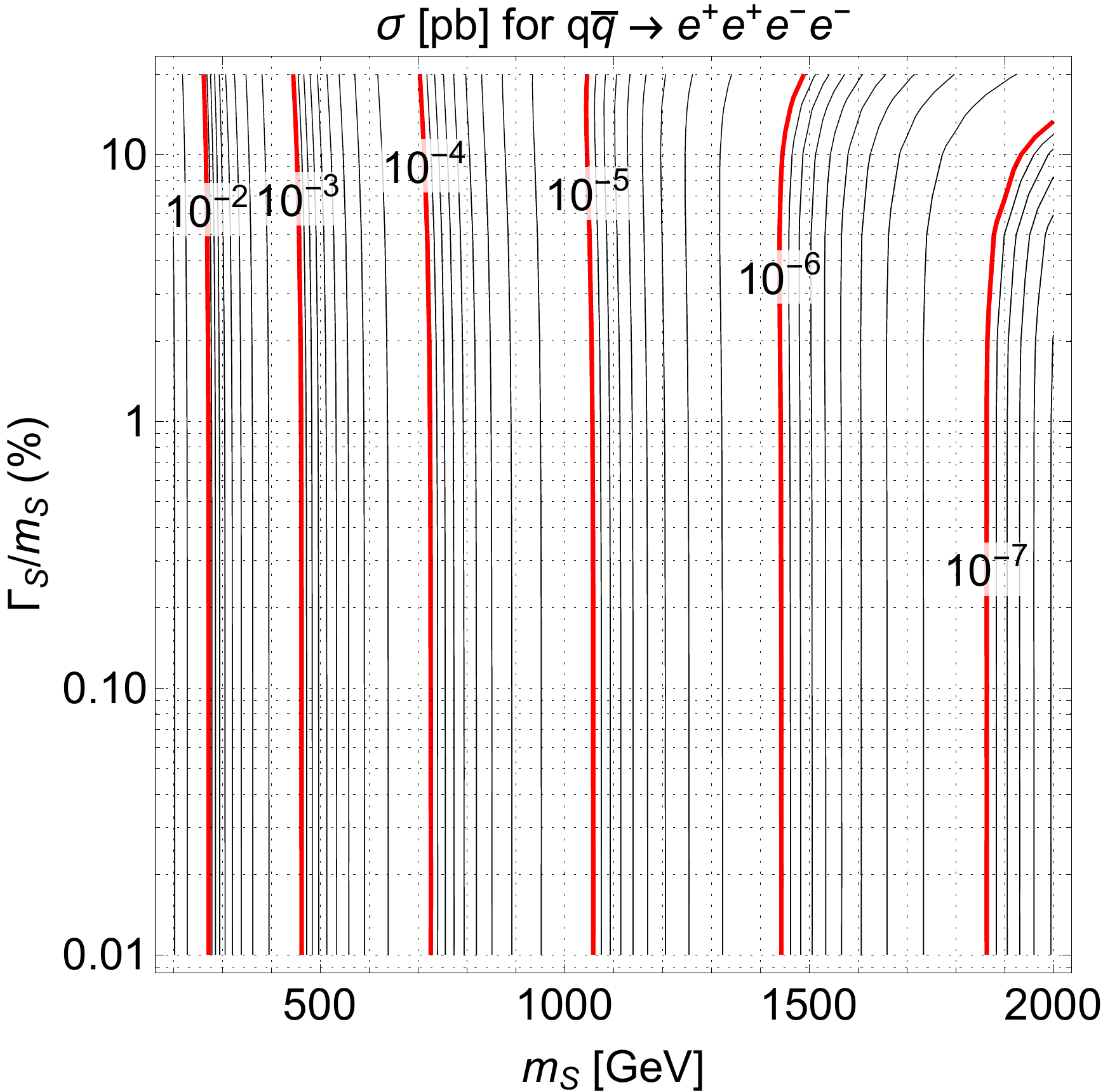}\hfill
\includegraphics[width=.33\textwidth]{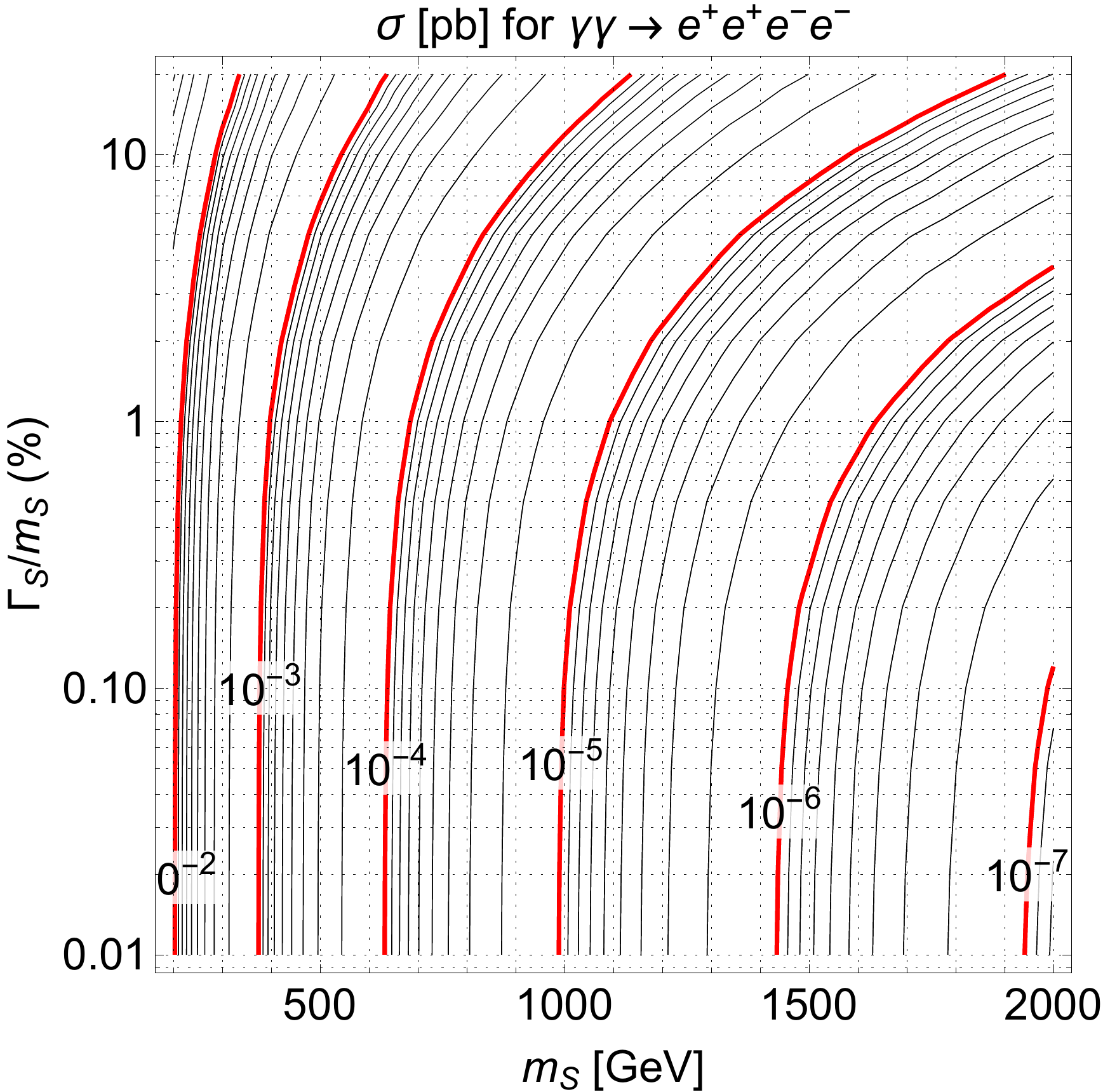}\hfill
\includegraphics[width=.33\textwidth]{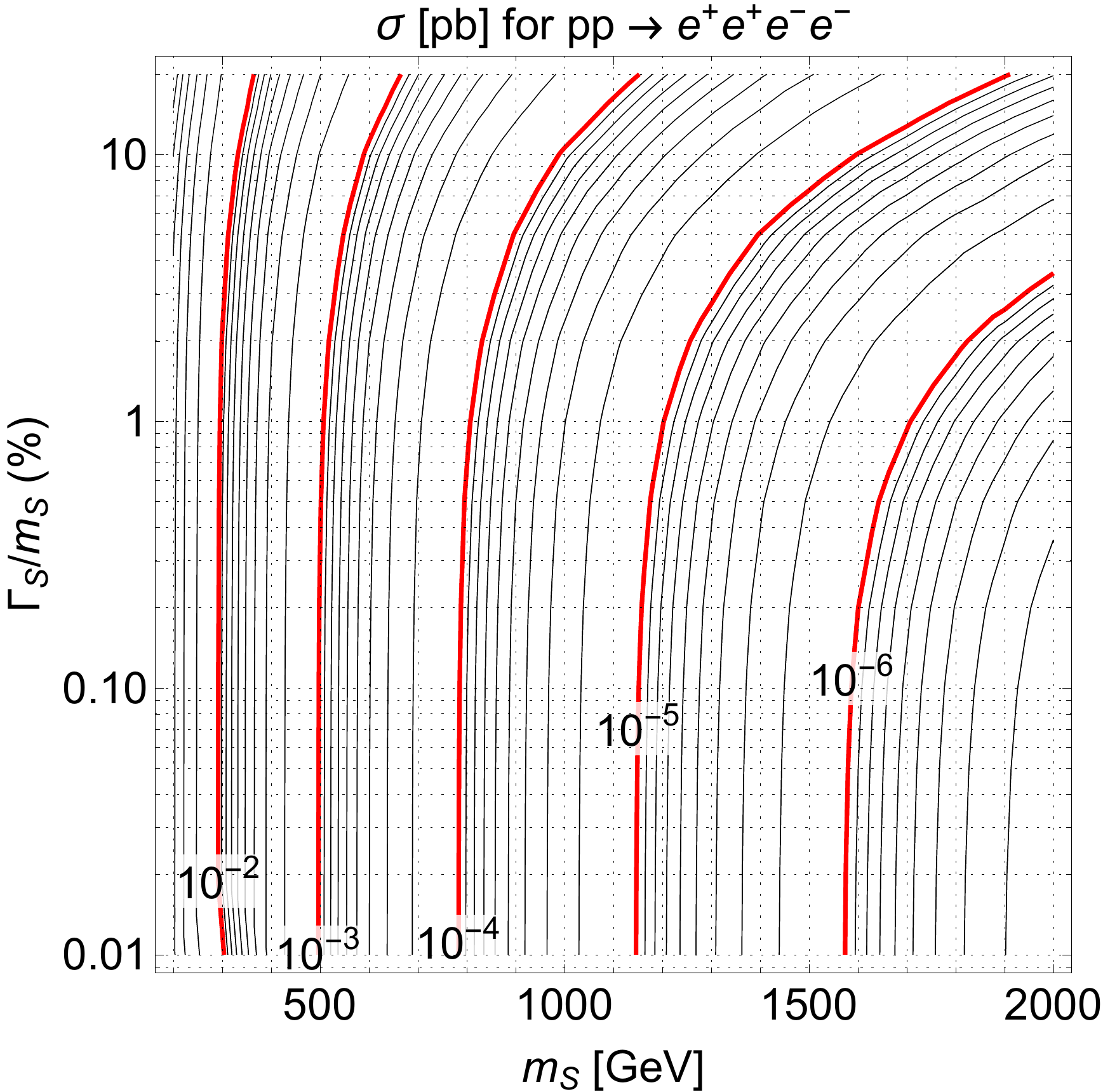}\\[5pt]
\includegraphics[width=.33\textwidth]{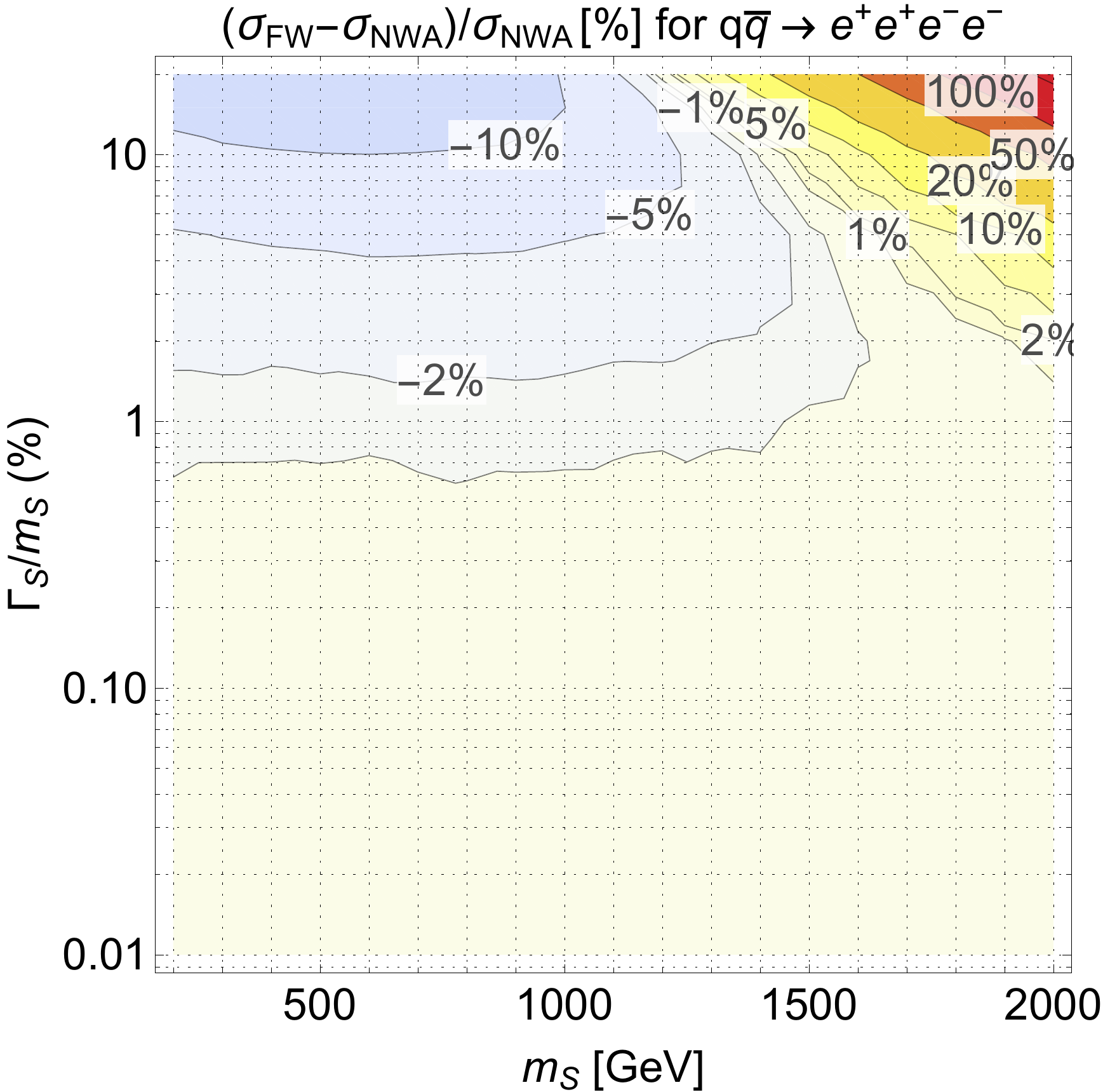}\hfill
\includegraphics[width=.33\textwidth]{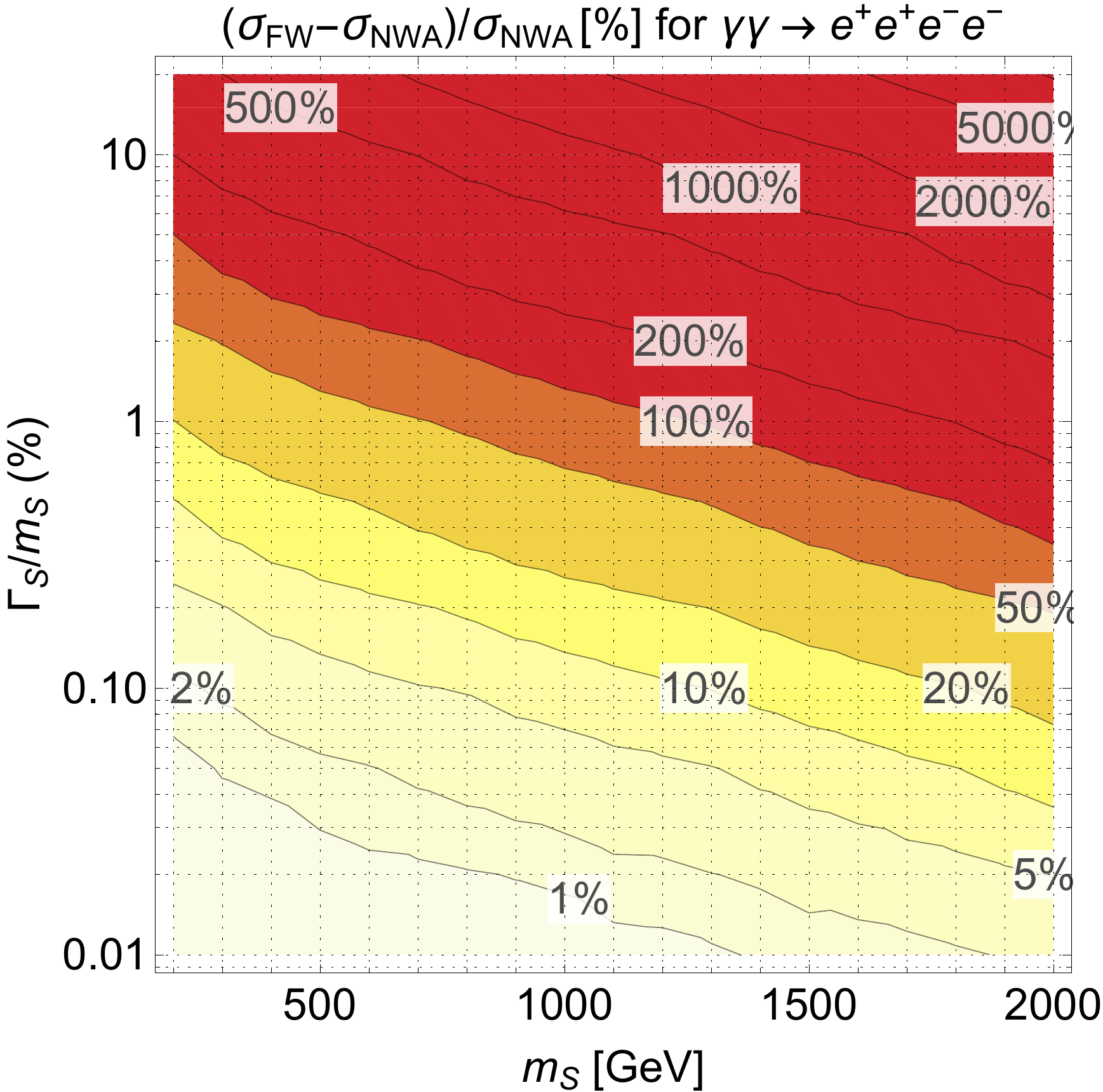}\hfill
\includegraphics[width=.33\textwidth]{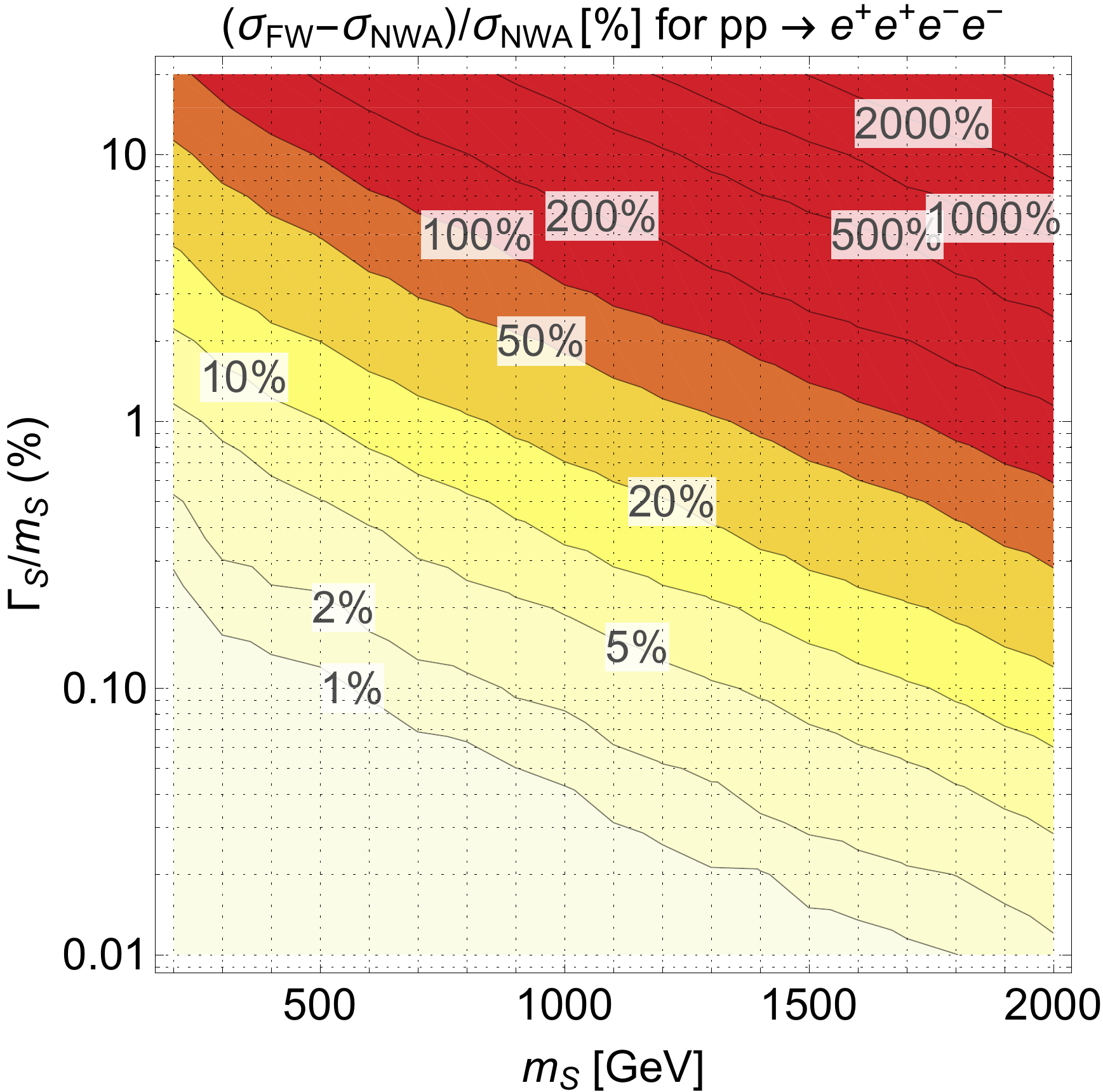}
\caption{\label{fig:FWvsNWA} Top row: cross section corresponding to saturating the Yukawa couplings to the maximum values associated to a given total width. Bottom row: relative ratio between cross sections in the large width regime and NWA. From left to right, here and in the following: quark-initiated process, photon-initiated process and total contribution.}
\end{figure}

Using this result, we can now compute the cross section corresponding to the maximum value of the coupling needed to obtain a given $\Gamma_S$, and compare it to the cross section in the NWA. Our results are reported in Figure~\ref{fig:FWvsNWA} for the $2e^+2e^-$ final state, as again, due to the large mass gap between $S$ and the SM leptons, all the other final states produce a qualitatively analogous result.

As expected, for relatively small values of the width (with respect to the mass), the relative differences are negligible, and the NWA can be used for the description of all processes. As the width increases, however, the relative differences become larger, though the dependence of the cross section on the $\Gamma_S/m_S$ ratio is much weaker for DY processes with respect to photon-initiated ones. Furthermore, in the DY processes, for values of $\Gamma_S / m_S$ above $\sim$1\% the relative difference is negative for $m_S\lesssim 1300$ GeV and positive for larger masses. Around $m_S\sim 1300$ GeV a cancellation between effects can be observed, due to a different scaling of the phase space and the PDFs with the transferred energy in the process depending on the width of the $S$. This effect has been observed and described in \cite{Moretti:2016gkr} for a different process. Of course, and analogously to what was found in ~\cite{Moretti:2016gkr}, for values of $m_S$ corresponding to a cancellation at the level of integrated cross section differences in the kinematics of the final state still appear at differential level and affects the efficiency of a specific set of experimental cuts. 

\begin{figure}[htbp!]
\centering
\includegraphics[width=.33\textwidth]{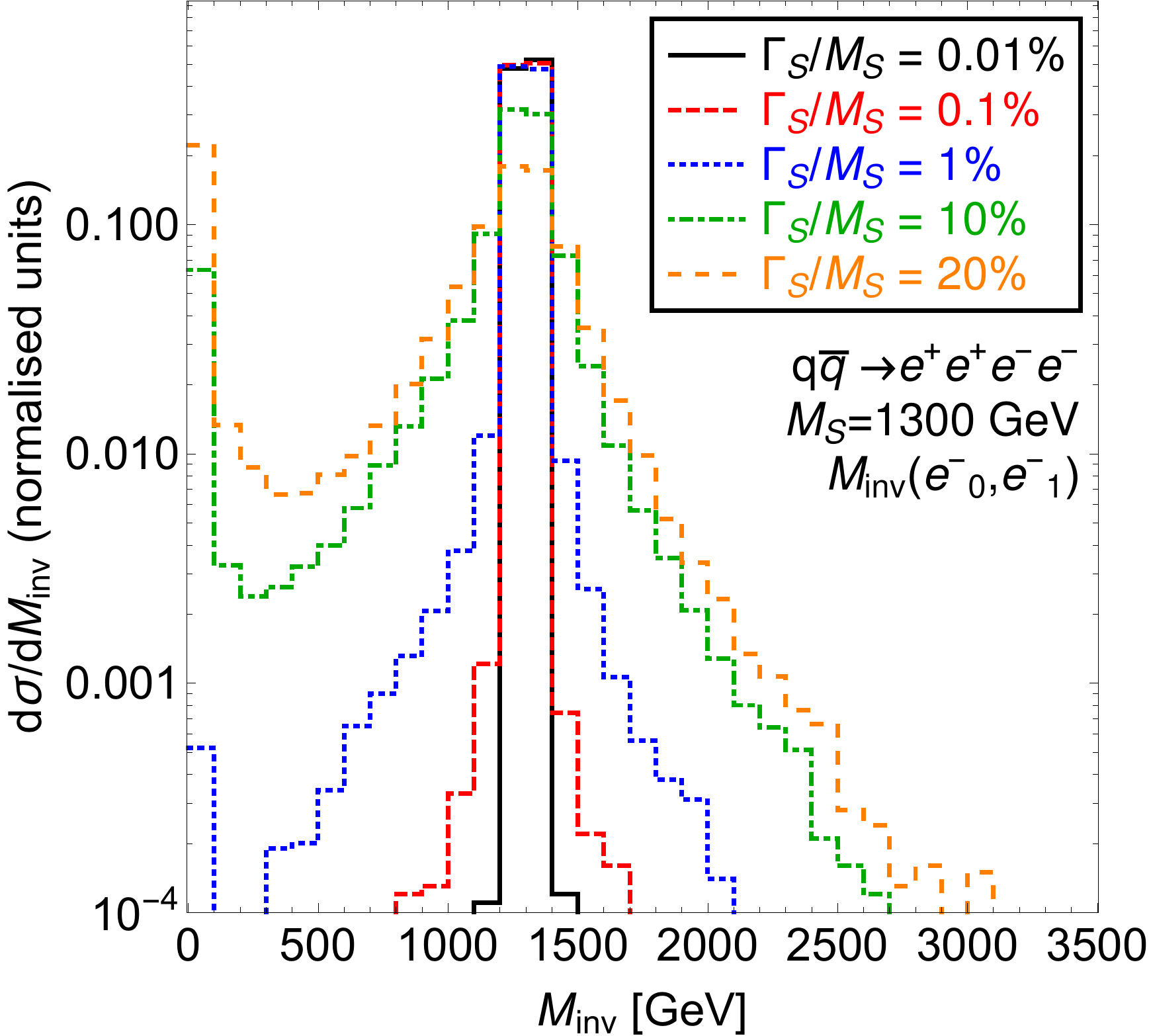}\hfill
\includegraphics[width=.33\textwidth]{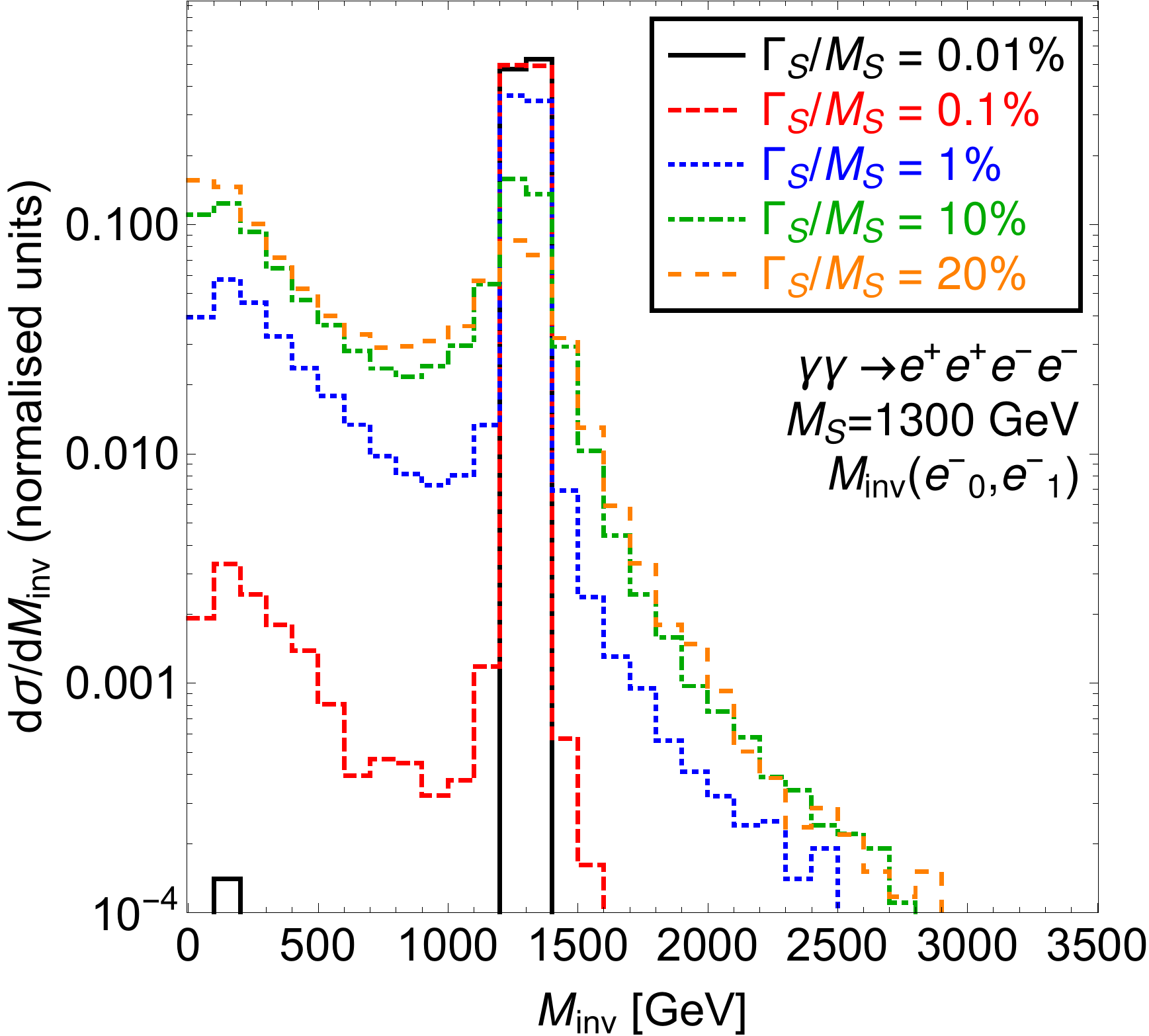}\hfill
\includegraphics[width=.33\textwidth]{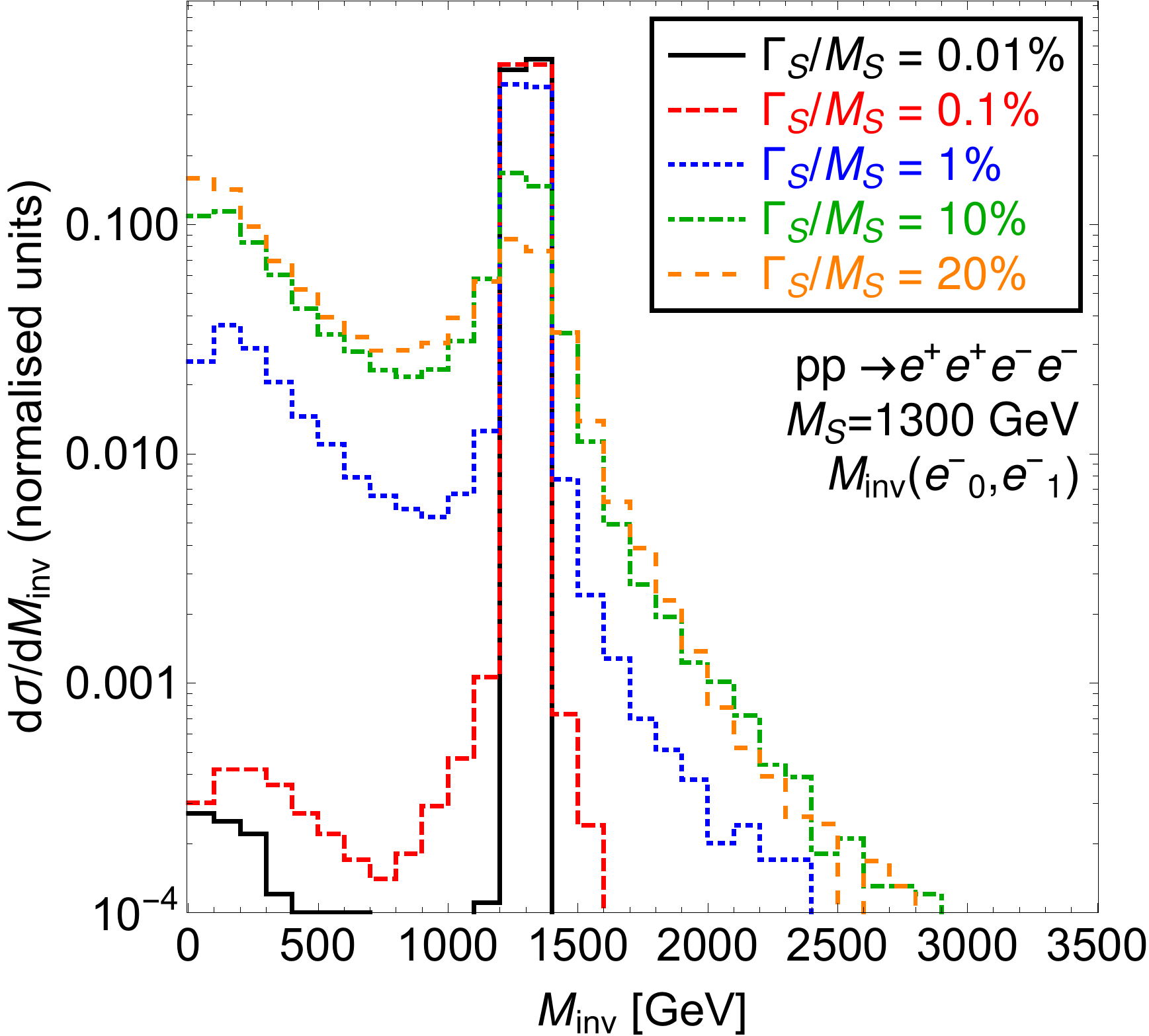}
\caption{\label{fig:distributions1300} Kinematic distributions of the invariant mass of the two same-sign electrons for the final state $2e^+2e^-$ with $m_S = 1300$ GeV and different $\Gamma_S/m_S$ ratios.}
\end{figure}

The kinematic distributions of the invariant mass of the two same-sign electrons for both individual components of the signal, \emph{i.e.} the DY and photon-initiated sub-processes, and for the total process are shown considering $m_S=1300$ GeV in Figure~\ref{fig:distributions1300}. 
The distributions show remarkable differences when the width is increased, and such differences appear in regions which largely contribute to the total cross section. As the $\Gamma_S/m_S$ ratio increases, the invariant mass distribution $M_{\rm inv}(e^-_0,e^-_1)$, which has a peak on the $S$ mass which broadens as the width increases, receives a contribution in the region below 500 GeV for larger widths, which is completely absent in the NWA. Such differences can thus strongly affect the efficiency of a given set of experimental cuts. It is also possible to notice that the distributions of the full process (right panel in Figure~\ref{fig:distributions1300}) reflect the fact that the $\gamma\gamma$ contribution largely dominates for large $\Gamma_S/m_S$.

\begin{figure}[htbp!]
\centering
\includegraphics[width=.33\textwidth]{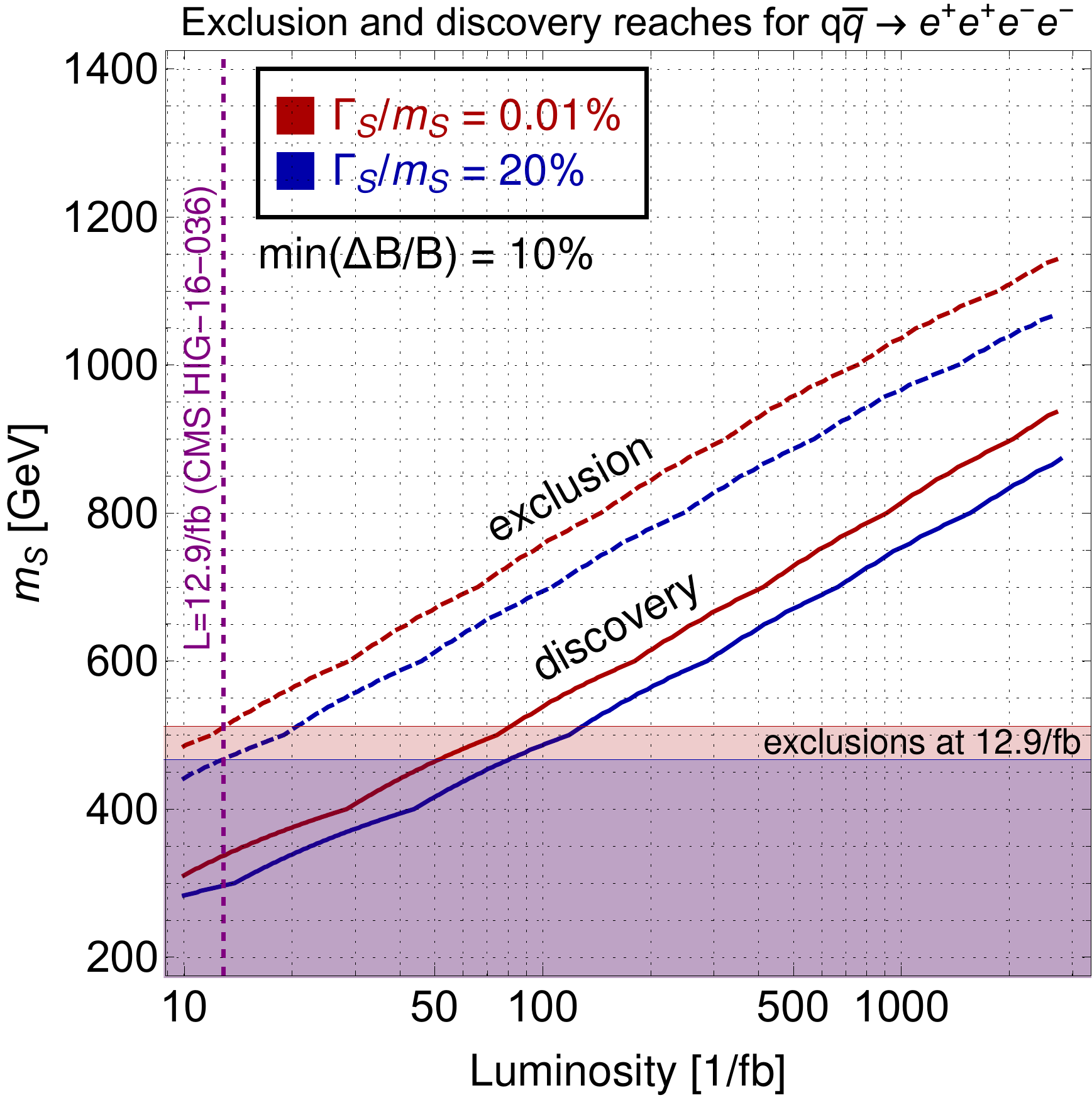}\hfill
\includegraphics[width=.33\textwidth]{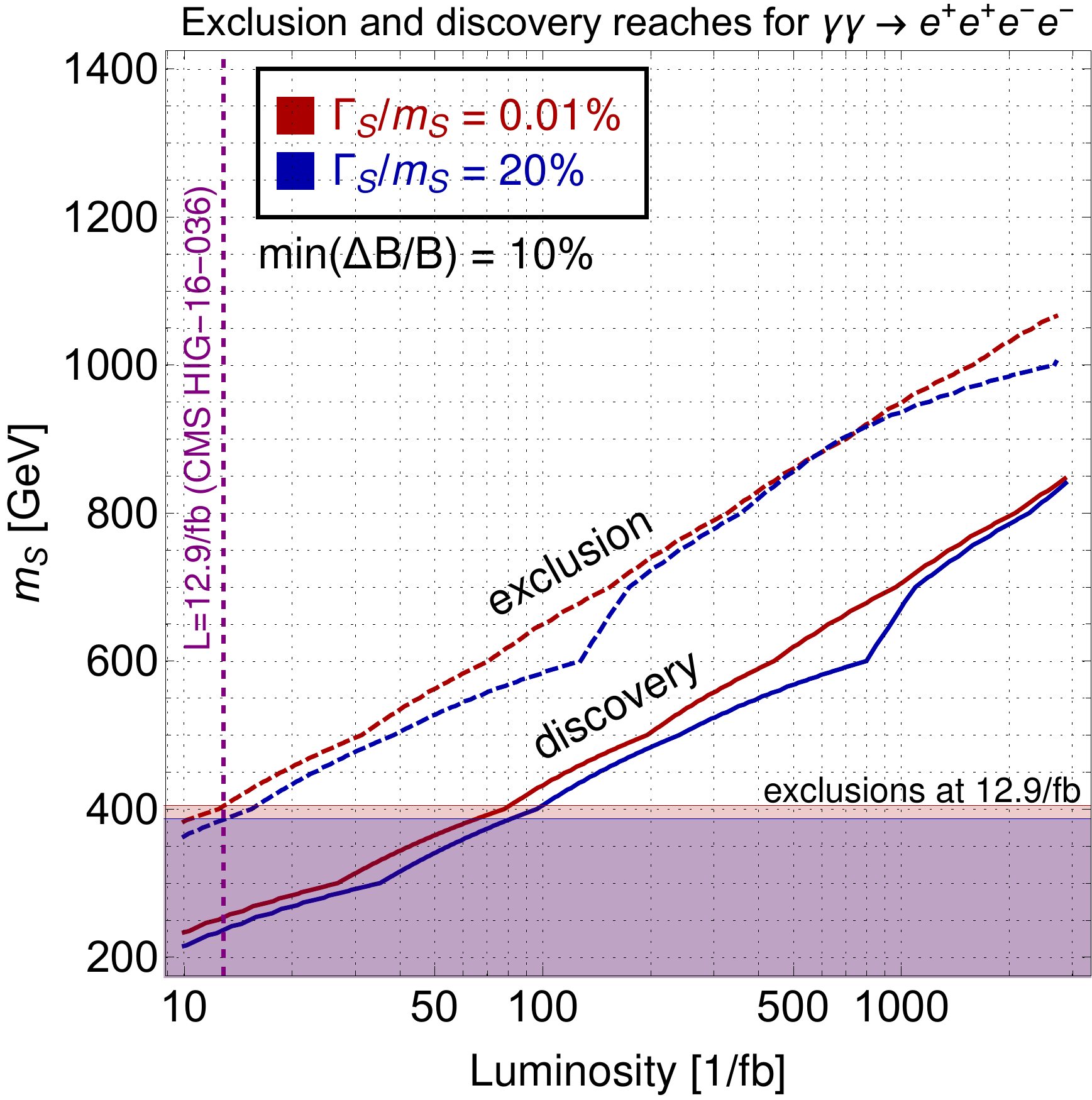}\hfill
\includegraphics[width=.33\textwidth]{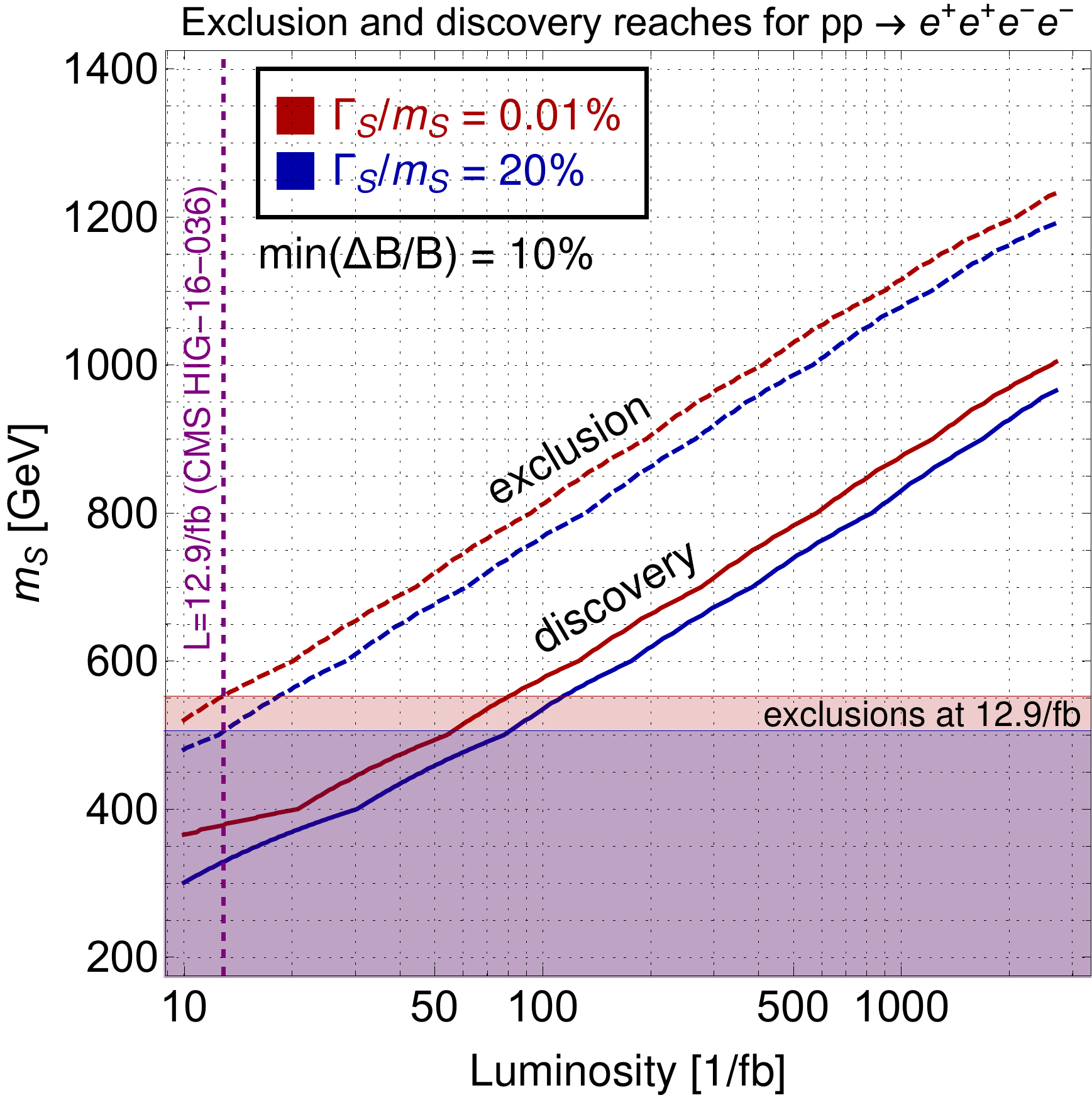}\\
\caption{\label{fig:significances} Exclusion ($\Sigma=2$) and discovery ($\Sigma=5$) reaches as function of the integrated luminosity.}
\end{figure}

The next step of the analysis is to evaluate the performance of experimental searches for doubly charged scalars when the width of $S$ is large. For this purpose we have recast a CMS search at 13 TeV~\cite{CMS:2017pet} within {\tt MadAnalysis5}. 
The cross section of the signal has been obtained considering the maximum value of the coupling which can produce a given total width. The exclusion and discovery reaches, corresponding to a significance $\Sigma=S/\sqrt{S+B+(\Delta B)^2}=2$ and 5 respectively, are summarised in Figure~\ref{fig:significances} for the $2e^+ 2e^-$ channel, considering luminosities from 12.9/fb, corresponding to the search~\cite{CMS:2017pet}, to 3000/fb, corresponding to the HL stage of the LHC. 
The projections have been obtained assuming that signal and background events scale linearly with the luminosity, while the uncertainty on the background scales like its square root down to a floor of systematic uncertainties corresponding to 10\% of the background events.\footnote{The background has been rescaled starting, for concreteness, from the value reported at 12.9/fb for $m_S=500$ GeV, {\emph i.e.} $B=(0.0523\pm0.0113)$, and as a simplifying assumption it is assumed constant over the whole $m_S$ range. The 10\% floor is reached at a luminosity of $\sim$60/fb.} With the same selection and cuts of the CMS search~\cite{CMS:2017pet} considered in this analysis, and under the assumptions above, the bound increases above the TeV. Given the current exclusion bounds, and with the same signal region defined in the experimental search, a discovery can only be made at luminosities larger than $\sim100$/fb.

\begin{figure}[htbp!]
\centering
\includegraphics[width=.33\textwidth]{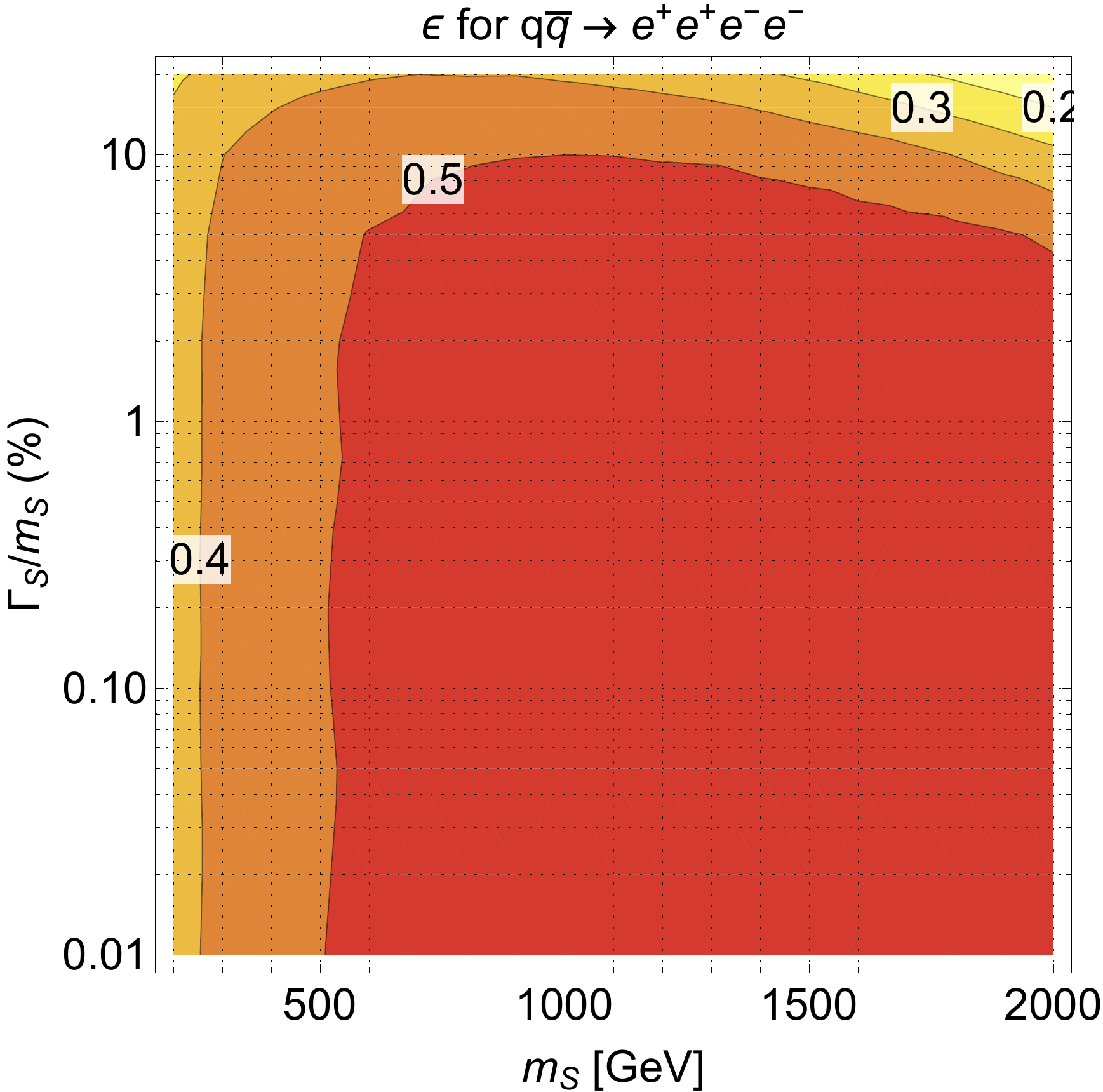}\hfill
\includegraphics[width=.33\textwidth]{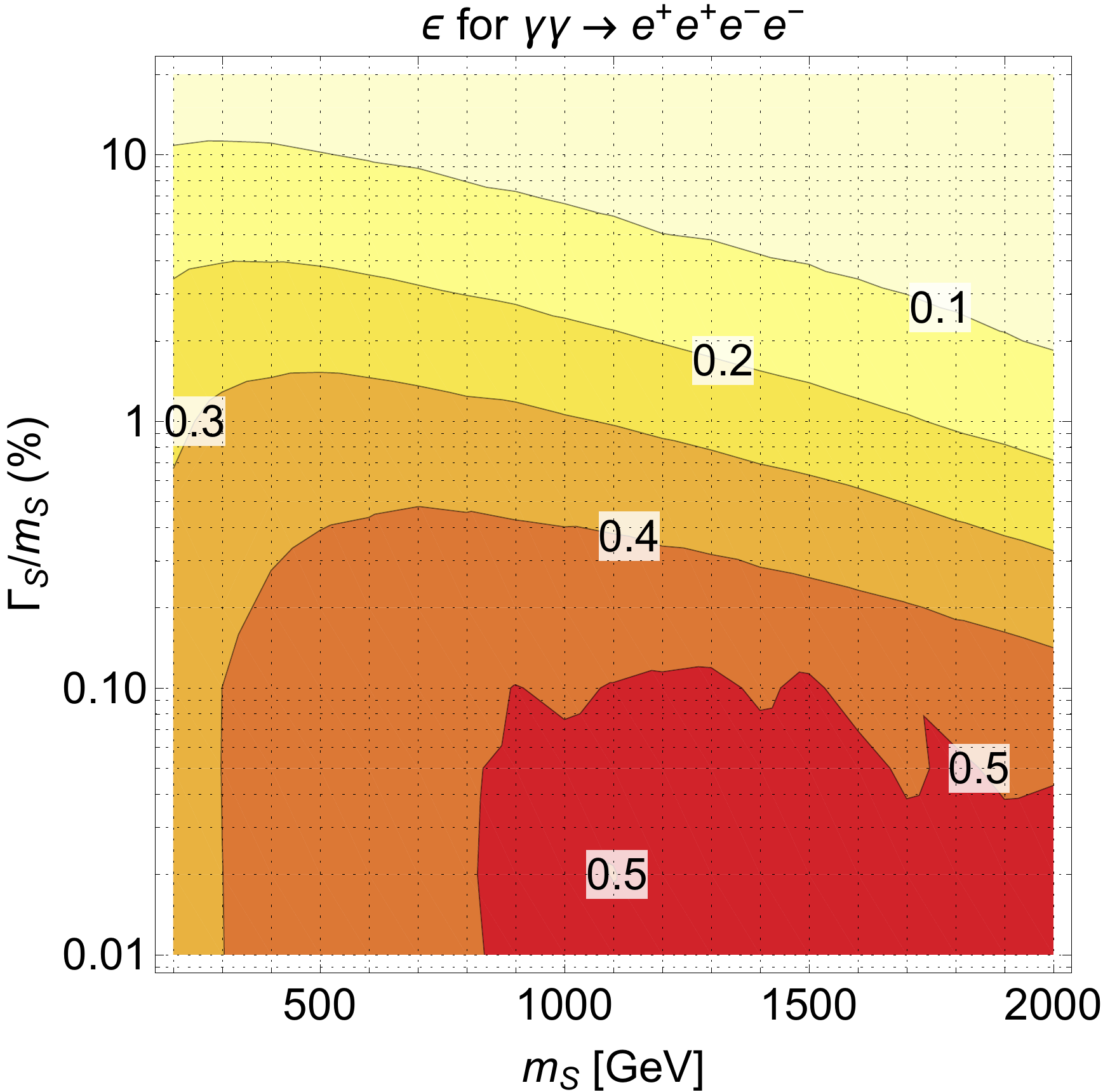}\hfill
\includegraphics[width=.33\textwidth]{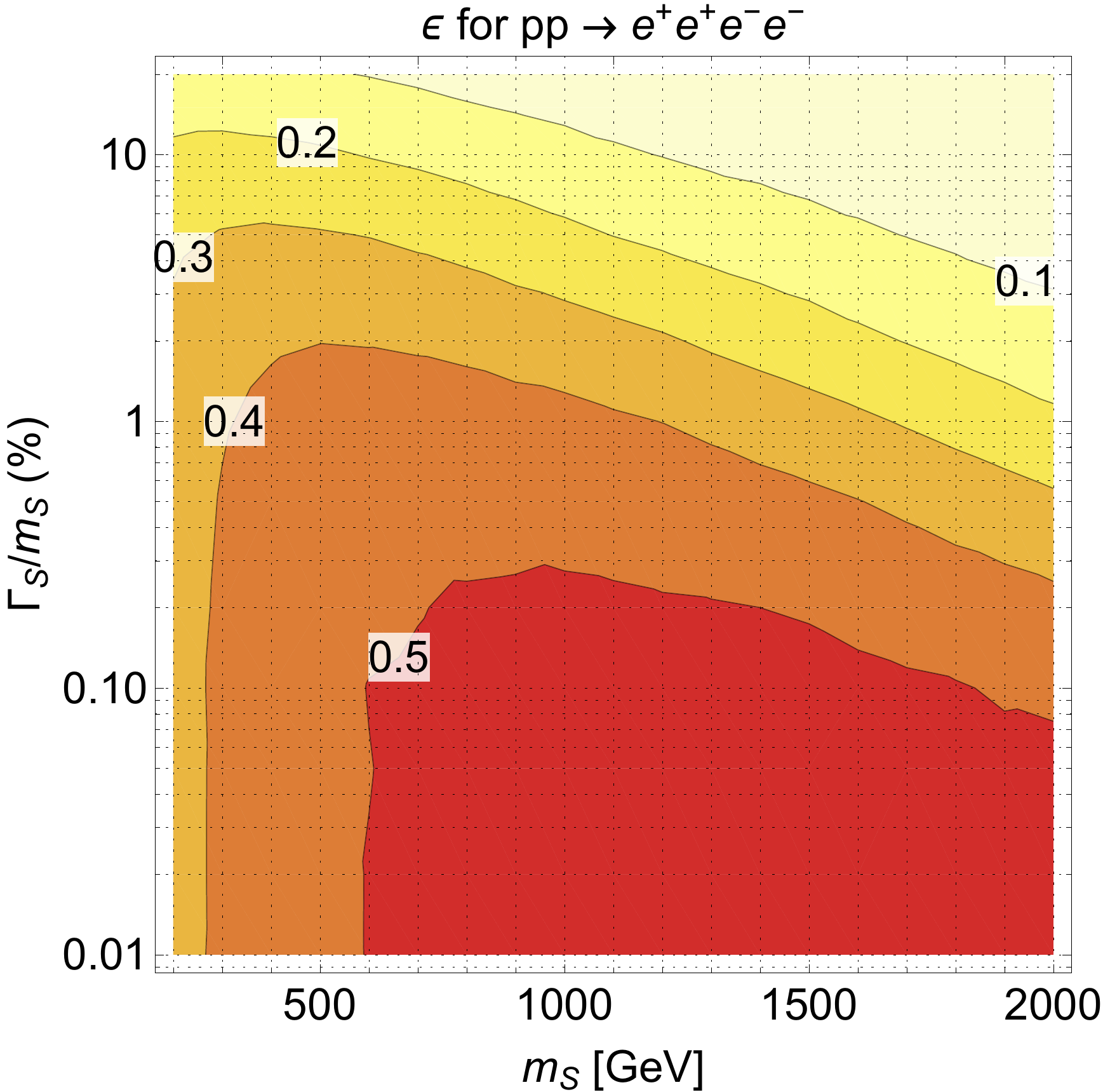}\\
\caption{\label{fig:efficiencies} Efficiency of the cuts in the four-lepton signal regions of the CMS search~\cite{CMS:2017pet}.}
\end{figure}

Under the assumption that the width is entirely generated by the decay channel under consideration, the dependence of the exclusion and discovery reaches on the total width is small, and the reason of this behaviour has to be found in the definition of the kinematic cuts of the CMS search. The signal region corresponding to the $2l^+ 2l^-$ channel (with $l=e,\mu$) selects events in a small invariant-mass window for same-sign dileptons in the region $\{0.9\times m_S,1.1\times m_S\}$. As the width increases, however, more and more events will fall outside such window, thus reducing the efficiency of the cut, as shown in Figure~\ref{fig:efficiencies}. The same figure also explains why the contribution from the $\gamma\gamma$ process is not large even when $\Gamma_S/m_S$ is sizeable: despite the rapidly increasing cross section, the cuts are efficient in filtering out events, compensating in this way the increase.

The bound obtained by the recast in the DY channel is different than the experimental one in the narrow-width limit. This is expected for two reasons: 1) our model contains a $S$ singlet, whereas the experimental bound was obtained for a doubly charged scalar belonging to a triplet with has a different $ZSS$ coupling; 2) our results are obtained at leading order, while the experimental ones have been corrected by a $k$-factor; as it is not possible (and beyond our goals) to apply the same $k$-factor outside the NWA, we have limited our analysis to the LO results.

Note that if the total width were affected by further decay channels, the discovery reach would depend on two factors which affect the results in opposite directions: the reduction in the cross-section due to smaller branching ratios and the potential increase in the number of signal events in the signal region due to contributions from new decay channels (see, {\it e.g.}~\cite{Alcaide:2017dcx}). Both contributions are clearly model-dependent and go beyond the scopes of our analysis.

\begin{figure}[htbp!]
\centering
\includegraphics[width=.33\textwidth]{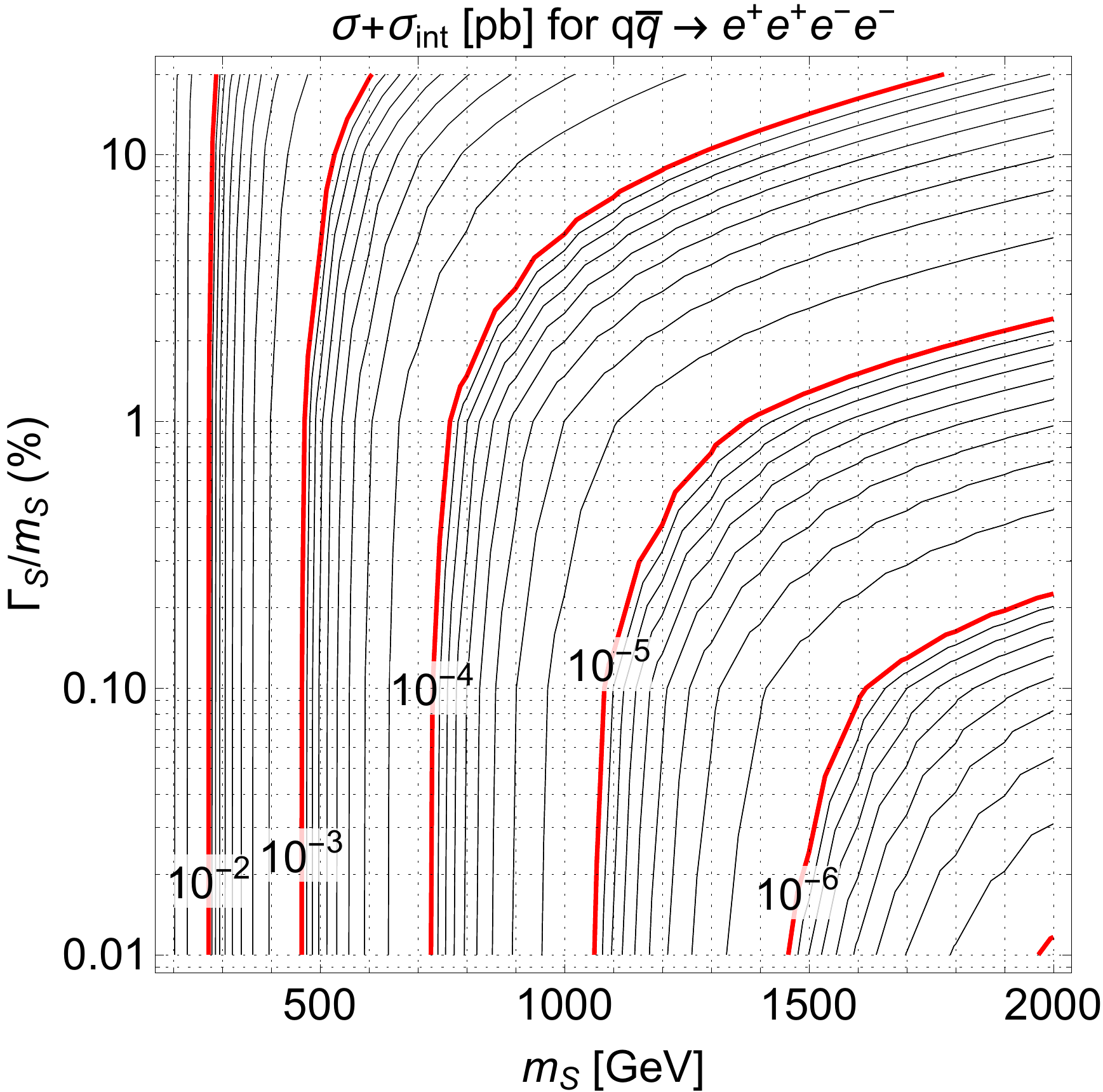}\hfill
\includegraphics[width=.33\textwidth]{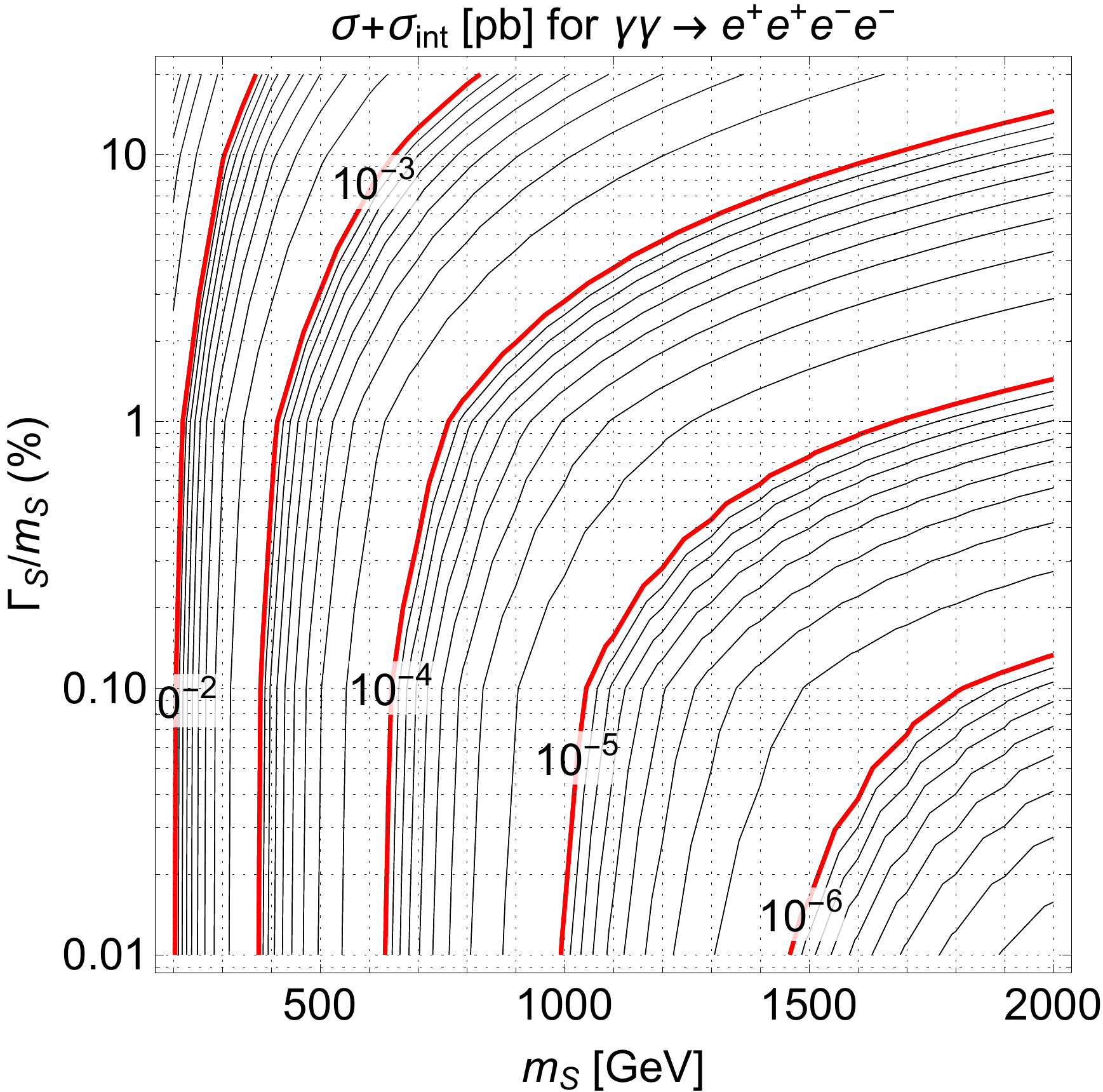}\hfill
\includegraphics[width=.33\textwidth]{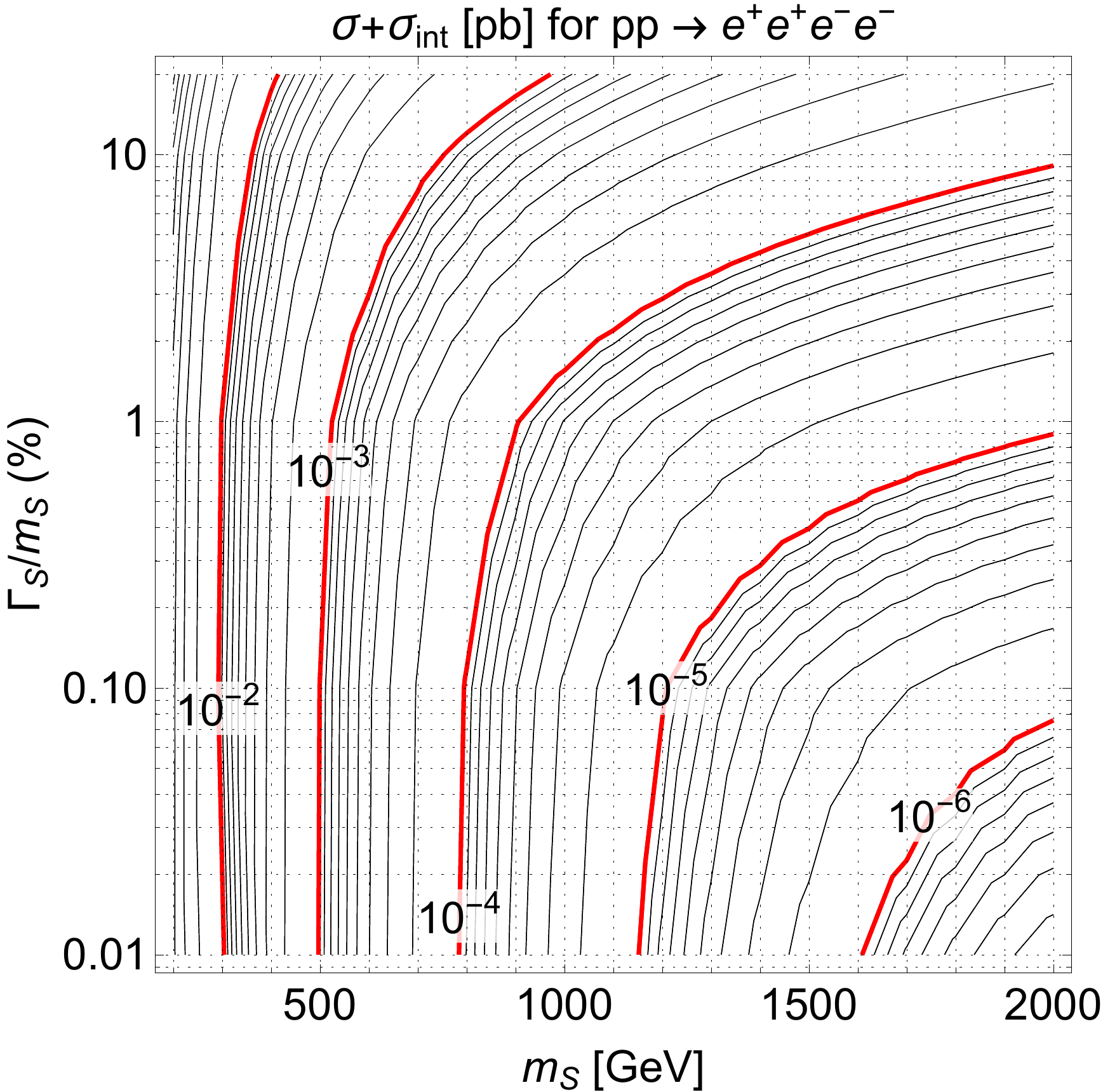}
\caption{\label{fig:sigmahatint} Cross section for signal and interference contributions corresponding to the maximum coupling values which can generate the given widths.}
\end{figure}

\begin{figure}[htbp!]
\centering
\includegraphics[width=.33\textwidth]{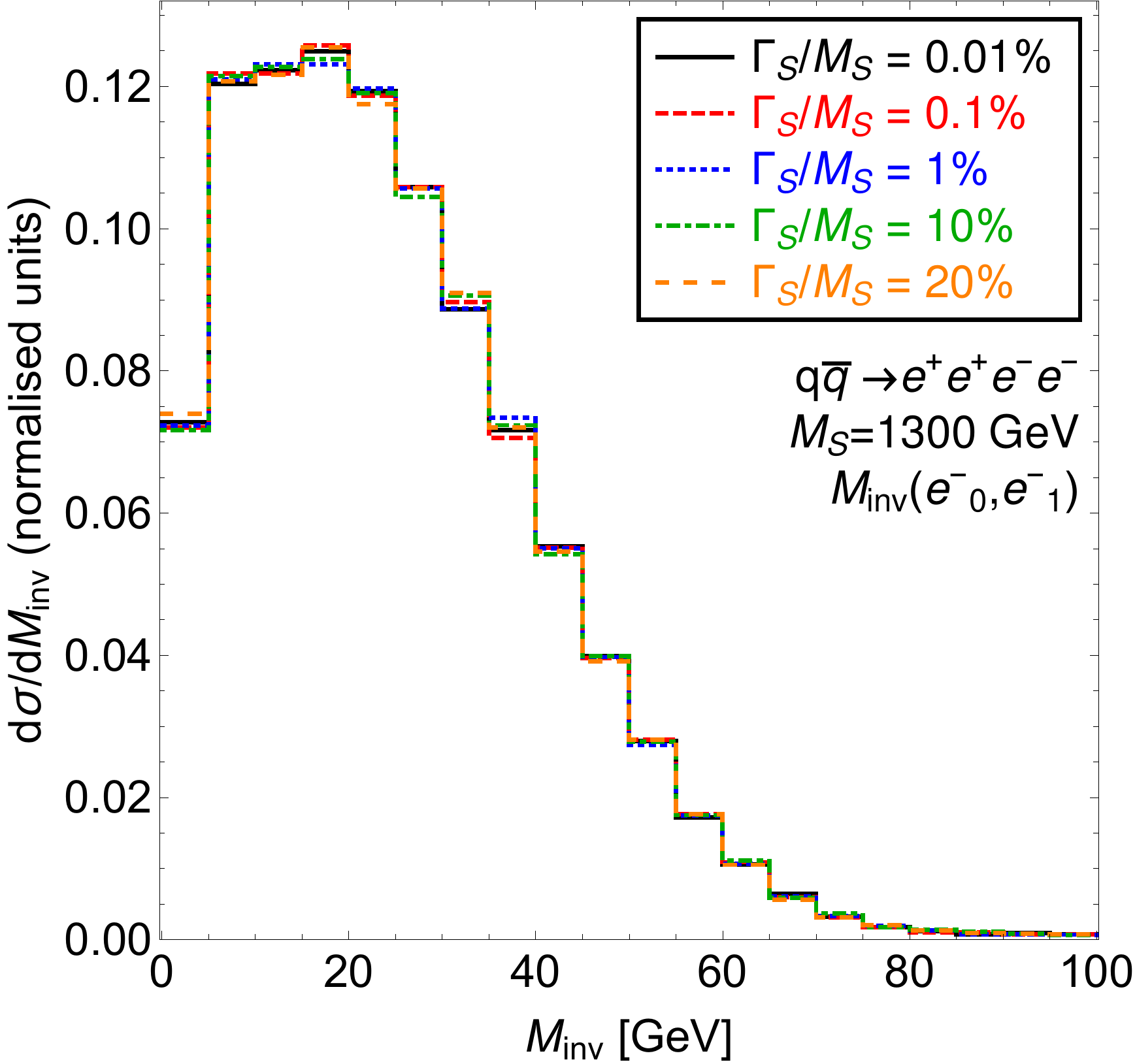}\hfill
\includegraphics[width=.33\textwidth]{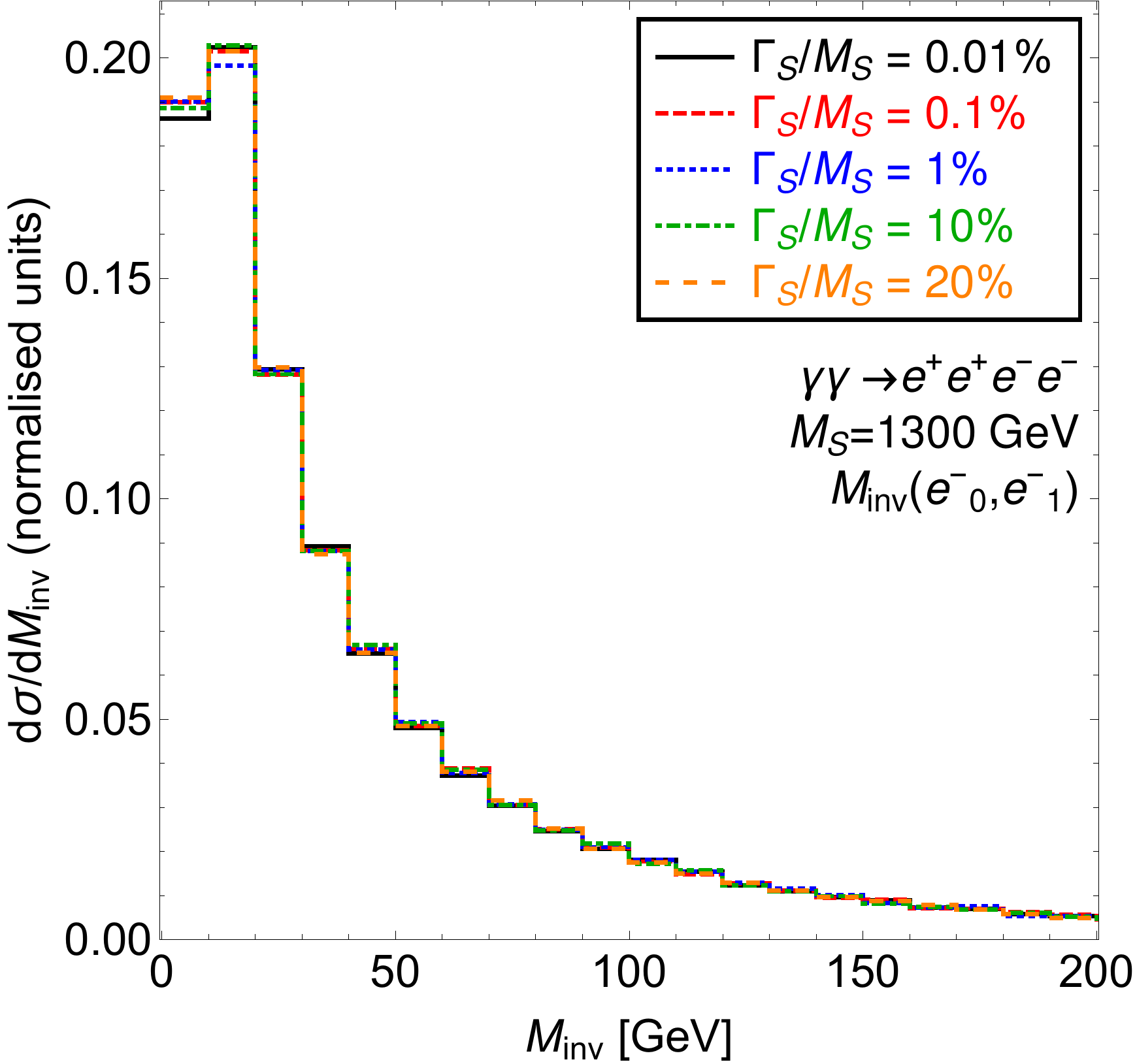}\hfill
\includegraphics[width=.33\textwidth]{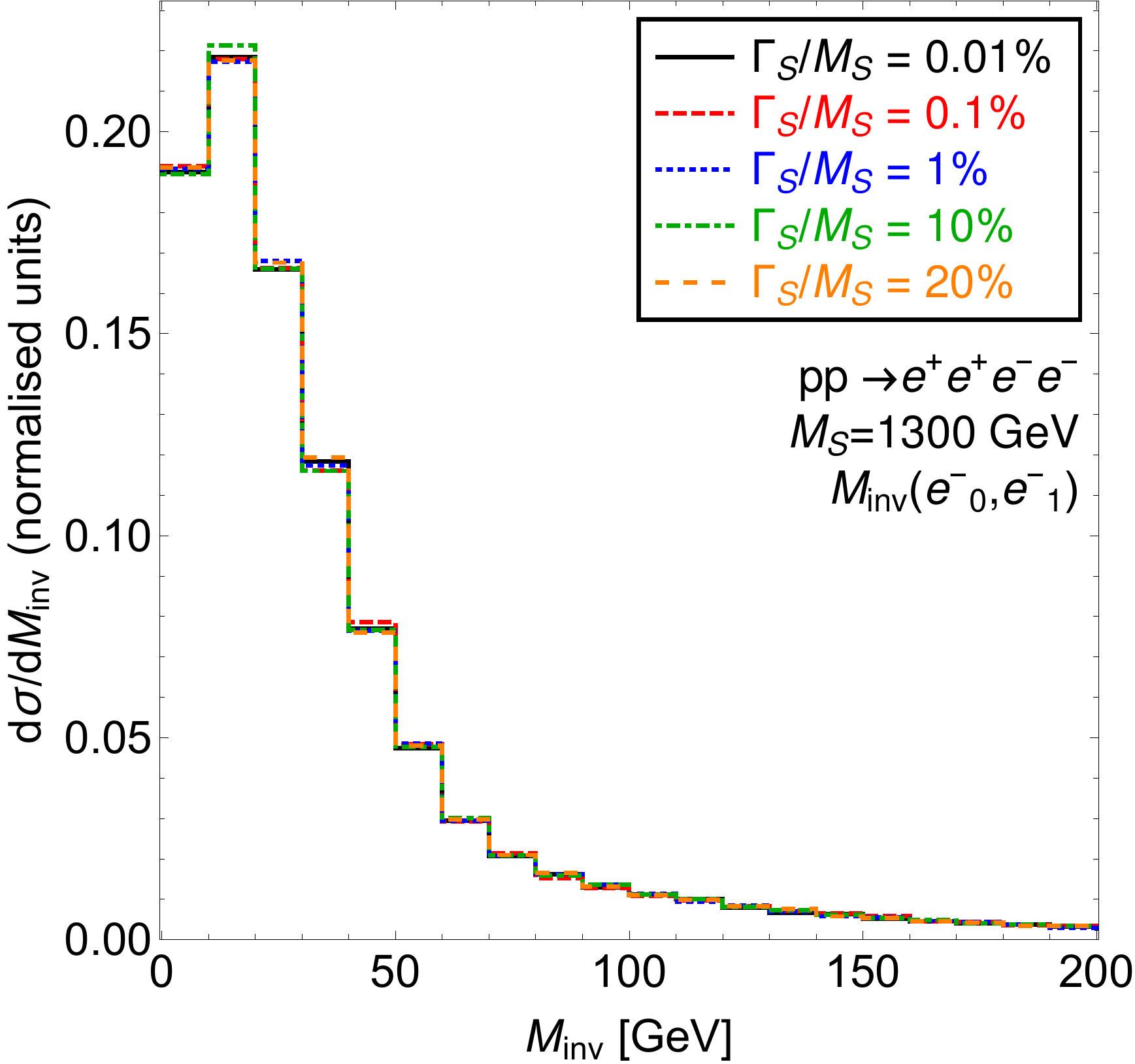}
\includegraphics[width=.33\textwidth]{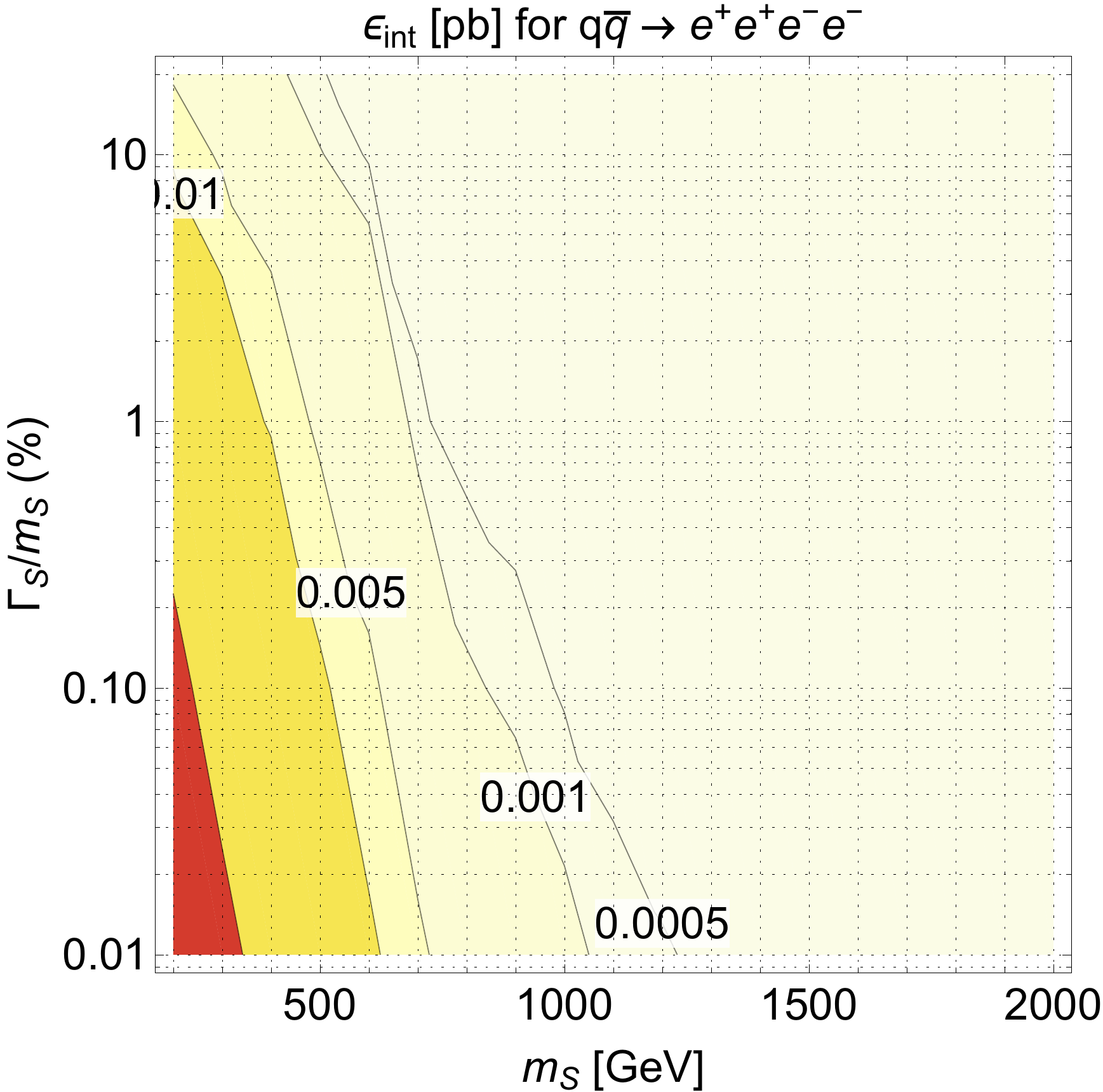}\hfill
\includegraphics[width=.33\textwidth]{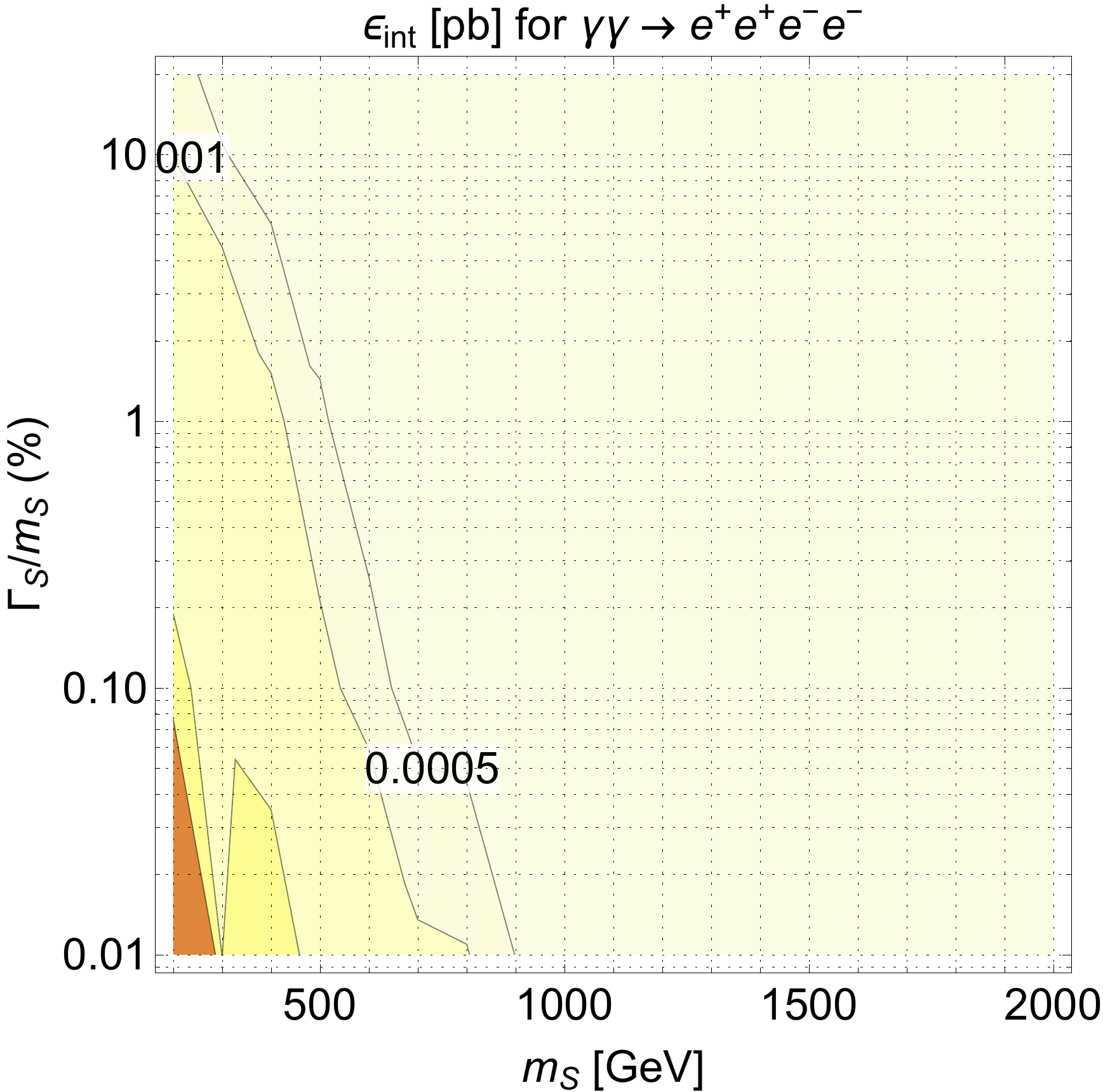}\hfill
\includegraphics[width=.33\textwidth]{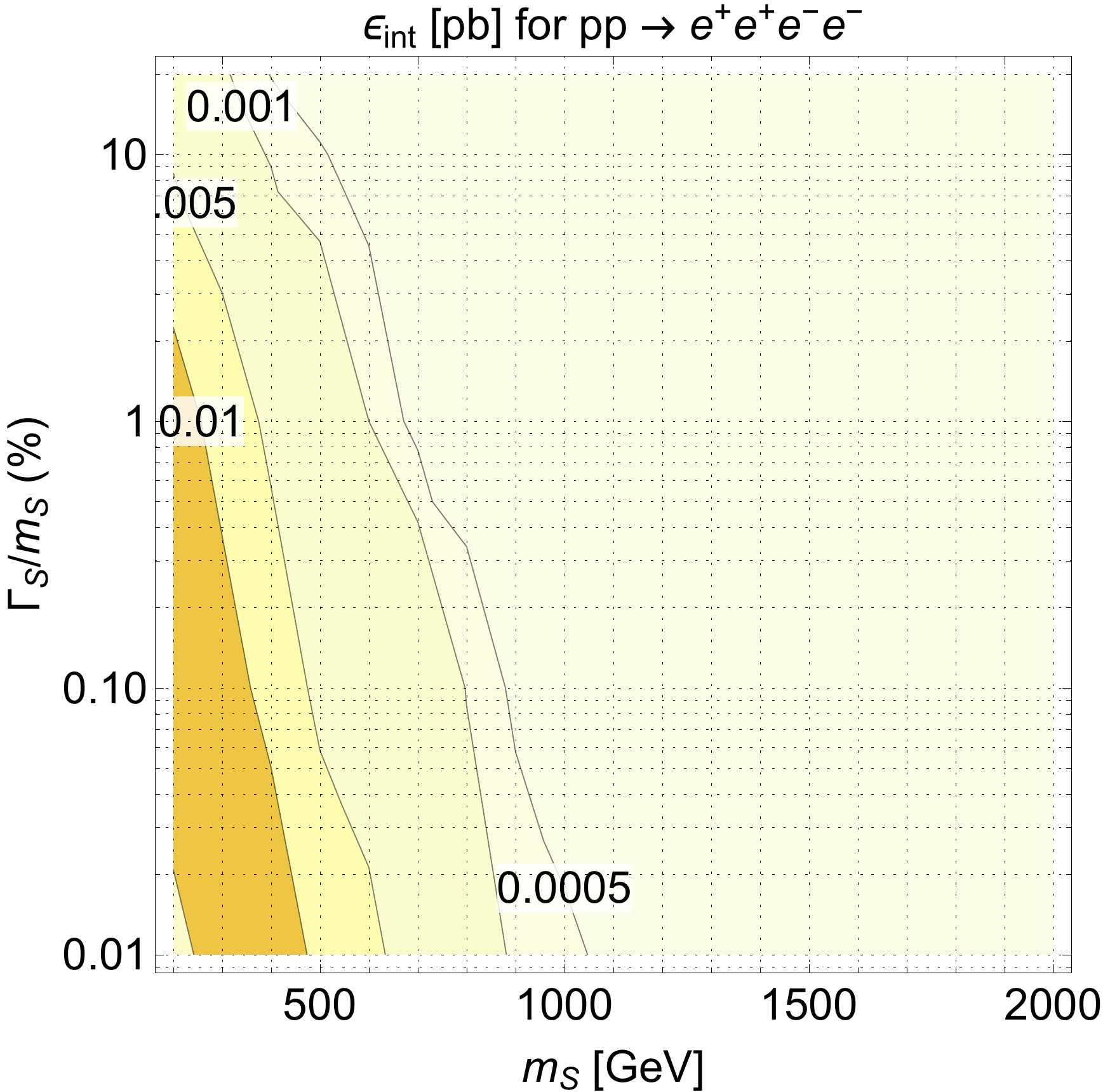}
\caption{\label{fig:interferencekinematics1300} Top row: kinematic distributions of the invariant mass of the two same-sign electrons for interference terms in the $2e^+2e^-$ final state with $m_S = 1300$ GeV and different $\Gamma_S/m_S$ ratios. Bottom row: efficiency of the cuts in the four-lepton signal regions of the CMS search~\cite{CMS:2017pet} for the interference terms.}
\end{figure}

The shape of the $M_{\rm inv}$ distribution in Figure~\ref{fig:distributions1300} shows a relatively large contribution in regions at low energy, where interference with the SM background can become sizeable in the large width regime. It is therefore important to assess the role of such contributions. Interference can arise for example from processes of $Z$ pair production, where the $Z$ boson subsequently decays into leptons. Of course, interference contributions depend on the final state: while final states characterised by two pairs of leptons of the same flavour and opposite charge (such as $2e^+2e^-$ or $e^+\mu^+e^-\mu^-$) can have interference with the SM background, final states with less than two pairs of leptons of same flavour and opposite charge (such as $2e^+2\mu^-$ or $2e^+e^-\mu^-$) do not interfere with the background at all. The combination of signal and interference cross sections obtained by maximising the value of the coupling is also shown in Figure~\ref{fig:sigmahatint}.
The size of interference terms is large enough to influence the cross section at large $S$ masses and widths. However, once the experimental cuts are taken into account, such an effect is completely removed. In Figure~\ref{fig:interferencekinematics1300} we show the distribution of the transverse momentum of the leading electron for the $2e^+2e^-$ final state: it peaks in the low $M_{\rm inv}$ region, which is completely filtered away by the cut on the invariant mass window of same-sign dileptons of the CMS search~\cite{CMS:2017pet}. 
This results in a negligible cut efficiency for interference contribution, shown in the bottom panels of Figure~\ref{fig:interferencekinematics1300}, which allows us to safely consider only the signal component for our phenomenological analysis. The contribution of interference in the large width limit should however be taken into account if considering selection cuts which do not require same-sign leptons to be in a mass window around the peak of the $S$ invariant mass, and if leptons with small transverse momentum are selected. Such information could indeed be used, in principle, for optimising the sensitivity of signal regions to probe final states generated by a $S$ with large width.


\section{Searches at future $e^+e^-$ colliders}
\label{sec_ILC}
\noindent
Future LCs such as the ILC~\cite{Behnke:2013xla,Baer:2013cma} and the CLIC~\cite{Aicheler:2012bya,Linssen:2012hp} have great potential to study BSM physics in the lepton sector. This section is devoted to the analysis of the sensitivity of these proposed colliders to the couplings  and the direct production of the $S$. For this purpose, our model has been implemented in {\tt FeynRules~v2.3}~\cite{Alloul:2013bka} to extract a model file for CalcHEP~\cite{Belyaev:2012qa}. The numerical simulations have been performed with {\tt CalcHEP~v3.6.29}, taking into account the initial state radiation and beamstrahlung. The former is implemented in CalcHEP using the expressions of Jadach, Skrzypek, and Ward~\cite{Jadach:1988gb, Skrzypek:1990qs}, while the latter is calculated by CalcHEP according to the parameters characterising the beams, which are given in the ILC Technical Design Report~\cite{Behnke:2013xla} and in the CLIC Conceptual Design Report~\cite{Aicheler:2012bya}. According to these documents, the expected centre-of-mass energies and integrated luminosities for the ILC and the CLIC correspond to the values reported in Table~\ref{tab_en-lum}. Furthermore, in the present analysis, standard acceptance cuts for a LC have been applied to the charged-lepton final state, namely
\begin{equation}
E_\ell >10\, \mbox{GeV}, \qquad\qquad |\cos(\theta_\ell)| < 0.95\,,
\end{equation}
where $E_\ell$ are the energies of the charged leptons ($\ell=e^\pm,\mu^\pm$) and $\theta_\ell$ are their angles with respect to the beam direction.

\begin{table}[htbp!]
\centering
\begin{tabular}{cccc}
\toprule
{\bf Stage} & I & II & III \\
\midrule
\boldmath{$\sqrt{s}_{\mbox{\tiny ILC}}$} & 250 GeV & 500 GeV & 1 TeV \\
\boldmath{$\mathcal{L}_{\mbox{\tiny ILC}}$} & 250 fb$^{-1}$ & 500 fb$^{-1}$ & 1 ab$^{-1}$ \\
\bottomrule
\end{tabular}
\hspace{0.1\textwidth}
\begin{tabular}{ccccc}
\toprule
{\bf Stage} & Ia & Ib & II & III \\
\midrule
\boldmath{$\sqrt{s}_{\mbox{\tiny CLIC}}$} & 350 GeV & 380 GeV & 1.5 TeV & 3 TeV \\
\boldmath{$\mathcal{L}_{\mbox{\tiny CLIC}}$} & 100 fb$^{-1}$ & 500 fb$^{-1}$ & 1.5 ab$^{-1}$ & 3 ab$^{-1}$ \\
\bottomrule
\end{tabular}
\caption{\small Centre-of-mass energies and expected integrated luminosities of ILC prototypes (left part) and CLIC prototypes (right part).}
\label{tab_en-lum}
\end{table}

The $e^+e^-$ colliders are sensitive to the product $\lambda_{1a}\lambda_{1b}$ since $S$ can be exchanged in the $t$-channel. Therefore, for flavour conserving final states a single coupling can be constrained while only combinations of different couplings are constrained by low-energy experiments (see Section~\ref{sec_2}).

Because $S$ in \eq{Lagrangian} only couples to right-handed leptons, correspondingly polarised beams can result in an enhancement of the production cross section. On the contrary, left-handed polarised beams decrease the sensitivity to our $S$, but would show the opposite trend if the doubly charged scalar were a component of $SU(2)_L$-triplet. Therefore, the beam polarisation can be useful to distinguish between the two scenarios and enhance the signal with respect to the background, to achieve a better sensitivity to the couplings~\cite{Nomura:2017abh}. 
In what follows, the polarisation features of both the ILC and the CLIC prototypes are exploited. In particular, ILC has the option to polarise the electron beam to $P_{e^-} = \pm 80\%$ and the positron beam to $P_{e^+} = \mp30\%$~\cite{Behnke:2013xla}, while CLIC has the option to polarise the electron beam to $P_{e^-} = \pm80\%$~\cite{Aicheler:2012bya}.

\begin{figure}
\begin{center}
\includegraphics[width=0.49\textwidth]{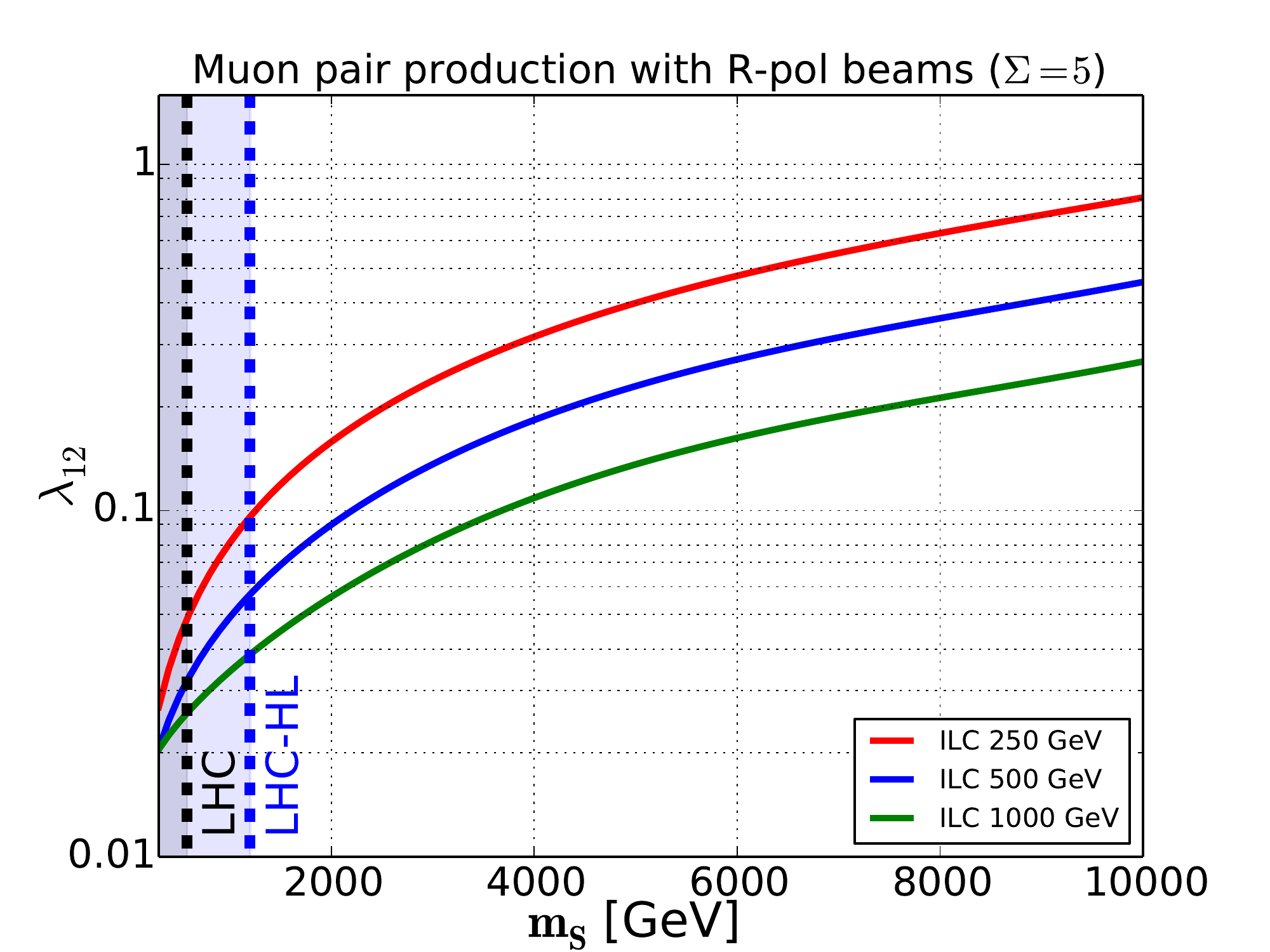}
\includegraphics[width=0.49\textwidth]{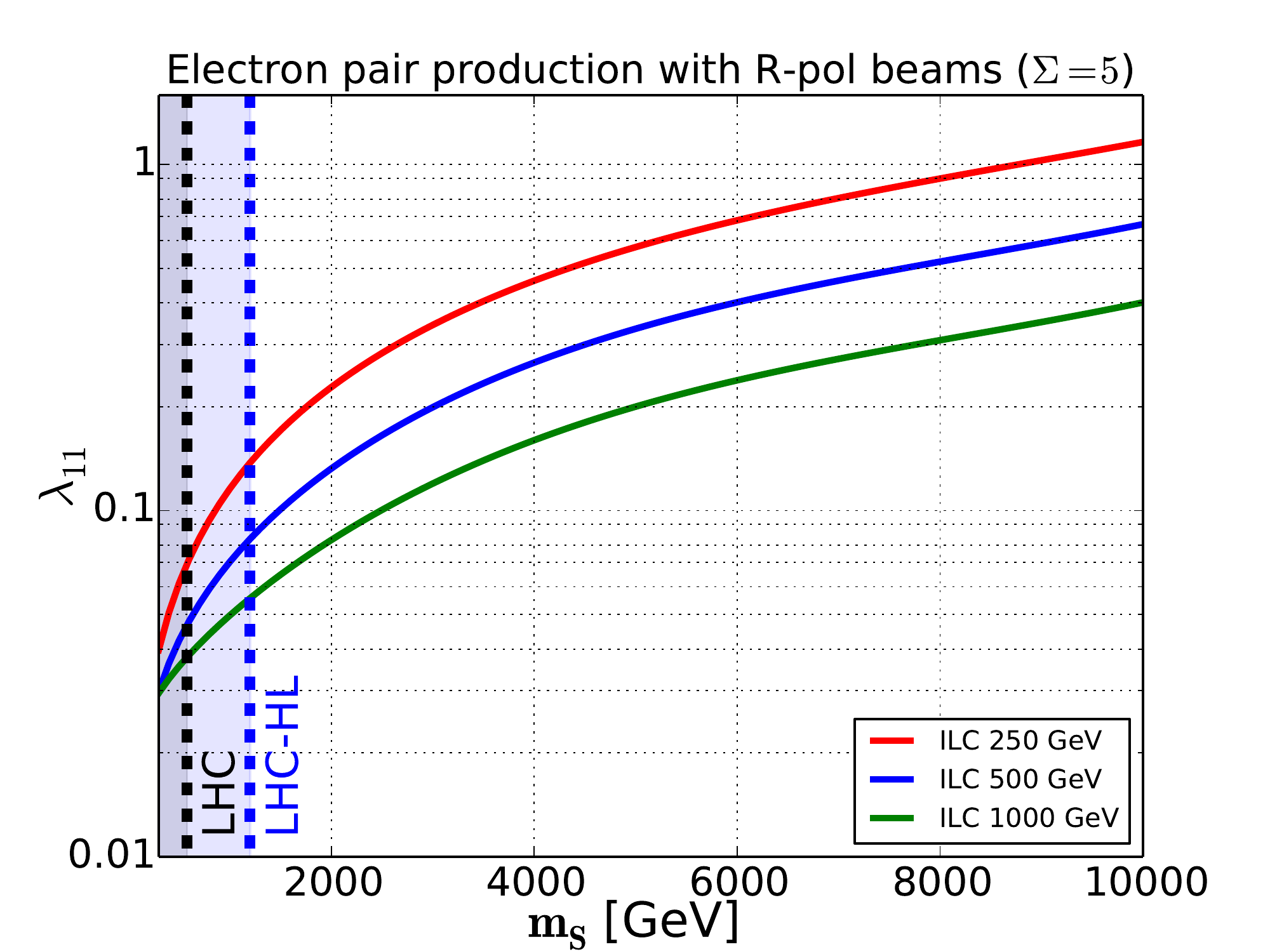} \\
\includegraphics[width=0.49\textwidth]{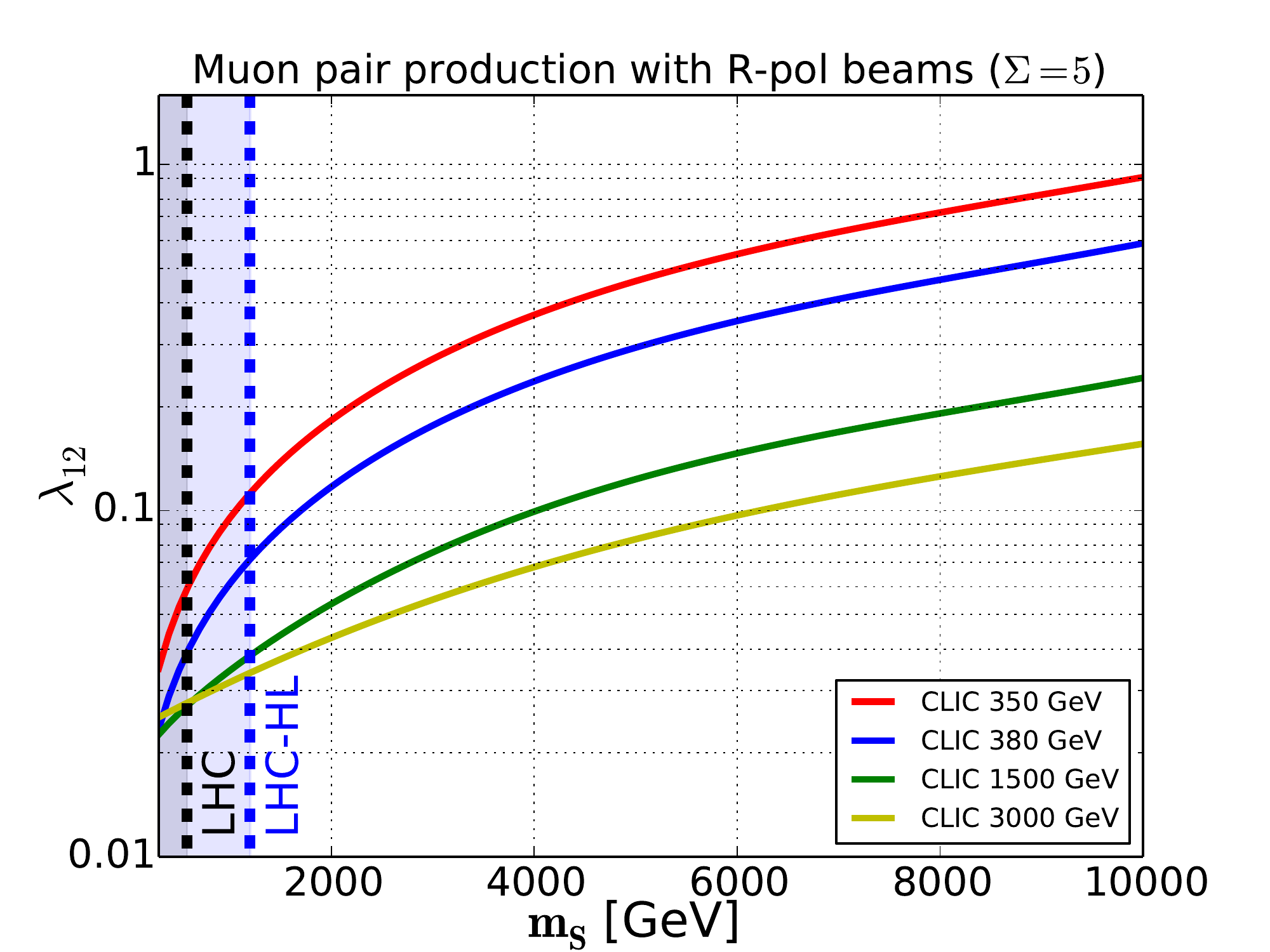}
\includegraphics[width=0.49\textwidth]{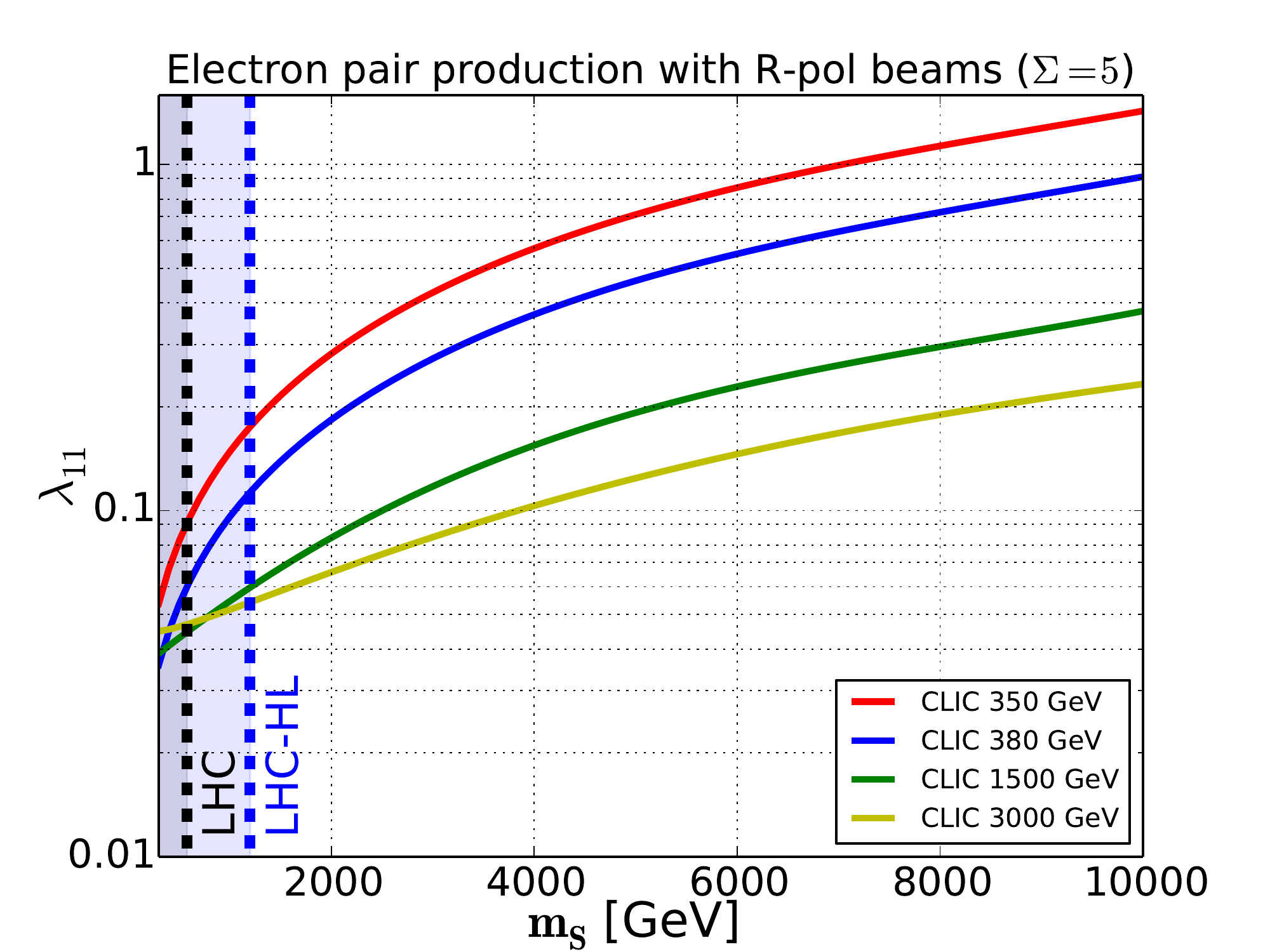}
\caption{Contours of the cross section of $e^+e^- \rightarrow \mu^+\mu^-$ (left panels) and $e^+e^- \rightarrow e^+e^-$ (right panels) with $\Sigma=5$ for different values of the coupling and the mass of the $S$, at ILC with right-handed polarised beams (upper panels) and CLIC with right-handed polarised electron beam (lower panels). For the electron-positron pair production, the restriction $|\cos\theta| < 0.5$ is also applied.}
\label{contours_eemumu-eeee}
\end{center}
\end{figure}

In Figure~\ref{contours_eemumu-eeee} the contours of the cross section of $e^+e^- \rightarrow \mu^+\mu^-$ and $e^+e^- \rightarrow e^+e^-$ for the discovery significance $\Sigma = 5$ are shown as a function of the mass $m_S$ and of the coupling $\lambda_{12}$ or $\lambda_{11}$ for the ILC and the CLIC prototypes, for the luminosities reported in Table~\ref{tab_en-lum}. The significance $\Sigma$, defined in Section~\ref{sec_LHC}, is calculated here assuming that the uncertainty on the background is negligible. As previously described, the beams are right-handed polarised in order to enhance the contribution from the $S$ exchanged in the $t$-channel. In the case of the electron-positron production, it is convenient to apply a stronger cut on the angle $\theta$, namely $|\cos\theta| < 0.5$~\cite{Nomura:2017abh}, to better cope with the large background. The better performance of the ILC compared to the CLIC is due to the positron beam polarisation. 
Sensitivity to the coupling $\lambda_{13}$ can be achieved via the process $e^+e^- \rightarrow \tau^+\tau^-$. Some benchmark points are reported in Table~\ref{tab_lam13ILCandCLIC}, where an efficiency rate of 70\% is assumed for the reconstruction of $\tau$ leptons decaying to hadrons.

\begin{table}
\begin{tabular}{c|ccc}
\toprule
 \boldmath{$\lambda_{13}$}& $m_S=500$ GeV & $m_S=1$ TeV & $m_S=2$ TeV \\
\midrule
ILC 250 & $  6.4\times 10^{-2}$ & $  1.2\times 10^{-1}$ & $  2.3\times 10^{-1}$  \\
ILC 500 & $  4.3\times 10^{-2}$ & $ 7.2\times 10^{-2}$ & $  1.4\times 10^{-1}$  \\
ILC 1000 & $  3.9\times 10^{-2}$ & $  5.0\times 10^{-2}$ & $  7.6\times 10^{-2}$  \\
\midrule
\midrule
CLIC 380 & $  6.7\times 10^{-2}$ & $  9.1\times 10^{-2}$ & $  1.8\times 10^{-1}$ \\
CLIC 1000 & $  3.9\times 10^{-2}$ & $  5.1\times 10^{-2}$ & $  8.1\times 10^{-2}$\\
CLIC 3000 & $  4.0\times 10^{-2}$ & $  4.7\times 10^{-2}$ & $  6.4\times 10^{-2}$\\
\bottomrule
\end{tabular}
\caption{\small Sensitivity of ILC (upper part) and CLIC (lower part) prototypes to $\lambda_{13}$ from the process $e^+e^- \rightarrow \tau^+\tau^-$.}
\label{tab_lam13ILCandCLIC}
\end{table}

\begin{figure}
\begin{center}
\includegraphics[width=0.49\textwidth]{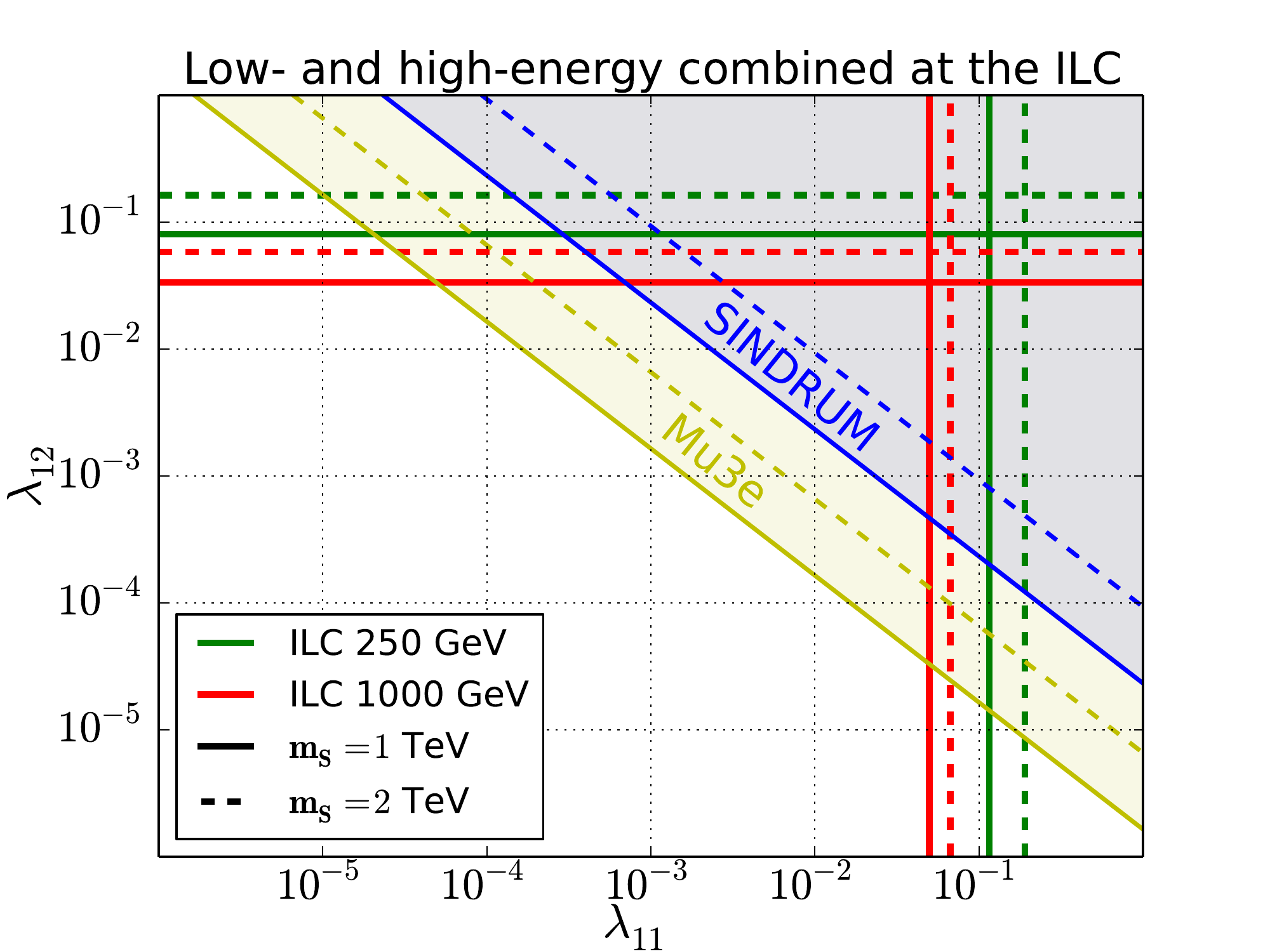}
\includegraphics[width=0.49\textwidth]{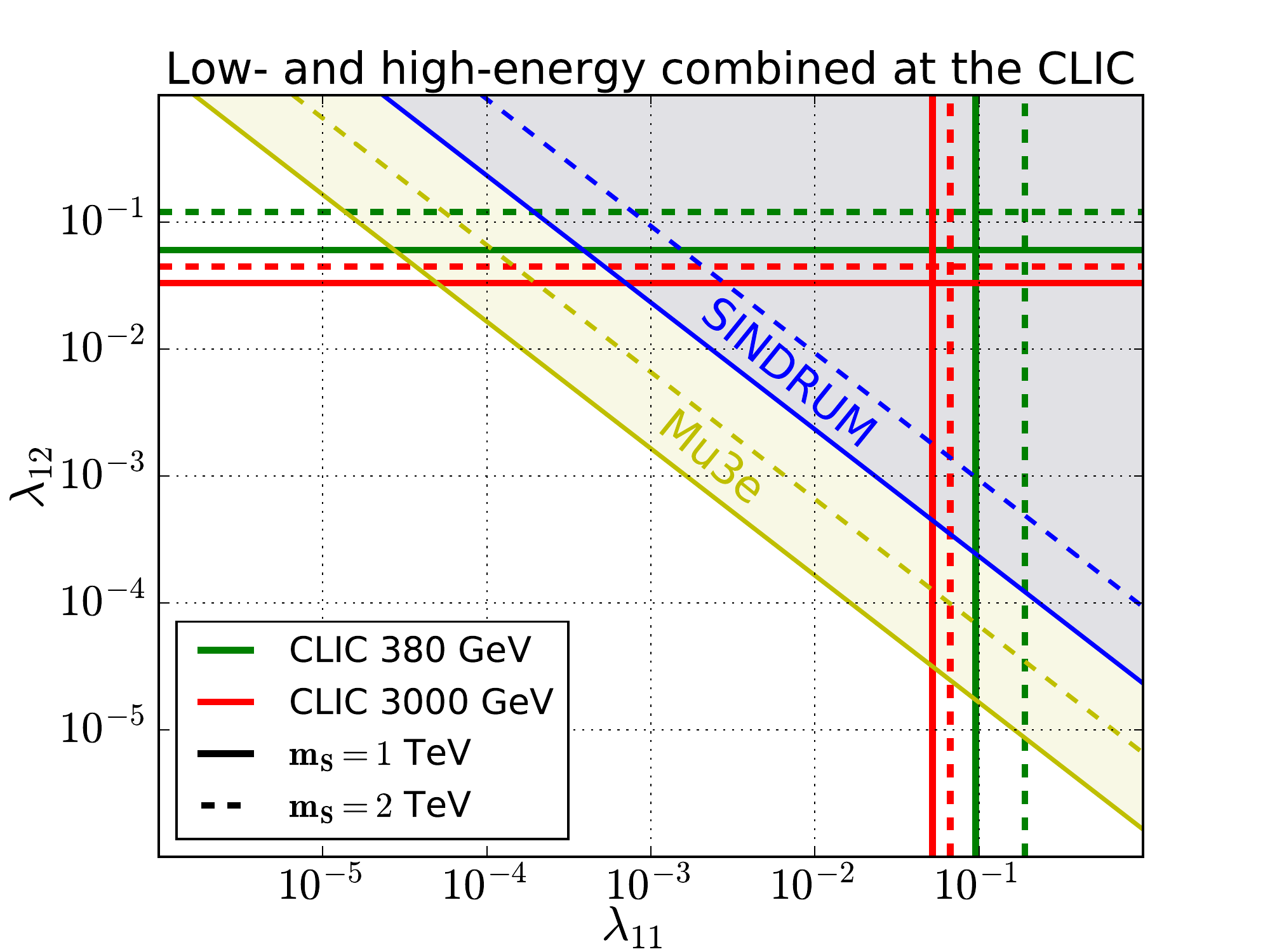}
\caption{Limits from SINDRUM and discovery power at the ILC (left panel) and CLIC (right panel) prototypes and at the Mu3e experiment (both panels).}
\label{ILC+SINDRUM}
\end{center}
\end{figure}

The discovery potential of future linear colliders has to be compared to the actual sensitivity of the low-energy experiments and to their planned future upgrades. The most important low-energy constraint on $\lambda_{11}$ and $\lambda_{12}$ comes from the three-body muon decay $\mu\to 3e$. The current limit is set to BR$\leq 10^{-12}$ by the SINDRUM experiment~\cite{Bellgardt:1987du} and is expected to be improved to BR$\leq 5\cdot 10^{-15}$ by the Phase I of the Mu3e experiment~\cite{Berger:2014vba,Perrevoort:2018cqi}. On the other hand, via the $S$ exchange in the $t$-channel the linear colliders can be sensitive to the couplings $\lambda_{11}$ and $\lambda_{12}$ independently and would be complementary to the low-energy experiments to this extent. Figure~\ref{ILC+SINDRUM} shows the combination of the sensitivities of ILC (left panel) and CLIC (right panel) to $\lambda_{11}$ and $\lambda_{12}$ with the current limit from the SINDRUM experiment and the expected limits from the Mu3e experiment. These limits on the product $\lambda_{11}\lambda_{12}$ are extracted assuming that the dominant contribution to the $\mu \rightarrow 3e$ decay comes from $\lambda_{11}$ and $\lambda_{12}$, while the other couplings are suppressed. In general, switching on the other $S$ couplings would result in more stringent bounds on $\lambda_{11}\lambda_{12}$, but fine-tuned regions of the parameter space where cancellations take place, relaxing the bounds, are also possible.

The leptonic colliders offer the opportunity to explore a new production channel, that is absent at the LHC: a single $S$, in association with two same-sign uncorrelated leptons, can be produced on-shell when the collider energy is compatible with the mass of the particle. The production proceeds via boson fusion and via radiation of the $S$ from initial or final leptonic states. These two sub-channels strongly interfere and cannot be separated at the level of the total cross section.
\begin{figure}
\begin{center}
\includegraphics[width=0.49\textwidth]{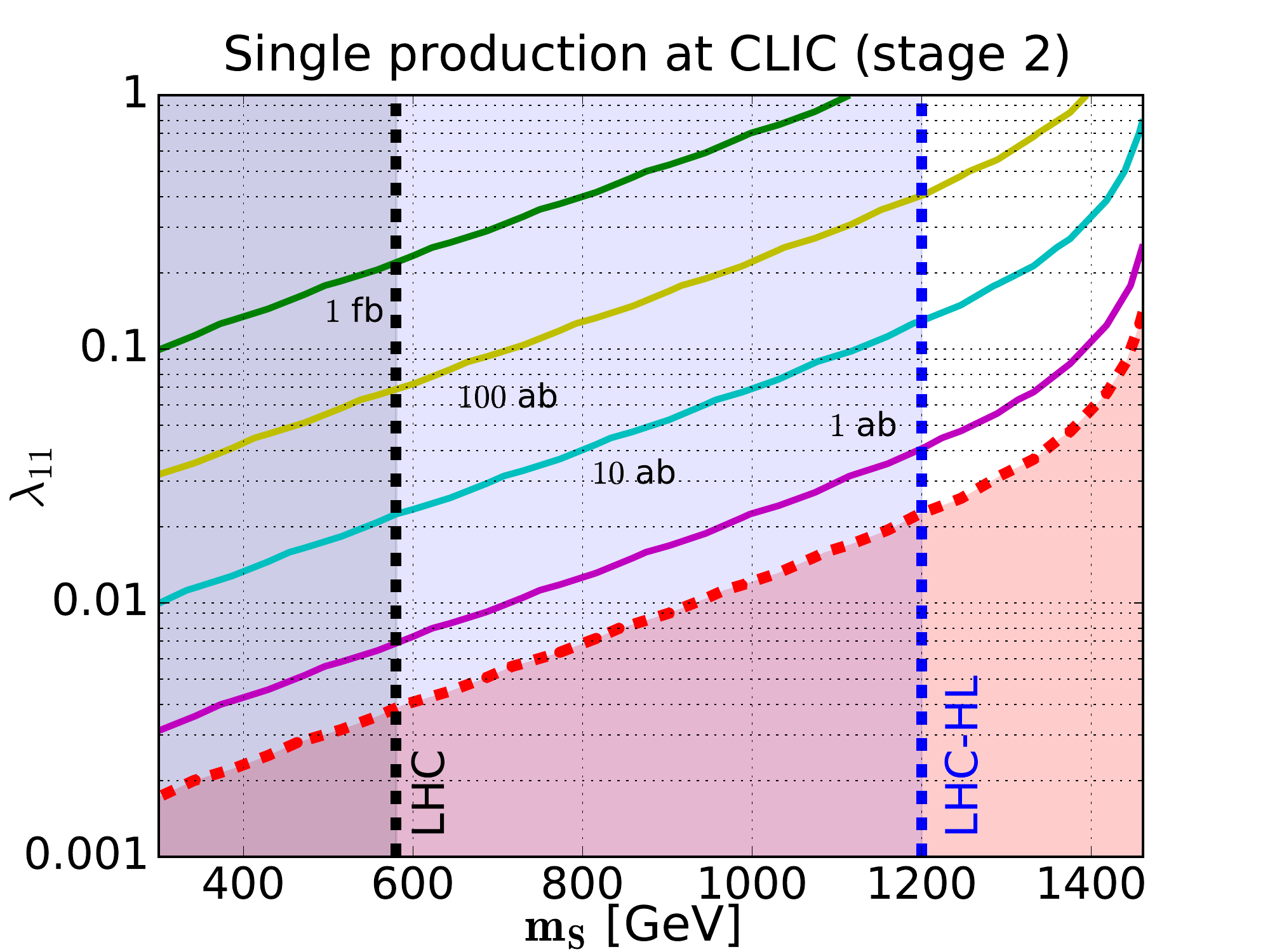}
\includegraphics[width=0.49\textwidth]{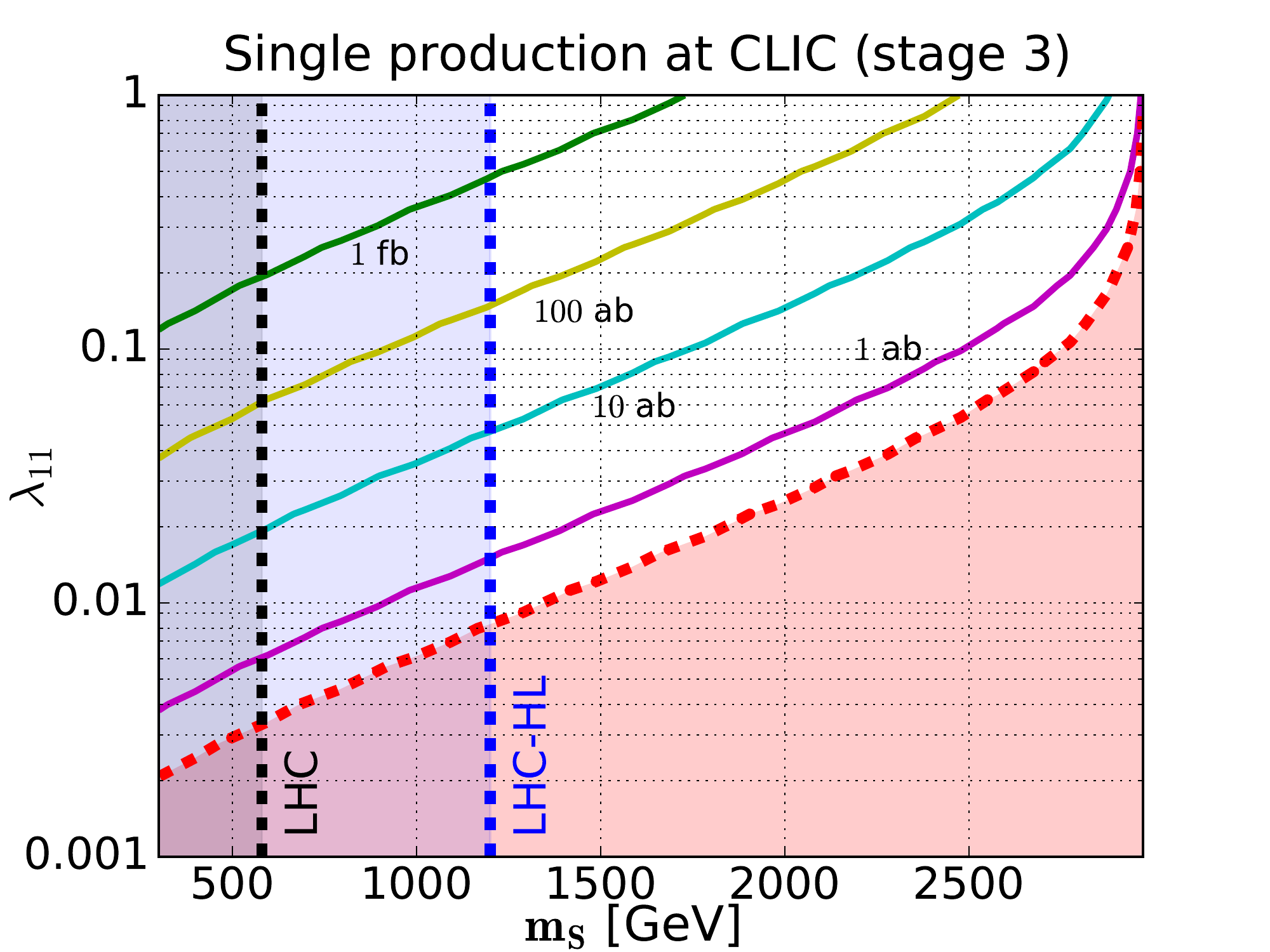}
\caption{Contours for $e^+e^-\to S^{++}e^-e^-$ cross section in the $m_s - \lambda_{11}$ plane for CLIC stage 2 (1.5 TeV) and stage 3 (3 TeV).}
\label{CS_ee-eeS_BS_1TeV}
\end{center}
\end{figure}
In Figure~\ref{CS_ee-eeS_BS_1TeV} the cross sections for the production of $2e^+2e^-$, of which at least a same-sign pair originated from the decay of a $S$, are shown as a function of $m_s$ and $\lambda_{11}$ for CLIC at 1.5 TeV and 3 TeV. In these plots, the width $\Gamma_S$ is entirely due to $\lambda_{11}$ and the electron beam is unpolarised. The red-dotted line represents the threshold for the production of a single event. The current LHC bound and the future LHC-HL bound are also shown for comparison.

\begin{figure}
\begin{center}
\includegraphics[width=0.8\textwidth]{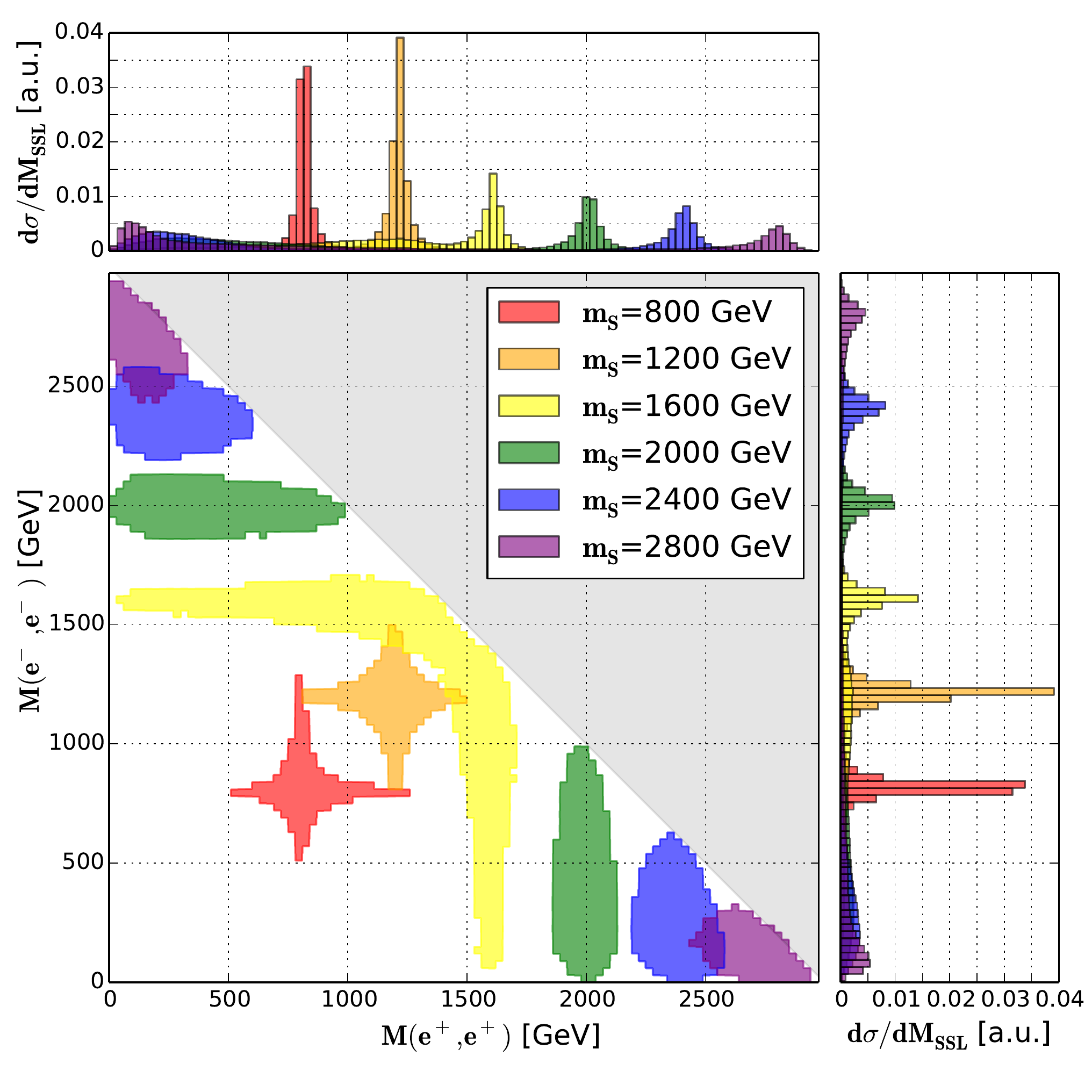}
\caption{Top and right panels: $e^+e^-\to 4e$ differential cross section plotted against the same-sign lepton invariant masses $M_{e^+e^+}$ (top) and $M_{e^-e^-}$ (right) in arbitrary units for several values of the doubly charged scalar mass $m_S$. Central panel: Corresponding contour plot for the differential cross section plotted in the $M_{e^+e^+}-M_{e^-e^-}$ plane.}
\label{single_prod_2D}
\end{center}
\end{figure}
In Figure~\ref{single_prod_2D}, the invariant mass distributions of both the electron and the positron pairs are plotted in arbitrary units for different values of the mass $m_S$, with a fixed $\Gamma_S$ corresponding to $\lambda_{11}=1$ and the other $\lambda$-couplings set to zero. The binning width has been conservatively set to $30$ GeV, corresponding to a factor of $\sim 2$ with respect to the value prescribed for $Z'$ searches~\cite{Basso:2009hf}. Notice that above the pair-production threshold, this production mode dominates and most of the production events contribute to the peak. On the contrary, below the pair-production threshold a shoulder appears beside the peak in the region of lower invariant masses. Contributions come mainly from the uncorrelated leptons associated to the lepton pair produced by the decay of the $S$, with a subleading contribution from topologies that acquire importance when the $S$ is (considerably) off-shell.

\begin{figure}
\begin{center}
\includegraphics[width=0.49\textwidth]{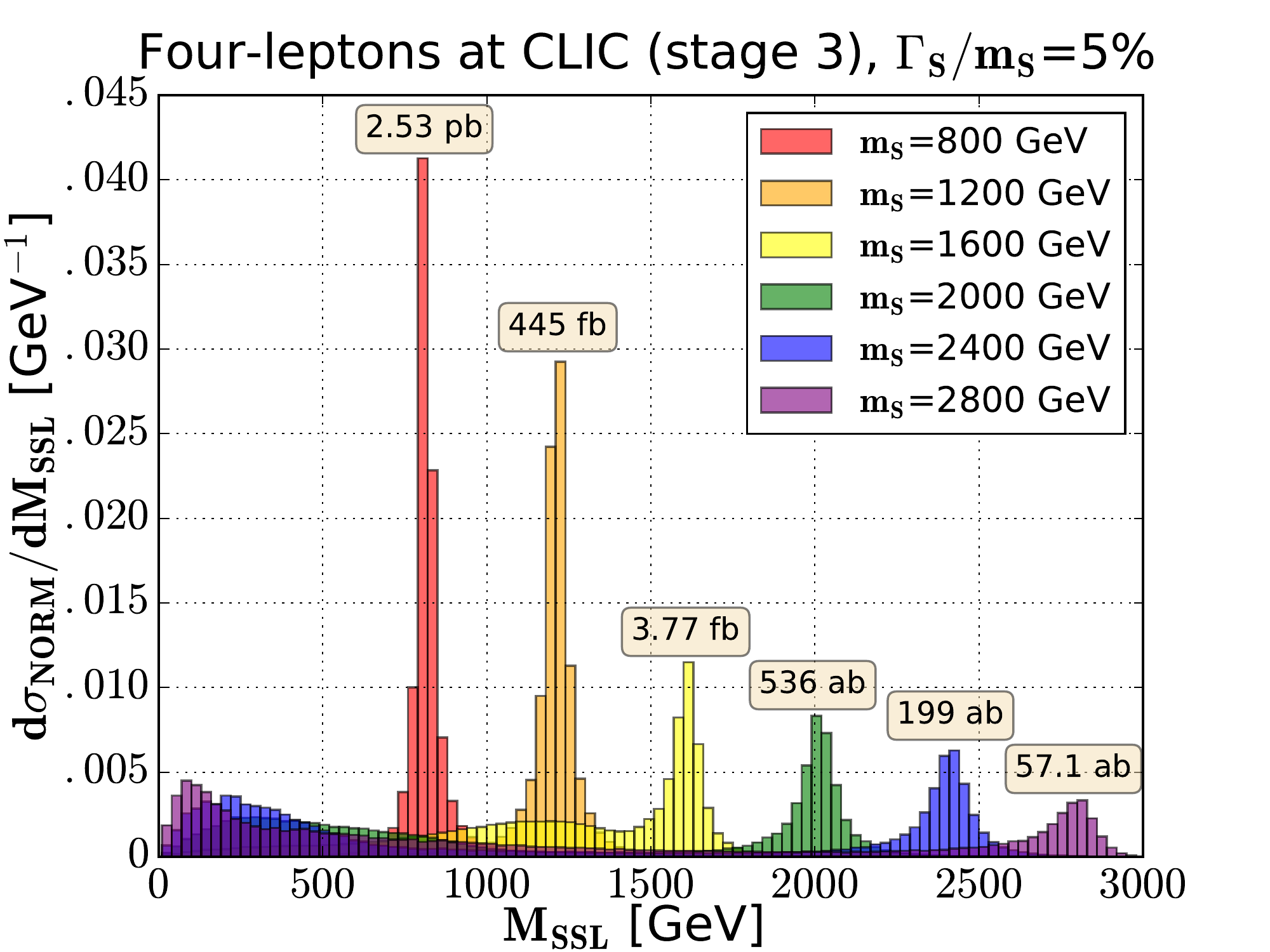}
\includegraphics[width=0.49\textwidth]{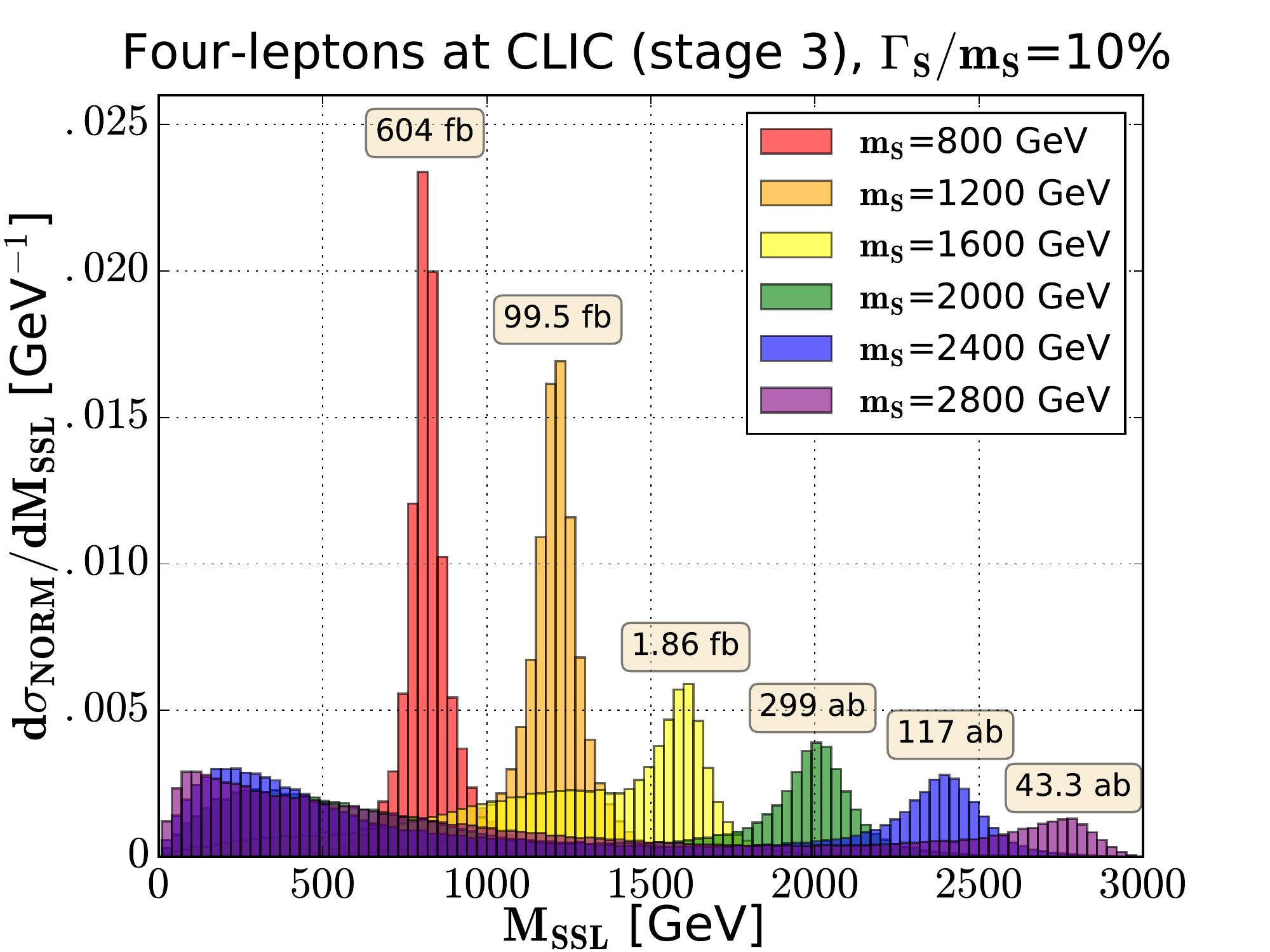}
\caption{Normalised distributions for different values of the mass of the $S$ are shown for $\lambda_{11}=1$ and $\Gamma_S/m_S=5\%,\, 10\%$  at the stage 3 of CLIC with 3 ab$^{-1}$ luminosity and unpolarised electron beam. The total cross section for each case is reported on the label of the corresponding peak.}
\label{single_prod_W05W10}
\end{center}
\end{figure}

In order to highlight the effects of a larger width, the shapes expected for different values of the $m_S$ are shown in Figure~\ref{single_prod_W05W10}, for the choice of parameters $\lambda_{11}=1$ and $\Gamma_S/m_S=5\%,\, 10\%$ at the stage 3 of CLIC with 3 ab$^{-1}$ luminosity and unpolarised electron beam. They are accompanied by the total cross sections, that can be rescaled to account for different values of $\lambda_{11}$.

\section{Coupling matrix textures and implications}
\label{sec_Pheno}
\noindent
As previously described in the Introduction section, models with a doubly charged scalar provide a natural mechanism for radiative neutrino mass generation~\cite{delAguila:2012nu,Gustafsson:2014vpa,Cai:2017jrq}. Even
without exploring the exact details, we know that this particle will produce an effective Majorana mass term
\begin{eqnarray}
m^\nu_{ab}\propto \lambda^{\phantom{l}}_{ab} \frac{m^l_a m^l_b}{\Lambda},
\label{mass-lambda}
\end{eqnarray}
where $m^l$ indicates the lepton mass, $a$ and $b$ are flavour
indices and $\Lambda$ represents some heavier UV completion scale with the ingredients that are required to trigger the Zee--Babu mechanism. Given the anarchic behaviour of the PMNS matrix~\cite{Esteban:2016qun}, we can classify two possible
scenarios related to $\lambda$-matrix patterns:
\begin{itemize}
\item pheno-inspired: the PMNS anarchic behaviour is caused by an anarchic behaviour in the $m^\nu$ mass matrix; this implies that $\lambda_{ab} \sim \left(y^l_a y^l_b\right)^{-1}$, where $y^l$ indicates the lepton SM Yukawa couplings; 
\item model-building-inspired: the PMNS anarchic behaviour is caused by a fine-tuning in the orthogonalisation of the $m^\nu$ mass matrix, but the $\lambda_{ab}$ entries shows only a mild hierarchical behaviour between diagonal and off-diagonal entries.
\end{itemize}
Furthermore, we can try to move from the neutrino-mass-generation logic and consider the hypothesis that the $\lambda$-matrix shows some alternative and nonetheless interesting behaviour. For illustrative purposes, we can adopt the following choice for the $\lambda_{ab}$:
\begin{itemize}
\item Yukawa-inspired: $\lambda_{ab}$ entries are disconnected from the
logic of the previous scenarios and reproduce a pattern that mimic the Yukawa matrix.
\end{itemize}
In what follows we will investigate these scenarios. We consider the impact of current (as listed in Table~\ref{tab:exp}) and future limits of low-energy experiments (for illustrative purposes we use ${\rm{BR}}\left[{{\mu^\mp} \to {e^\mp}{e^\pm}{e^\mp}} \right] \leq 1.0 \times10^{-16}$ and ${\rm BR}_{\mu \to e}^{\rm Al} \leq 1.0 \times
10^{-16}$, \emph{i.e.} the limit that will be reached in the ultimate phase of the Mu3e experiment~\cite{Berger:2014vba,Perrevoort:2018cqi} and the limit expected by the future experiments probing muon conversion in nuclei~\cite{Carey:2008zz,Cui:2009zz,Kutschke:2011ux}, respectively) and confront them with the limits that can be obtained from a future $e^+ e^-$ collider.

\subsection*{Pheno-inspired scenario}

Taking $\lambda_{ab} \sim \left(y^l_a y^l_b\right)^{-1}$ as input, the
matrix $\lambda_{ab}$ parametrically takes the form
\begin{align}
\label{lambda-pheno}
\lambda_{ab} = \lambda \left(
\begin{array}{ccc}
   \pm 1 & \nu^{2} & \nu^{3} \\
   \nu^{2} & \nu^{4} & \nu^{5} \\
   \nu^{3} & \nu^{5} & \nu^{6}
   \end{array}
   \right)
\end{align}
with $\nu \sim \mathcal{O}(m^\mu/m^\tau)$. The couplings of the $S$ to the lightest families are the largest. At the same time, processes involving these couplings also have the strongest experimental constraints. As a result, processes involving $\tau$ leptons play virtually no role in constraining the model in this scenario.

\begin{figure}
\begin{center}
\includegraphics[width=0.49\textwidth]{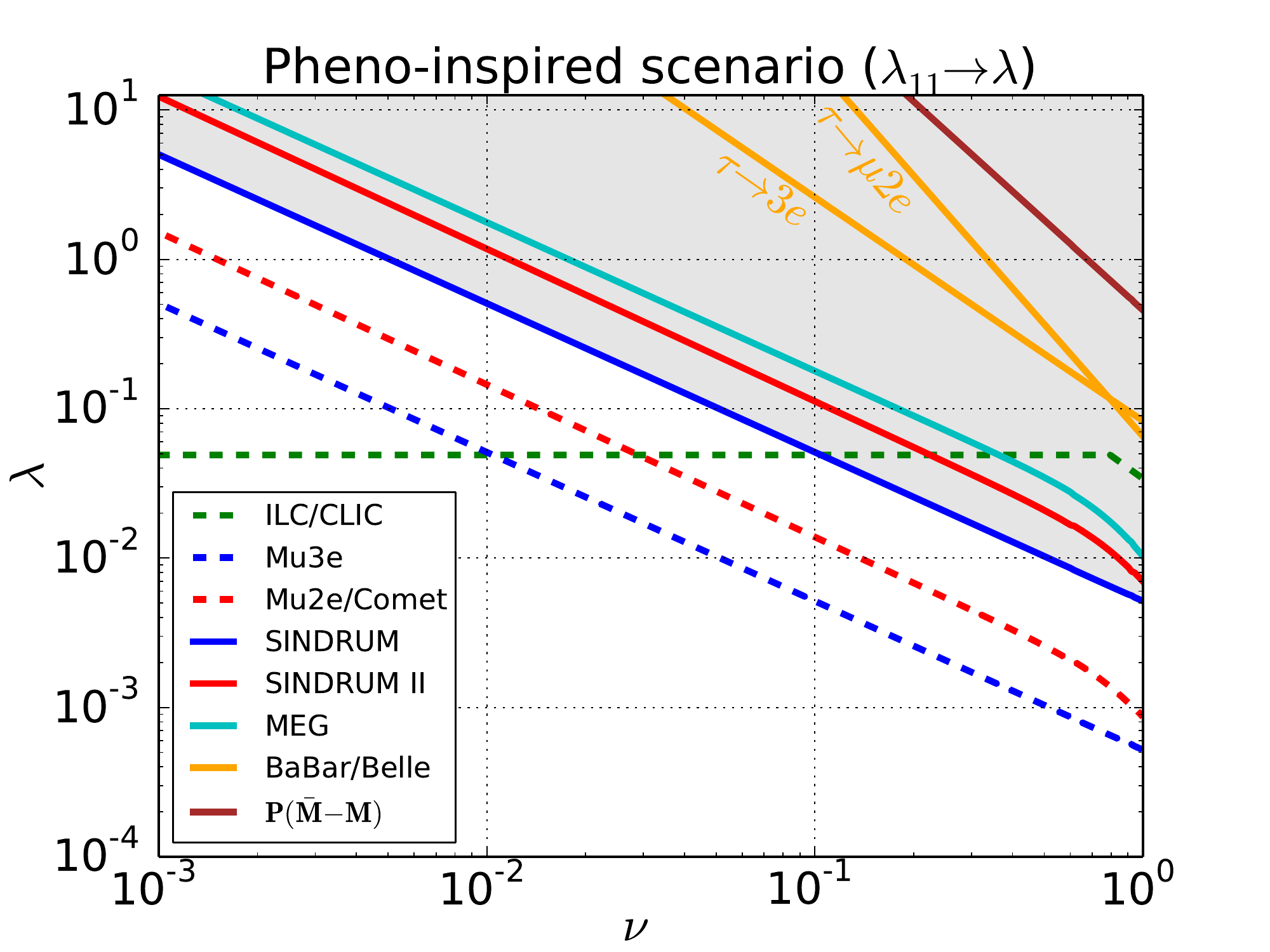}
\includegraphics[width=0.49\textwidth]{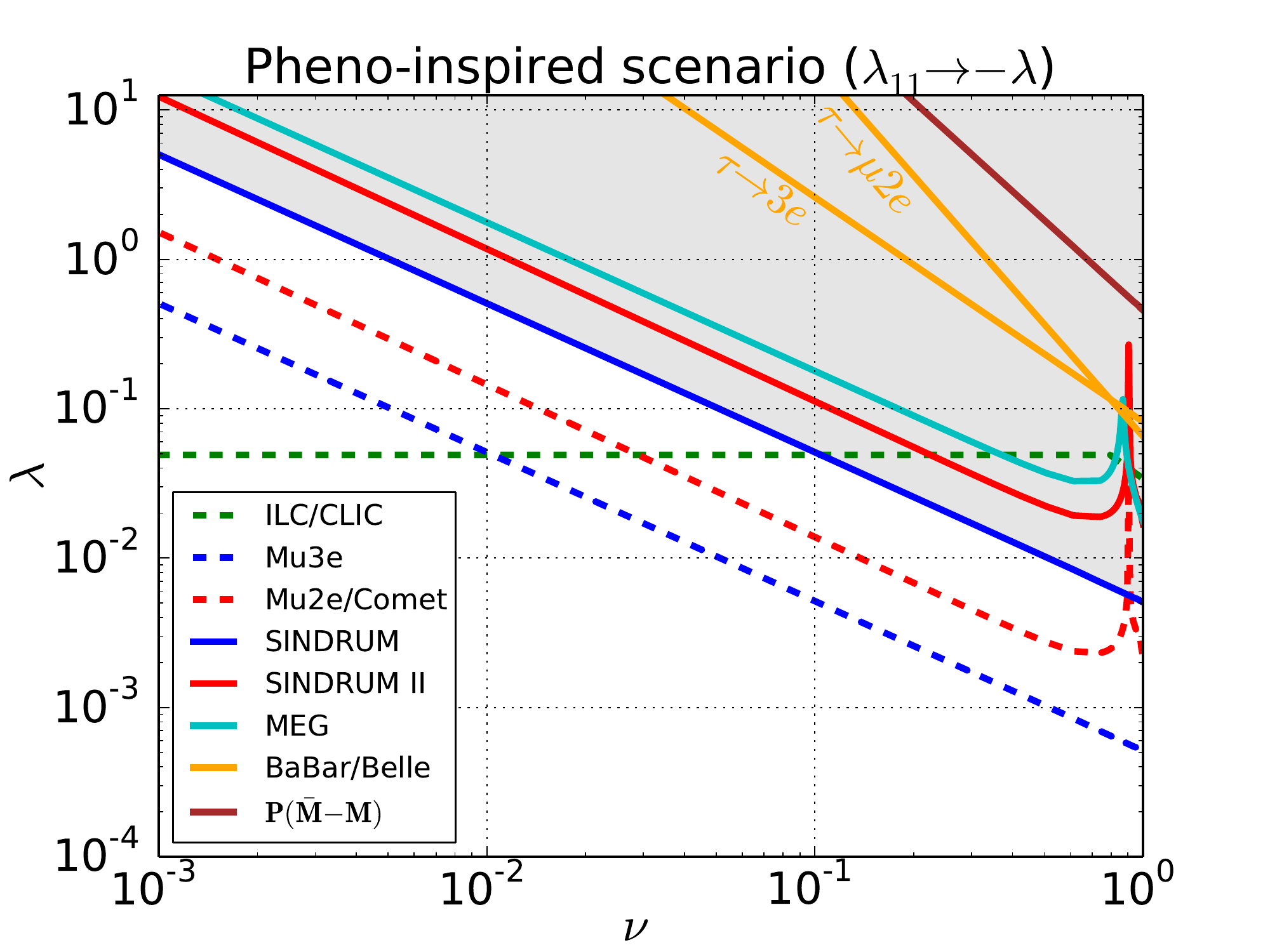}
\caption{Limits assuming couplings given in \eqref{lambda-pheno} (left) and changing $\lambda_{11} \to
-\lambda$ (right) for $m_S=1$ TeV.}
\label{pheno1}
\end{center}
\end{figure}

In order to illustrate this, we make the simplifying assumption that the coupling matrix of the $S$ takes precisely the form given in \eqref{lambda-pheno}. We choose a fixed mass $m_S = 1$~TeV and compare the limits from various processes in the $\lambda$-$\nu$ plane. The results are shown in Figure~\ref{pheno1}, where the region on the top-right of the various curves is excluded. Not surprisingly, the strongest limits are due to the SINDRUM result for $\mu\to 3 e$
(solid-blue line). The MEG limit (light-blue line) and muon conversion in gold (solid-red line) result in somewhat weaker limits. Future improvements to $\mu$-$e$ conversion (dashed-red line) and $\mu\to 3 e$ (dashed-blue line) will have a large impact. On the other hand, the best limits involving $\tau$ leptons are from the processes $\tau\to 3
e$ and $\tau\to \mu (e^+ e^-)$ (shown as orange lines) and are considerably weaker. The same is true for $\overbar{M}$-$M$ oscillation (brown line).  For comparison the limits obtainable at the latest stage of both the ILC and the CLIC are also depicted (green-dashed line). For very small values of $\nu$ the limit on $\lambda_{11}$ is competitive. The reason is that at the ILC/CLIC a limit on $\lambda_{11}$ can be obtained even if all other couplings tend to zero. 

Let us stress that we do not allege that the strict equality in \eqref{lambda-pheno} is a realistic scenario. Typically it is expected that the values of the couplings vary. We just use \eqref{lambda-pheno} to facilitate the presentation of the salient features of the constraints obtained from low-energy and future $e^+ e^-$ collider experiments. In fact, in the right panel of Figure~\ref{pheno1} we show the limit using Eq~(\ref{lambda-pheno}) but changing the sign of $\lambda_{11}$. This sign change induces cancellations in the matching of the dipole operators and can have strong effects in certain regions of parameter space. Thus the low-energy limits presented in this section have to be taken as generic indications and do not replace a proper check of the validity of a certain point in the parameter space.

\subsection*{Model-building-inspired scenario}

Let us turn now to the model-building scenario, where it is assumed that all couplings are of the same order, possibly with a small hierarchy between diagonal and off-diagonal elements. Again, we fix $m_S = 1$~TeV and use the drastically simplified version for the couplings
\begin{align}
\lambda_{ab} = \lambda \left(
\begin{array}{ccc}
   \pm 1 & \nu & \nu^{2} \\
   \nu & 1 & \nu \\
   \nu^{2} & \nu & 1
   \end{array}
   \right)
   \label{lambda-model}
\end{align}
Note that original motivation would suggest $\nu < 1$, but we also
consider $\nu > 1$.

\begin{figure}
\begin{center}
\includegraphics[width=0.49\textwidth]{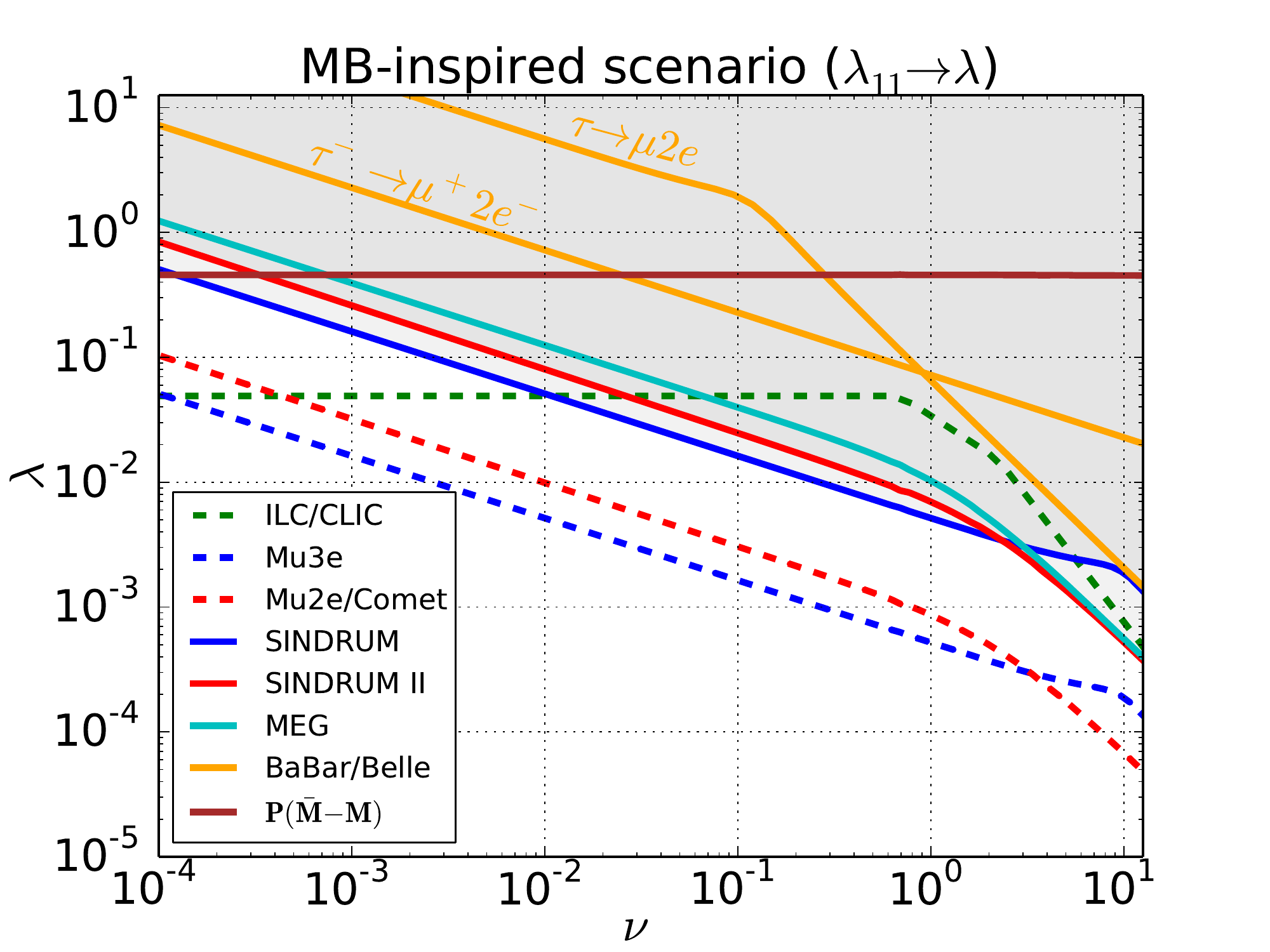}
\includegraphics[width=0.49\textwidth]{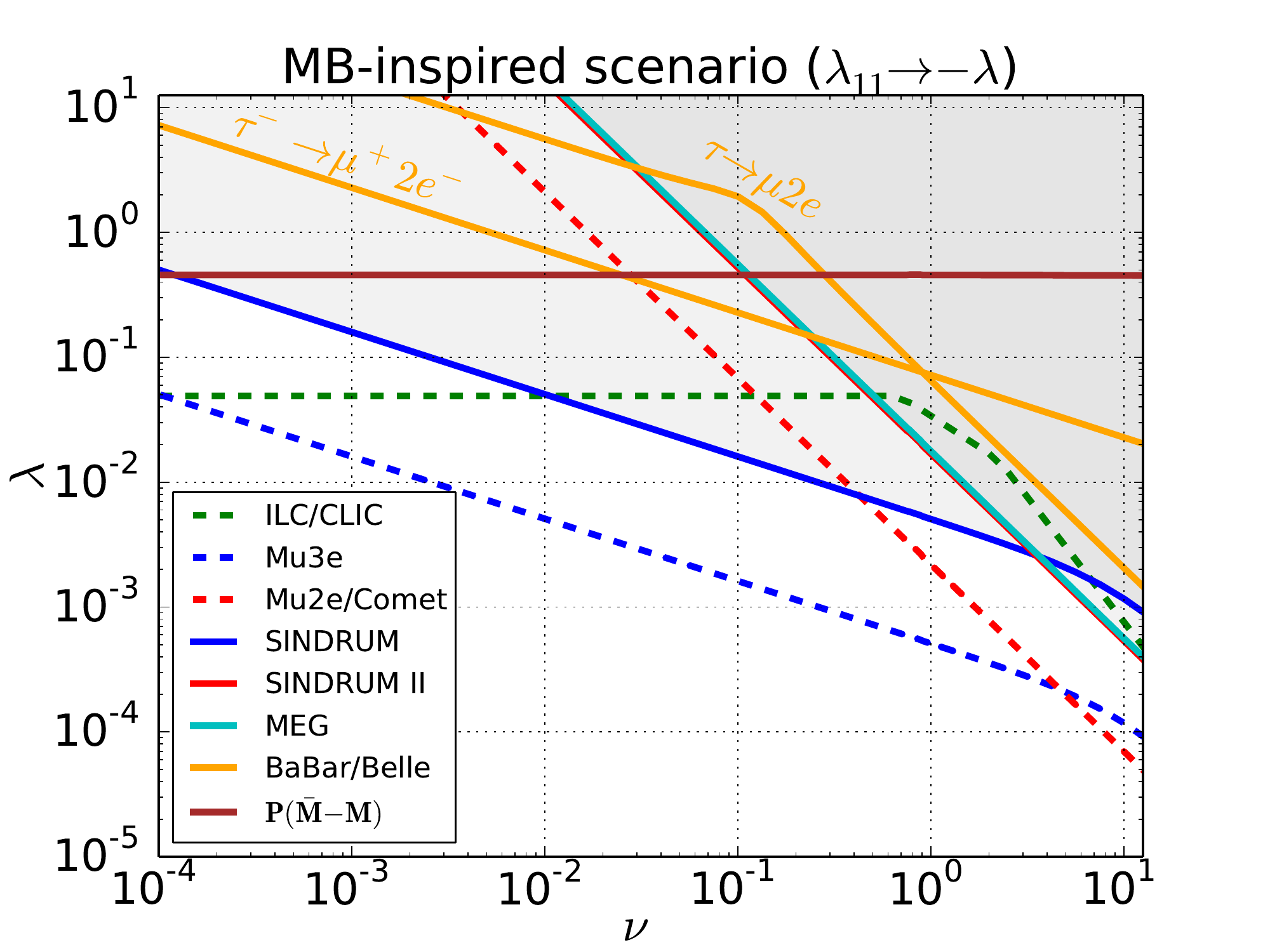}
\caption{Limits from various processes, as Figure~\ref{pheno1}, but
using \eqref{lambda-model} with the plus sign (left panel) and
minus sign (right panel) for $m_S=1$ TeV.}
\label{model1}
\end{center}
\end{figure}

If all couplings are of the same order, generally speaking it is still
the case that low-energy processes involving $\tau$ leptons are less
constraining. However, as can be seen in Figure~\ref{model1},
processes like $\tau\to \mu (e^+ e^-)$ start to serve as a useful
cross check. The kink in the limit for $\tau\to \mu (e^+ e^-)$ is due
to RGE effects. For $\nu \gtrsim 0.1$ the branching ratio for
$\tau\to \mu (e^+ e^-)$ is dominated by the single operator that is
present at the EWSB scale, $C_{VRR}^{3211}$. For smaller values
of $\nu$ the operators induced by the RGE become numerically important
though and substantially modify the limits.

For small values of $\nu$ (\emph{i.e.} $\lambda_{ab}$ approaching a diagonal
matrix) $\overbar{M}$-$M$ oscillation becomes increasingly competitive. But
a substantial improvement of the experimental bound would be required
to be competitive with limits from ILC/CLIC. There are two kinks in the ILC/CLIC limit around $\nu=1$. In the first horizontal part the bound comes from $\lambda_{11}$ whereas after the first and second kink the bound originates from $\lambda_{12}$ and  $\lambda_{13}$, respectively.

Once more, the limits depend on the precise values of the
couplings. To illustrate this, in Figure~\ref{model1} we compare the
limits due to the most important processes using
\eqref{lambda-model} with the plus sign (left panel) and the minus
sign (right panel). The strongest effect of the sign change is in $\mu$-$e$
conversion and $\mu\to e \gamma$, again due to cancellations in the
Wilson coefficient of the dipole operator.

\subsection*{Yukawa-inspired scenario}

In the examples considered so far, processes with $\tau$ leptons played a minor part, since their experimental bounds are
weaker. However, if we consider a scenario where couplings to the first (and second) generation are suppressed, these processes will be much more important. In this spirit we consider couplings that follow a  pattern similar to the Yukawa couplings, and write
\begin{align}
\lambda_{ab} = \lambda \left(
\begin{array}{ccc}
   \nu^{2d} & \nu^{d+1} & \nu^{d} \\
   \nu^{d+1} & \nu^{2} & \nu \\
   \nu^{d} & \nu & 1
   \end{array}
   \right)
   \mbox{ with }
   d \in \{2, 4 \}
   \label{lambda-box}
\end{align}
assuming $\nu < 1$.

\begin{figure}
\begin{center}
\includegraphics[width=0.49\textwidth]{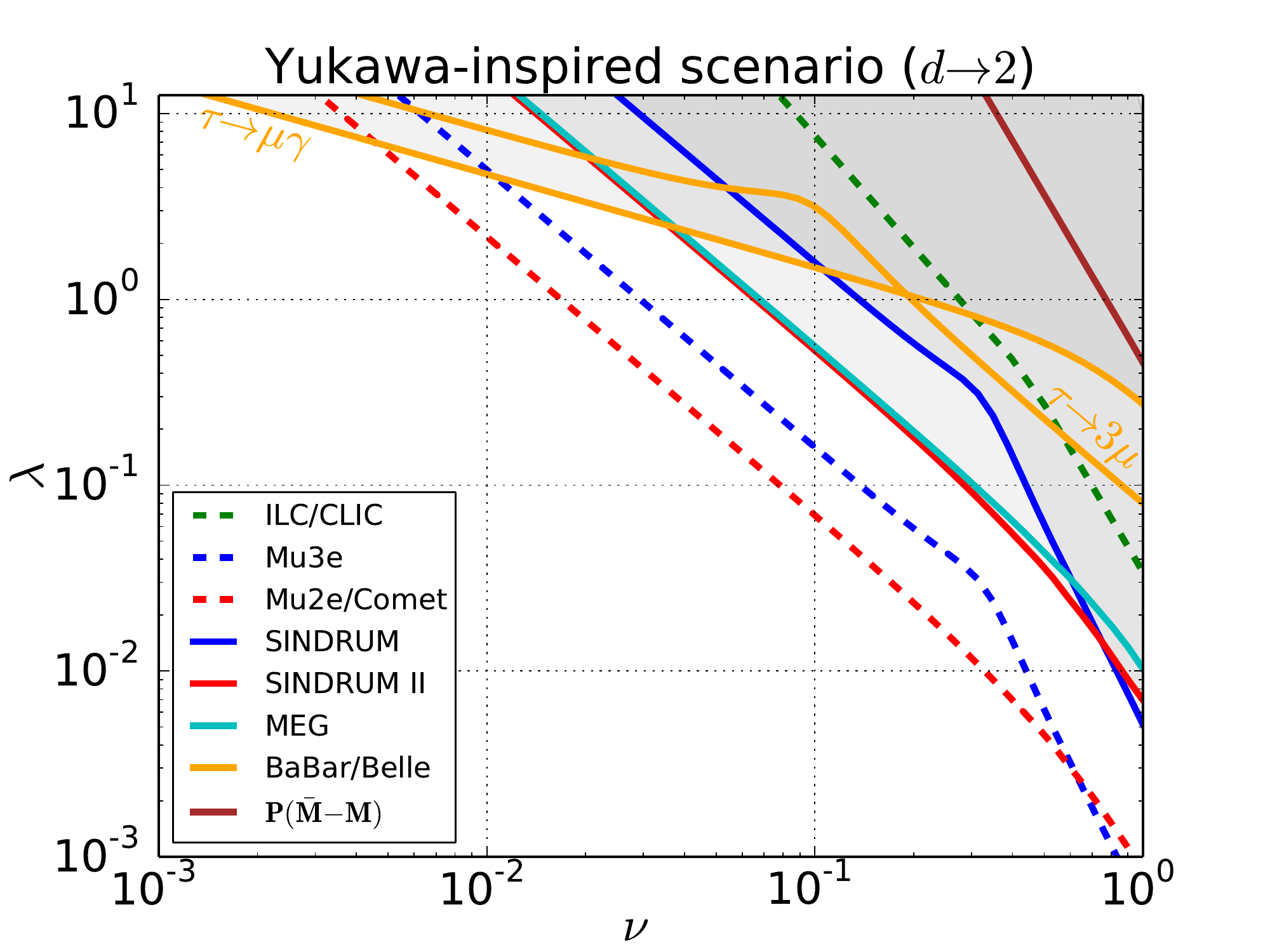}
\includegraphics[width=0.49\textwidth]{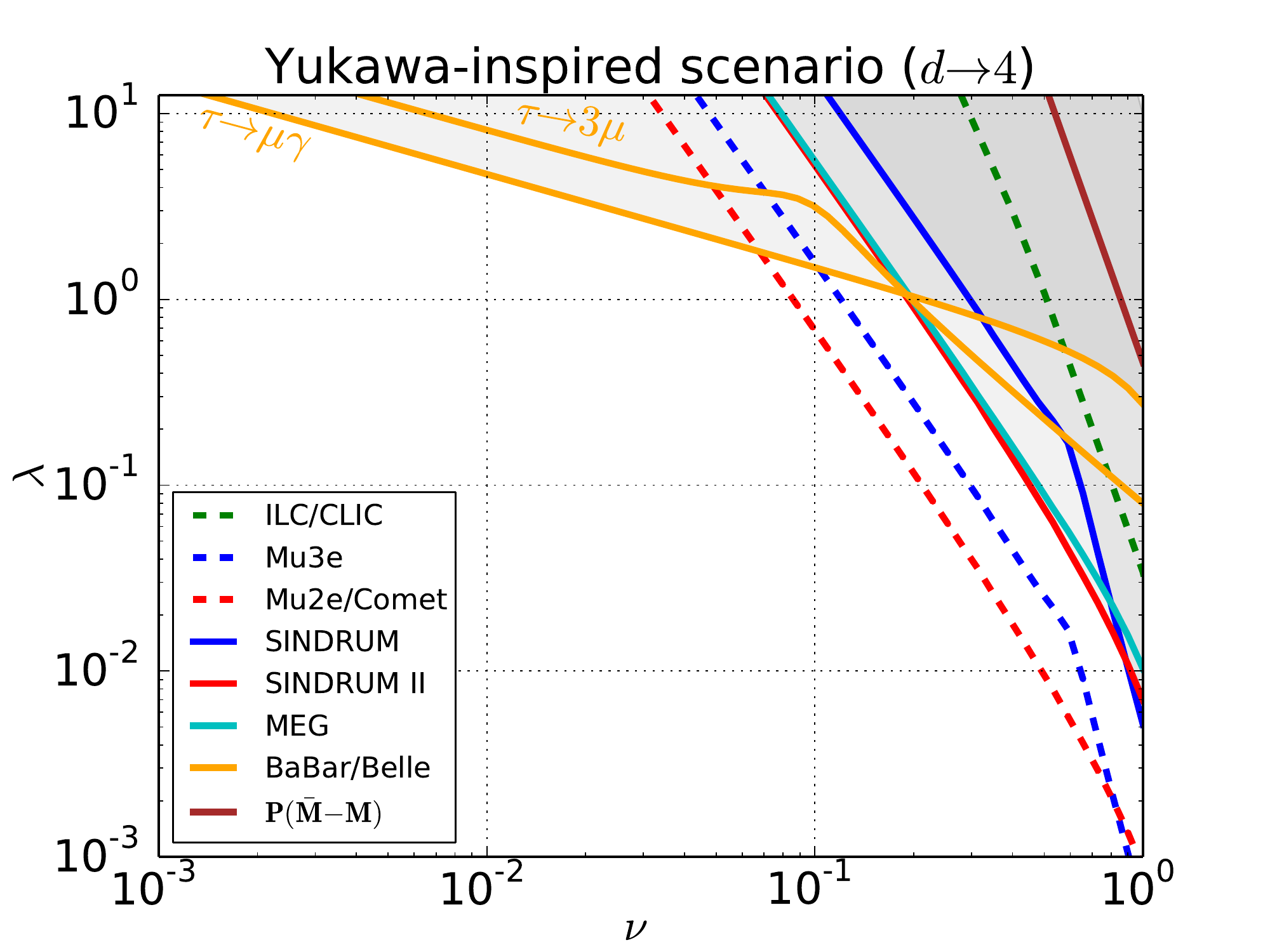}
\caption{Limits from various processes, as Figure~\ref{pheno1}, but
using \eqref{lambda-box} with $d=2$ (left panel) and
$d=4$ (right panel) for $m_S=1$ TeV.}
\label{box1}
\end{center}
\end{figure}

As shown in Figure~\ref{box1}, for small values of $\nu$ the
processes $\tau\to\mu\gamma$ and $\tau\to\mu (e^+ e^-)$ become even
more competitive, in particular for $d=4$. But except for very small
values of $\nu$, the stringent limits on $\mu\to e \gamma$ and in
particular future limits on $\mu\to 3 e$ and $\mu$-$e$ conversion keep
playing a decisive role. An extreme hierarchy is required to
compensate for the weaker experimental bounds. Thus, charged LFV processes
with only muons and electrons keep playing a crucial role, even if
third-generation couplings are strongly enhanced.

\subsection*{General remarks}

In the three scenarios considered above only a small number of processes enter. However, allowing the couplings to take arbitrary values, for virtually any of the processes listed in
Table~\ref{tab:exp}, it is possible to find a corner in parameter space
where it provides the dominant constraint. It is thus
imperative not to focus the experimental activity on a few observables,
but to take all of them into consideration. We also stress it is
virtually impossible to make statements that are generally valid concerning the
allowed region of a single coupling since all points in the six-dimensional parameter space of $\lambda_{ab}$ need to be
considered independently.


\section{Conclusions}
\label{sec_4}
\noindent
Doubly charged scalars appear in popular extensions of the SM, mostly motivated by left-right symmetry, neutrino mass generation, non-minimal EWSB mechanisms and grand-unified theories.

In this article we have investigated the phenomenology of a $SU(2)$-singlet doubly charged scalar. We considered the impact of low-energy precision experiments and current searches at the LHC on the constraints on its mass and couplings. Moreover, we studied the scope of future low- and high-energy experiments to probe the surviving parameter space for such particles in the light of specific benchmark scenarios.

The new particle violates explicitly both lepton flavour and lepton number, thus triggering low-energy processes that are not allowed in the SM and therefore very constrained by experimental searches. In this paper we have analysed the impact of the doubly charged scalar on LFV observables by means of a systematic dimension-six EFT approach. Then, we have interpreted the experimental limits as bounds on the parameter space of the effective coefficients and converted them into constraints at higher energies by means of one-loop RGE corrections. This approach is crucial to describe correctly the experimental limits from $\mu$-$e$ conversion in terms of bounds on mass and couplings.

At the LHC, besides the main partonic channel $q\bar{q}\rightarrow \gamma^*(Z^*)\rightarrow S^{++}S^{--}$ we have also included corrections due to photon-initiated processes, and reinterpreted the current experimental limits, also considering large width effects. Even though the precise value of these limits depends on the assumptions, the effects of a large width on the bounds are found to be mild: they decrease the NWA mass limit of $\sim 50$ GeV at most, for both current and future integrated luminosities.  The limits will improve with the HL run from the current limit of $\sim 500$ GeV to the ultimate HL limit of $\sim 1200$ GeV, unless a discovery is made.

Future searches at $e^+e^-$ colliders are also very promising. Specific signatures can be investigated both at the ILC and CLIC. Lepton scattering with a doubly charged scalar exchanged in the $t$-channel can be studied in order to explore much higher scales than the centre-of-mass energy of the collider. For couplings $\lambda\sim 0.1$, the discovery potential reaches up to masses of to several TeV. Such a range can be extended by one order of magnitude for couplings $\lambda \sim 1$, making the linear collider the only available option to single out the contribution of specific couplings with a reach of $\mathcal{O}(10)$ TeV. Furthermore, LC machines can display a unique power in determining the line shapes in case of resonant production, especially in presence of large width effects. For values of the $\lambda$-couplings close to the unit, cross sections above $\sim 10$ ab and $\sim 1$ fb can be reached for $m_S\sim 1.5$ TeV by the CLIC at stage II and stage III, respectively.

In order to obtain comprehensive constraints on the full coupling matrix, a combined approach involving as many observables as possible is the only possible option. Specific setups should be analysed case by case. We have considered several $\lambda$-matrix textures inspired by various theoretical approaches. We have shown explicitly that each experimental observable was found to be the most relevant in some specific portion of the parameter space. Consequently, no observable can be discarded from the analysis without losing crucial information on some region of the parameter space.

In conclusion, we stress the importance of the complementarity in low- and high-energy searches for doubly charged scalars, especially in the light of the promising future experimental plans for linear collider facilities and high-intensity experiments.

\acknowledgments
\noindent
The authors are grateful to A. Akeroyd, D. Barducci, M. Hoferichter, U. Langenegger, A. Pukhov and M. Zaro  for their useful comments during the development of this paper and to T. Geib and A. Merle for their important contributions in the initial stage of the project. LP acknowledges the use of the IRIDIS HPC Facility at the University of Southampton. The work of AC is supported by an Ambizione Grant No. PZ00P2\_154834 of the Swiss National Science Foundation. MG and GMP are partially supported by the SNSF under contract No. 200021\_160156. 

\bibliography{BIB}

\end{document}